\documentclass[%
 reprint,
superscriptaddress,
 amsmath,amssymb,
 aps,
floatfix,
]{revtex4-2}

\usepackage{graphicx}
\usepackage{dcolumn}
\usepackage{bm}
\usepackage{float}
\usepackage{subcaption}
\usepackage{placeins}

\usepackage[document]{ragged2e}
\usepackage{caption}
\usepackage{amssymb}
\usepackage{color}
\DeclareUnicodeCharacter{2212}{-}
\usepackage{textgreek}
\DeclareCaptionJustification{justified}{\justifying}
\makeatletter
\def\blx@err@patch#1{}
\makeatother
\usepackage{url}


\begin{document}


\title{Elastic and structural anisotropy in silica thin films for gravitational-wave detectors}

\author{Brenda Bracco}
\thanks{These authors contributed equally to this work.}
\affiliation{Dipartimento di Chimica, Biologia e Biotecnologie, Universit\`a di Perugia, 06100 Perugia, Italy}
\affiliation{Istituto Nazionale di Fisica Nucleare, Sezione di Perugia, 06100 Perugia, Italy}

\author{Michele Magnozzi}
\affiliation{OptMatLab, Dipartimento di Fisica, Universit\`a di Genova, 16146 Genova, Italy}
\affiliation{Istituto Nazionale di Fisica Nucleare, Sezione di Genova, 16146 Genova, Italy}

\author{Stefano Colace}
\author{Maurizio Canepa}
\affiliation{OptMatLab, Dipartimento di Fisica, Universit\`a di Genova, 16146 Genova, Italy}

\author{Giulio Favaro}
\affiliation{CNR - Istituto di Fotonica e Nanotecnologie (IFN), 35131 Padova, Italy}
\affiliation{Dipartimento di Fisica e Astronomia, Universit\`a di Padova, 35131 Padova, Italy}

\author{Marco Bazzan}
\affiliation{Dipartimento di Fisica e Astronomia, Universit\`a di Padova, 35131 Padova, Italy}

\author{Massimo Granata}
\author{David Hofman}
\affiliation{Laboratoire des Mat\'eriaux Avanc\'es - IP2I, CNRS, Universit\'e de Lyon, Universit\'e Claude Bernard Lyon 1, 69100 Villeurbanne, France}

\author{Alessandro Di Michele}
\affiliation{Dipartimento di Fisica e Geologia, Universit\`a di Perugia, 06100 Perugia, Italy}
\affiliation{Istituto Nazionale di Fisica Nucleare, Sezione di Perugia, 06100 Perugia, Italy}

\author{Laura Silenzi}
\affiliation{School of Science and Technology - Physics Division, Università di Camerino, 62032 Camerino, Italy}
\affiliation{Istituto Nazionale di Fisica Nucleare, Sezione di Perugia, 06100 Perugia, Italy}

\author{Gianpietro Cagnoli}
\affiliation{Universit\'e de Lyon, Universit\'e Claude Bernard Lyon 1, CNRS, Institut Lumière Mati\`ere, 69100 Villeurbanne, France}

\author{Giovanni Carlotti}
\affiliation{Dipartimento di Fisica e Geologia, Universit\`a di Perugia, 06100 Perugia, Italy}

\author{Paola Sassi}
\email{paola.sassi@unipg.it}
\affiliation{Dipartimento di Chimica, Biologia e Biotecnologie, Universit\`a di Perugia, 06100 Perugia, Italy}
\affiliation{Istituto Nazionale di Fisica Nucleare, Sezione di Perugia, 06100 Perugia, Italy}

\author{Silvia Corezzi $^{*,}$}
\email{silvia.corezzi@unipg.it}
\affiliation{Dipartimento di Fisica e Geologia, Universit\`a di Perugia, 06100 Perugia, Italy}
\affiliation{Istituto Nazionale di Fisica Nucleare, Sezione di Perugia, 06100 Perugia, Italy}


\begin{abstract}
	\justify
The thermal noise of mirror coatings for gravitational-wave detectors critically depends on the elastic properties of the constituent materials. Data analyses and theoretical models typically assume each material is homogeneous and isotropic, but isotropy has never been explicitly verified. Using Brillouin light scattering (BLS), we demonstrate for the first time that ion-beam-sputtered SiO$_{2}$ --- a material still viable for future mirror coatings --- exhibits cylindrical elastic symmetry, with in-plane isotropy but a notable $6\%$ compressive anisotropy along the film normal. This anisotropy remains unchanged after the post-deposition heat treatment currently used in ground-based detectors (500 °C, 10 h) but is nearly eliminated at 900 °C. Infrared reflectivity experiments support these findings by directly revealing heterogeneities in the distribution of bridging and non-bridging oxygen structures along the growth axis. While BLS measures the real part of the elastic constants at GHz frequencies, the data reveal negligible contributions from mechanical relaxations in the kHz--GHz range, making BLS a valid substitute for low-frequency properties obtained from standard anisotropy-insensitive techniques. Our results highlight that restoring isotropy through heat treatment --- by softening the material, enabling more than $7\%$ out-of-plane expansion, and smoothing out structural heterogeneities --- may play a key role in reducing thermal noise. This proof-of-concept study extends beyond silica, providing critical insights for the design of future coatings.
\end{abstract}

\maketitle


\justify

\section{\label{Intro}Introduction}

Gravitational wave (GW) detectors are precision instruments designed to measure minute displacements caused by astrophysical signals. The current ground-based detectors are kilometer-scale laser interferometers with Fabry-Pérot cavities, where mirrors are made of high-purity substrates coated with high-reflectivity Bragg reflectors, which consist of multilayer thin-film stacks made of alternating high- and low-refractive-index materials \cite{Aasi_CQG2015, Acernese_CQG2015, Akutsu_PTEP2020}. Surface fluctuations of the mirrors, arising from thermally activated relaxations in the coatings, are a major limitation to detector sensitivity. Therefore, to enhance the astrophysical reach of future detectors, it is crucial to reduce this thermal noise by minimizing elastic energy dissipation in the constituent materials.

The coatings of aLIGO (Advanced Laser Interferometer Gravitational-Wave Observatory) and AdV (Advanced Virgo) detectors, produced by LMA (Laboratoire des Mat\'eriaux Avanc\'es) \cite{LMA} via ion beam sputtering (IBS), consist of alternating layers of amorphous SiO$_{2}$ (low $n$) and TiO$_{2}$:Ta$_{2}$O$_{5}$ (high $n$) \cite{PinardAO2017}. Several strategies are being explored to reduce coating thermal noise, including variations in deposition conditions \cite{GranataAO2020, DuranteOptMat2024}, use of nanolayering \cite{YangACSANM2020}, optimization of post-deposition treatments \cite{GranataPRM2018, AmatoJVSTB2019, ColaceCQG2024}, and, importantly, the search for less dissipative high-refractive-index materials to replace TiO$_{2}$:Ta$_{2}$O$_{5}$. Promising candidates include silicon nitrides (SiN$_{\mathrm{x}}$) \cite{GranataAO2020, AmatoPRD2025} and amorphous oxide mixtures (TiO$_{2}$:GeO$_{2}$ \cite{VajentePRL2021} or TiO$_{2}$:SiO$_{2}$ \cite{McGheePRL2023}). Meanwhile, SiO$_{2}$ remains the amorphous material with the lowest known room-temperature elastic losses in the acoustic frequency range, and as such, it continues to be the unmatched low-refractive-index candidate for room temperature applications.

The elastic properties of the materials composing the coating are key factors in determining the coating thermal noise. It is important to note that, up to now, all simulations used to derive elastic properties from experimental data \cite{GranataCQG2020, GranataPRD2016, MalhaireJVSTA2023} and all theoretical models developed to predict the spectral density of thermal noise incorporate varying levels of simplification \cite{HongPRD2013, YamPRD2015, TaitPRL2020, FejerLIGO2021, VajentePRL2021, LovelaceCQG2018}. Specifically, despite deposition via IBS introduces stress \cite{MalhaireJVSTA2023}, they all rely on the assumption that each individual material can be treated as elastic isotropic and homogeneous. In reality, the hypotheses of isotropy and homogeneity have never been verified, and the measured thermal noise of coatings still exceeds theoretical expectations \cite{GranataCQG2020, GranataPRD2016}. This highlights the need for a more detailed and accurate representation of the behavior of ion-beam sputtered materials, going beyond the current literature studies.

In this work, we investigate anisotropic signatures in ion-beam sputtered SiO$_{2}$ thin films produced by LMA, a material still viable for future mirror coatings. Brillouin light scattering (BLS) and infrared (IR) reflectivity experiments are used to probe the elastic properties at the macroscopic scale and examine structural characteristics ---such as defects and inhomogeneities--- at the atomic scale, while monitoring their sensitivity to thermal exposure. BLS spectroscopy is the most powerful tool for accessing the elastic constants of thin films at GHz frequencies, though it has not been previously applied to characterize materials for GW detectors. Meanwhile, IR spectroscopy remains underexploited in thin film applications. These surface-sensitive techniques, combined with supplementary measurements, demonstrate considerable potential for improved characterization of materials for next-generation high-performance GW mirror coatings.

\section{\label{Methods}Materials and Methods}

         \subsection{Samples}

IBS SiO$_{2}$ films were deposited on single-side-polished (100)-p-doped Si wafers ($\varnothing=3''$, thickness=0.5 mm) at LMA-IP2I (\url{http://lma.in2p3.fr/}) using a commercial Veeco Spector system with accelerated Ar$^{+}$ ions as sputtering particles. Before deposition, the base pressure in the vacuum chamber was reduced to below 10$^{-5}$ mbar. During the coating process, the total pressure was maintained at approximately 10$^{-4}$ mbar, with 18 sccm of Ar supplied to the ion source and 15 sccm of O$_{2}$ injected inside the vaccum chamber. The energy and current of the sputtering ion beam were set to 1.25 keV and 0.6 A, respectively. SiO$_{2}$ films were deposited with thicknesses of approximately 720 nm (runs S19175 and S21088) and 2.5 $\mu$m (run S21089). Each coated wafer was laser-engraved and cleaved into several 1 cm $\times$ 2 cm rectangular pieces, to study both the as-deposited material and the material after heat treatments at various temperatures. All annealing cycles were performed in air, with the temperature ramped up at 100 °C per hour, held at the target value for 10 hours, and then ramped down at 100 °C per hour. The target annealing temperatures included 500 °C and 900 °C. 

       \subsection{\label{Char}Characterizations}
       
The mass density ($\rho$) of the films was determined using x-ray reflectivity (XRR) measurements performed with a Philips MRD diffractometer equipped with a Cu tube operating at 40 kV and 40 mA. The instrument was monochromatized to the Cu K-$\alpha$ line ($\lambda$=1.5406 \AA), and the reflected x-rays were detected with a Xe counter detector whose angular acceptance was defined by a Soller slit, a parallel plate collimator, and an additional 0.01 rad slit. Data analysis was carried out with the REFLEX software \cite{VignaudJApplCry2019}, modeling the sample as a silicon substrate with a thin SiO$_{2}$ layer on top. Figure~\ref{fig:XRR} in Appendix~\ref{AppA} presents the spectra of the sample S21088 in both the as-deposited state and after annealing at 900 °C. The film density was extracted from the critical angle of the sample, determined by the condition of total external reflection. 
       
The refractive index ($n$) and thickness ($h$) of the films were determined using spectroscopic ellipsometry (SE) with a J.A. Woollam Co. Variable Angle Spectroscopic Ellipsometer. Data were acquired over the 300--2500 nm spectral range at incidence angles of 60°, 65°, and 70° to optimize the data analysis accuracy \cite{WoollamSPIE1999}. The analysis was performed using the WVASE software, modeling the sample as two distinct layers (substrate and coating) with sharp interfaces and defined dielectric functions \cite{MagnozziOptMat2018, AmatoACSAOM2023, DuranteOptMat2024, AmatoPRD2025}. The optical properties of the Si substrate were characterized independently via SE, while the SiO$_{2}$ coating was modelled both as a isotropic material (see Fig.~\ref{fig:SE} in Appendix~\ref{AppA}) and as a material uniaxially birefringent along the direction perpendicular to the surface. While the anisotropic model slightly improves agreement in the visible portion of the spectrum, both models accurately reproduce the experimental data for all samples. In Appendix~\ref{AppA}, Fig.~ \ref{fig:SE} shows the measured ellipsometric angles for the as-deposited sample S21088, along with the corresponding fit generated by the isotropic model. In the same Appendix, Fig.~\ref{fig:n} presents the wavelength-dependent refractive index of the film in its as-deposited state and after annealing at 500 °C and 900 °C. In the thicker sample S21089, interference peaks in the UV region restricted analysis to the 400--2500 nm range, though thickness accuracy ($\sim 0.3 \%$) remained uncompromised. However, due to higher depolarization, the refractive index at 532.3 nm ---while fully consistent with sample S21088--- was estimated with greater uncertainty.
       
The film thickness was additionally measured using scanning electron microscopy (SEM) with a FE-SEM LEO 1525 ZEISS. Samples were mounted vertically on conductive carbon adhesive tape to acquire cross-sectional images of the coating on the substrate (Appendix~\ref{AppA}, Fig.~\ref{fig:SEM}). The coating thickness for each sample was estimated with an accuracy of $\sim 0.3 \%$, similar to that of SE, based on the average of 30--40 SEM measurements taken at various points. Elemental composition was confirmed using a Bruker Quantax EDX system.

     \subsection{Brillouin light scattering experiments}

Brillouin light scattering (BLS) measurements were performed in air at room temperature using a high-resolution, high-contrast Sandercock-type 3+3-pass tandem Fabry-P\'erot interferometer. The setup employed a solid-state laser with a wavelength $\lambda$=532.3 nm (power on the sample: $\sim$160 mW). Light was focused and collected in a backscattering geometry using a 50 mm camera objective, with the sample mounted vertically on a goniometer rotatable in the horizontal scattering plane. The incident light was either s-polarized (perpendicular to the plane of incidence) or p-polarized (parallel to the plane of incidence), and the scattered light was analyzed in either s- or p-polarization. A 2 mm-wide vertical slit placed behind the polarization analyzer restricted the angular acceptance of the interferometer around the backscattering direction. 
The three samples, S21089, S21088, and S19175, were measured in their as-deposited state and after annealing at 500 °C and 900 °C. Polarized (s-s; p-p) and depolarized (s-p) spectra were acquired over a frequency range up to 40 GHz at various angles of incidence ($\theta$=10°, 12°, 20°, 30°, 40°, 45°, 50°, 55°, 60°, 65°, 70°, 75°). At the lowest incidence angle ($\theta$=10°), used to prevent reflected elastic light from entering the spectrometer, the refracted light inside the film propagated at an angle $\alpha<7^{\circ}$, as determined by Snell’s law, which is nearly perpendicular to the film surface. Measurements were repeated at two or three different points on each sample, and with the sample plane rotated around the normal axis for additional data. Acquisition times ranged from 2 to 46 hours, depending on the scattering cross-section of the acoustic modes, which varies with sample thickness, annealing state, and scattering configuration. S-s and s-p spectra were recorded at room temperature and $\theta$= 45° on a fused-silica platelet (Suprasil 300, Heraeus), sputter-coated on one side with several hundred nanometers of chromium.

      \subsection{Infrared spectroscopy experiments}

Infrared (IR) measurements were performed in both specular reflection (SR) and attenuated total reflection (ATR) modes.

SR-IR spectra were acquired using a Bruker INVENIO-R interferometer equipped with a PIKE Technologies VeeMAX III variable-angle specular reflection accessory. The three samples (S21089, S21088, and S19175) were analyzed in their as-deposited state and after annealing at 500, 600, and 900 °C. Spectra were recorded with unpolarized and p-polarized radiation at incidence angles $\theta$=30°, 40°, 45°, 50°, 55°, 60°, 65°, 70°, 75°, 80°, as well as at a fixed incidence of $\theta$=67° with s-polarized radiation. Polarization was selected using a ZnSe manual polarizer inserted into the dedicated slot of the VeeMAX III accessory. Each spectrum was acquired with a resolution of 4 cm$^{-1}$ across the wavevector range 100--9000 cm$^{-1}$, by averaging 60 scans. A gold mirror served as the standard reference for all measurements. Unpolarized and p-polarized spectra, acquired at the same incidence angles as for the IBS films, were recorded at room temperature on a fused-silica sample supplied by Neyco.

ATR-IR spectra were recorded for the same set of samples using the Bruker INVENIO-R system equipped with the ATRMAXII variable-angle ATR accessory and a Ge ATR crystal with a 60° face angle. Measurements were taken at effective incidence angles of $\theta$=50°, 52.5°, 55°, 57.5°, with a resolution of 6 cm$^{-1}$, by averaging 150 scans. A sample press ensured close contact between the sample and the ATR crystal surface. For each acquisition, a background spectrum was collected under identical conditions using the clean Ge ATR crystal.

\section{\label{Background}Background on acoustic and optical modes in thin films}

        \subsection{\label{Acoustic}Acoustic modes and selective determination of elastic constants}
        
\begin{figure}
	\includegraphics[width=7 cm]{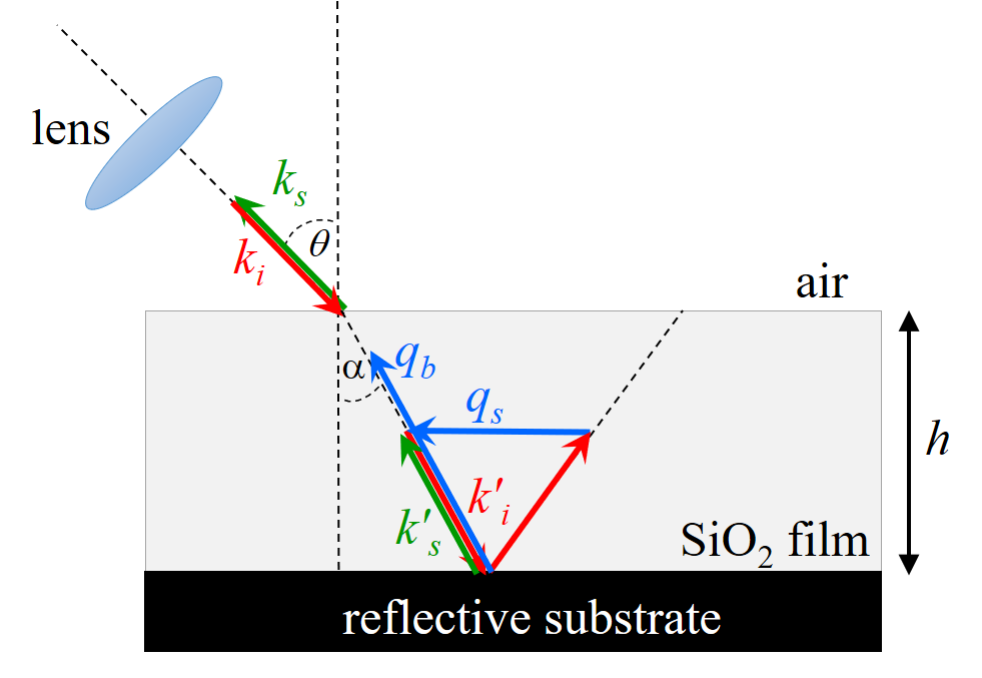}
	\centering
	\caption{\small Schematic diagram of the backscattering geometry used for BLS measurements. The wavevectors of incident, refracted, and reflected light are indicated in red, the wavector of scattered light in green, and the wavevectors of bulk-like and surface-like acoustic waves responsible for backscattered light in blue. All wavevectors lie within the scattering plane. In the experimental setup, the sample is mounted vertically, the scattering plane is horizontal, and s- or p-polarized incoming light is used.} \label{fig:Immagine1}
\end{figure}

BLS is the inelastic scattering of photons by thermally-induced acoustic phonons naturally present in a material \cite{Brillouin1922}. This phenomenon arises from the interaction between the electric field of the monochromatic incident light and the periodic modulation of the material's dielectric constant caused by acoustic waves. The modulation can occur through two primary mechanisms: the photoelastic effect, and surface corrugations (also known as the rippling effect). Whatever the mechanism, from a quantum-mechanical perspective, a Brillouin scattering event is described as a photon-phonon collision in which energy and momentum are conserved, through phonon creation (Stokes process) or phonon annihilation (anti-Stokes process). It follows that in a typical BLS experiment, the probed phonons have wavelengths comparable to that of light, placing their frequencies in the GHz range \cite{BLSreview1, BLSreview2}.\\
The configuration with a transparent amorphous SiO$_{2}$ film deposited on a reflective Si substrate enables two distinct backscattering channels to be accessed within a single BLS experiment. As illustrated in Fig.~\ref{fig:Immagine1}, the incident light (wavelength $\lambda$; wavevector $k_{i}=2\pi/\lambda$; angle of incidence $\theta$) passes through the objective lens, refracts at the air-film interface ($k'_{i}=n k_{i}$ where $n$ is the refractive index of the film) at an angle $\alpha=\arcsin(\sin\theta/n)$, is focused onto the film-substrate interface and then reflected by the substrate. For momentum conservation, only interactions with specific acoustic waves travelling parallel to the surface (\textit{surface-like} phonons) and within the medium at an angle $\alpha$ (\textit{bulk-like} phonons) originate backscattered light ($k'_{s}=n k_{s}\approx k'_{i}$). On one hand, after reflection from the substrate, light is backscattered by acoustic waves having parallel-to-surface  wavevector $q_{s}=2 k'_{i}\sin\alpha = 2 k_{i}\sin\theta$. On the other hand, when the film thickness $h$ exceeds both the phonon mean-free-path $\ell$ and the light wavelength $\lambda$, acoustic waves traveling within the medium at an angle $\alpha$ with a wavevector $q_{b}=2nk_{i}$, also contribute to the backscattered light, as they do in an infinite bulk material. The frequency of the scattered light is shifted by $\pm f_{BLS}$ relative to the incident light and, for energy conservation, $f_{BLS}$ is connected to the phase velocity ($v_{b,s}$) of the corresponding acoustic wave as: 
\begin{equation}
	v_{b}=\frac{2\pi f_{BLS}}{q_{b}}=\frac{\lambda f_{BLS}}{2 n}; \,
	v_{s}=\frac{2\pi f_{BLS}}{q_{s}}=\frac{\lambda f_{BLS}}{2\sin\theta}
	\label{eq:vbvs}
\end{equation}
for bulk and surface modes, respectively. Note that the velocity of surface waves, unlike bulk ones, can be directly determined from $f_{BLS}$ without requiring prior knowledge of the material’s refractive index.   

By using various angles of incidence for the incoming ligth and a careful polarization analysis of the incident and scattered light, it is possible to detect and identify several acoustic waves in BLS spectra: the Rayleigh surface wave (RW), the surface-like longitudinal mode (LM), the bulk-like longitudinal mode (LB), all of which contribute to polarized scattering, and the surface-like shear horizontal mode (SHM), which contributes to depolarized scattering. A schematic representation of these modes is provided in Fig. 2. 
\begin{figure}
	\includegraphics[width=8.5 cm]{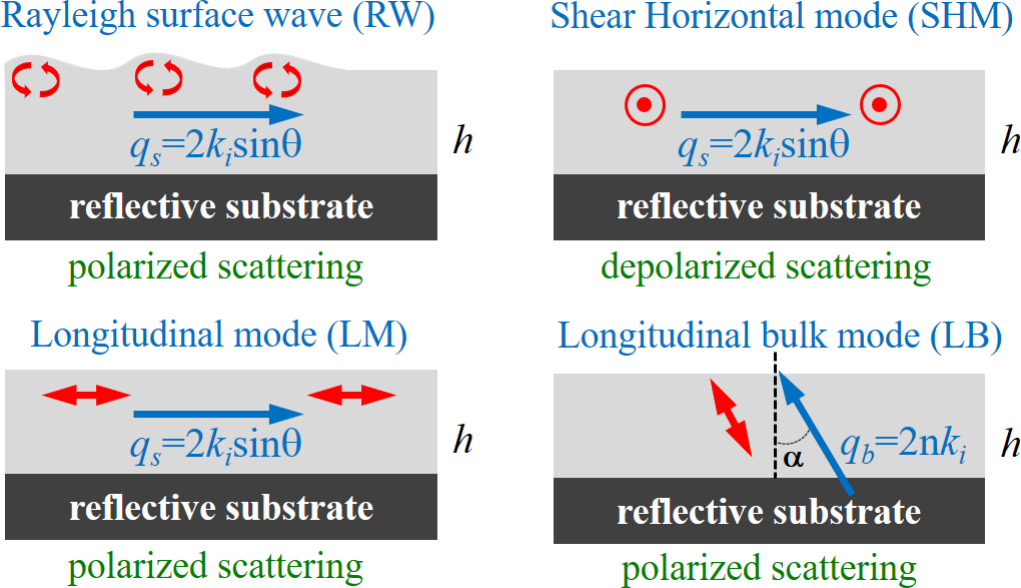}
	\caption{\small Schematic view of the four types of acoustic waves detectable in BLS spectra from SiO$_{2}$ thin films deposited on a reflective substrate. For each acoustic wave, the propagation direction is indicated by a blue vector, and the particle motion is represented by a red symbol. The polarization of the scattered light is indicated below each image.} \label{fig:modi}
\end{figure} 

It has to be noted that, in thin films ($h < \lambda, \ell$) of a homogeneous lossless material supported on a hard substrate, bulk-like phonons propagating perpendicular to the surface ($q_{\parallel}=0$) give rise to standing modes resulting from the constructive interference of waves reflecting back and forth at the film surfaces, having nodes at the interface with the substrate and antinodes at the free surface --- as occurs in organ pipes \cite{ZhangPRB2001}. To satisfy the boundary conditions, the wavevectors along the direction perpendicular to the surface must have discrete values $q_{\perp}=(2m-1)\pi/2h$, where $m=1,2,3 \ldots$ is an integer \cite{ZhangPRB2001, ZhangPRB2003}. The interaction of light at normal incidence with these standing modes produces in the BLS spectrum a series of equally spaced peaks, with a frequency separation $\Delta f=v_{LB}/2h$, where $v_{LB}$ is the velocity of surface-normal longitudinal bulk-like phonons. The intensity of these peaks is modulated by a $\textnormal{sinc}^{2}f$ envelope centered around the frequency $f_{LB}$, corresponding to the single peak produced by longitudinal phonons traveling normal to the surface under bulk conditions \cite{SandercockPRL1972, Sandercock1982BOOK, GomopoulosMACRO2009}. Within this theoretical framework, only when the film thickness increases such that $\Delta f$ becomes smaller than the peaks half-width, the discrete multipeak structure is no longer resolved, and the spectrum merges into a single symmetric peak centered at $f_{LB}$ \cite{ElAboutiCRY2022, PasseriBioAdv2023}.

The BLS signal integrates the contributions from the entire illuminated volume, which is defined by the laser spot size of $\sim$30 $\mu$m and a lens focal depth of the order of 100 $\mu$m, significantly larger than the film thickness. Consequently, at the macroscopic length scale probed by BLS, the sample behaves as a homogeneous medium. From an elastic perspective, thin films grown via IBS are reasonably expected to exhibit transverse isotropy, characterized by an extraordinary axis along the growth direction and equivalence among all directions parallel to the surface. Transverse isotropy implies cylindrical symmetry, resulting in five independent elastic constants in the elasticity tensor $c_{ij}$: $c_{11}$, $c_{33}$, $c_{13}$, $c_{44}$, and $c_{66}$ (see Appendix, Sec.~\ref{B1}). Importantly, the phase velocity of the detectable acoustic modes generally depends on the elastic constants of both the film and the substrate. Nevertheless, for sufficiently large film thicknesses, the influence of the substrate becomes negligible, and some of the film's elastic constants can be selectively extracted from the measured acoustic velocities, as in the case of a semi-infinite medium \cite{CarlottiApplSci2018}:
\begin{equation}
	v_{LM}=\sqrt{\frac{c_{11}}{\rho}} ; \, 	v_{SHM}=\sqrt{\frac{c_{66}}{\rho}} ; \, v_{LB}(\alpha=0^{\circ})=\sqrt{\frac{c_{33}}{\rho}} ,
	\label{eq:cij}
\end{equation}
with $\rho$ the material density; the remaining constants can be determined from $v_{RW}$ and $v_{LB}(\alpha \neq 0^{\circ})$, which both depend on $c_{11}$, $c_{33}$, $c_{13}$, $c_{44}$, and $\rho$ (see details in the Appendix, Sec.~\ref{B2} and \ref{B3}). Numerical simulations show that acoustic-wave propagation in SiO$_{2}$ films on silicon is well approximated by that in a semi-infinite medium --- and thus effectively substrate-independent --- when the film thickness exceeds the phonon wavelength $\Lambda$. Remarkably, this also holds for much harder substrates such as sapphire (Fig.~\ref{14} in Appendix~\ref{AppC}). 

Notice that elastic moduli are frequency-dependent complex quantities, and, from sound velocities, BLS spectroscopy determines their real part at the single frequency $f_{BLS}$, located in the tens of GHz range. 
Only in isotropic materials, there are two instead of five independent elastic constants, for example the longitudinal modulus $M \equiv c_{11}=c_{33}$ and the shear modulus $G\equiv c_{66}=c_{44}$, whose real parts can be determined by combining the velocity equations (\ref{eq:vbvs}) with the elasticity relations (\ref{eq:cij}). The isotropic elastic response can alternatively be described using the bulk modulus $K$, the Young’s modulus $Y$, and the Poisson’s ratio $\nu$, which are analytically related to $M$ and $G$ as follows: 
\begin{equation}
	Y=M-\frac{(M-2G)^{2}}{(M-G)} \quad \nu=\frac{2G-M}{2(G-M)} \quad  K=M-\frac{4}{3}G
	\label{eq:Ynu}
\end{equation}  
Another critical property is the internal elastic energy dissipation, characterized by the loss angle $\Phi = \tan^{-1} \left(\operatorname{Re}(Y)/\operatorname{Im}(Y)\right)$, determined by the ratio of the imaginary to the real part of the elastic modulus.

       \subsection{\label{Optical}Optical modes and local structure}

\begin{figure}
	\includegraphics[width=9 cm]{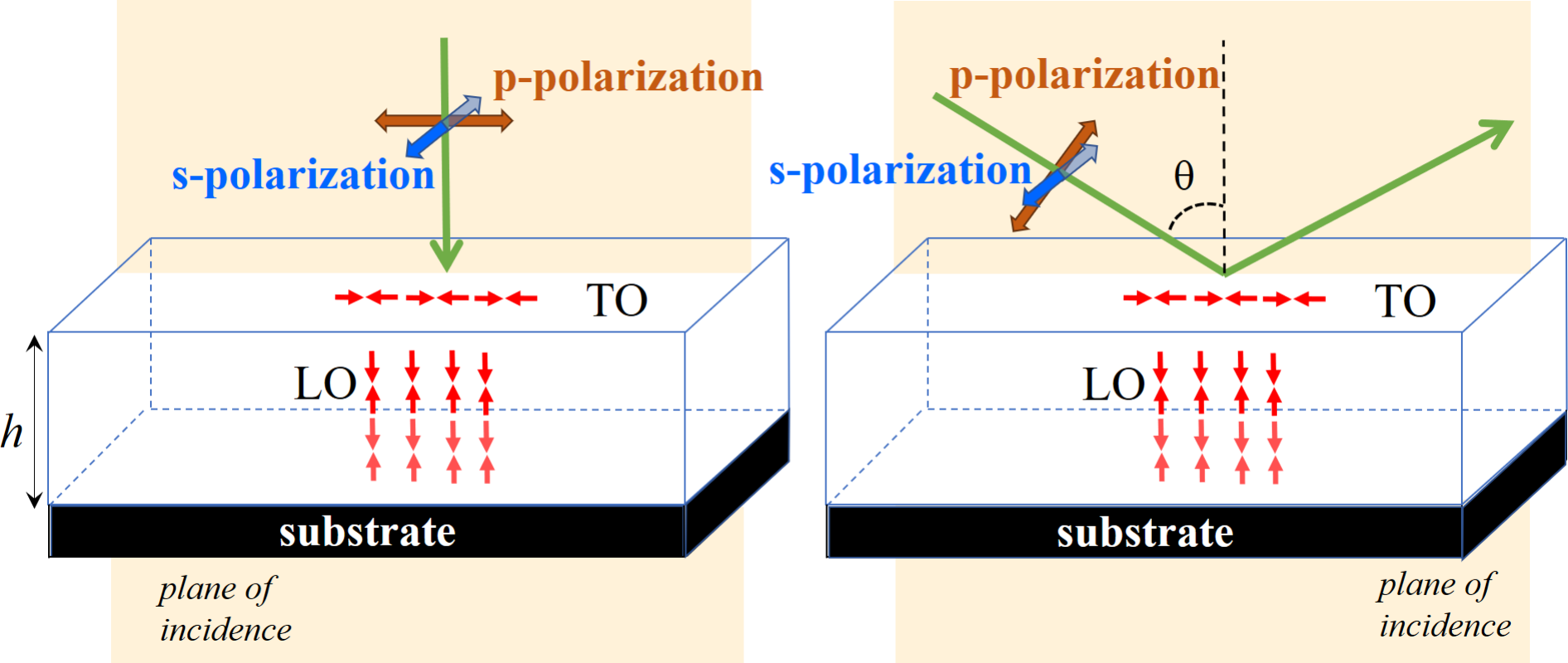}
	\caption{\small Schematic representation of TO and LO modes (red arrows) probed by s- and p-polarized IR radiation. At normal incidence (left panel), IR spectroscopy detects only the transverse (TO) component of the antisymmetric Si-O-Si stretching vibration. At oblique incidence (right panel), both the transverse (TO) and longitudinal (LO) components can be observed.} \label{fig:TOLO}
\end{figure}

In both crystalline and amorphous silica, oxygen atoms that link the SiO$_{4}$ tetrahedra, which are characteristic of the structure, are referred to as bridging oxygens (BOs). In fused silica, the angle formed by the Si-O-Si bonds, with a BO at the vertex, typically averages around 140° \cite{HosonoJAP1991, DeLosArcosVibSpect2021, EfthimiopoulosJAppPhys2018}. This angle can vary under conditions of stress or chemical inhomogeneity. The Si-O-Si stretching vibration reflects these variations, as its force constant depends on the arrangement (distance and relative orientation) of the SiO$_{4}$ units, thus serving as a probe for the local structure. In SiO$_{2}$ films, the long-range Coulomb interactions inherent to this polar material result in anisotropic vibrations: atomic displacements along directions parallel and perpendicular to the surface give rise to two distinct modes, the longitudinal-optic (LO) and the transverse-optic (TO) modes \cite{DeLosArcosVibSpect2021, EfthimiopoulosJAppPhys2018, AmmaJNCS2015}. Additionally, the stretching vibrations of groups containing non-bridging oxygens (NBOs), such as Si-O$^{-}$ and Si-OH, can be detected, providing further insights into the material's local environment.  

The IR signal of the antisymmetric Si-O-Si vibration is particularly intense, but the information obtained from this signal depends on the selection rules for IR absorption, which vary with the polarization and incidence angle of the radiation. For s-polarized radiation, the electric field is always parallel to the sample surface and thus interacts only with vibrations that exhibit a change in dipole moment with a surface-parallel component, such as the TO mode. Conversely, for p-polarized radiation at oblique incidence, the electric field has both parallel and perpendicular components relative to the surface, enabling excitation of both the TO and LO modes, as illustrated in Fig.~\ref{fig:TOLO}.

The spectral signals of the LO and TO modes are observed in the ranges 1200--1250 cm$^{-1}$ and 1100–1120 cm$^{-1}$, respectively, while the NBO signal is located in the range of 1000--1020 cm$^{-1}$. The configuration with a SiO$_{2}$ film deposited on a substrate enables to exploit IR reflectivity to monitor these vibrations while avoiding substrate absorption contributions. This is achieved in either SR or ATR configurations, which utilize the external or internal reflection of the incoming radiation, respectively. The penetration depth of radiation into the film, defined as the distance where the electric field amplitude drops to $1/e$ of its value at the surface, differs significantly in the two configurations. In SR-IR mode, the penetration depth depends on the wavelength $\lambda$ and the incidence angle $\theta$ of the incoming radiation, according to \cite{TanOPE2005}: 
\begin{equation}
	d_{p}^{SR}(\lambda, \theta)=\frac{\lambda}{4\pi k} \sqrt{1-\frac{\sin^{2} \theta}{n^{2}+k^{2}}}
	\label{eq:dpSR}
\end{equation} 
\\ where $n(\lambda)$ and $k(\lambda)$ are the real and imaginary parts of the material's complex refractive index at wavelength $\lambda$.      
In ATR-IR mode, a crystal (e.g., Ge) is used as internal reflection element due to its high refractive index and superior IR transmission. Here, the incident radiation udergoes total internal reflection within the crystal, and the sample, in close contact with it, absorbs the evanescent wave that propagates outside the crystal at each reflection. This configuration results in reduced radiation penetration into the sample, with a penetration depth given by \cite{HosonoJAP1991}: 
\begin{widetext}
	\begin{equation}
		d_{p}^{ATR}(\lambda, \theta)=(\frac{\lambda}{2\pi}) \left[ \frac{\sqrt{(n_{ATR}^{2}\sin^{2}\theta - n^{2} + (\ln 10 k)^{2} )^{2} +(2n_{ATR}\ln 10 k)^{2} }+(n_{ATR}^{2}\sin^{2}\theta - n^{2} + (\ln 10 k)^{2}  )}{2} \right]^{-1/2}
		\label{eq:dpATR}
	\end{equation}
\end{widetext}
where $n_{ATR}$ is the refractive index of the ATR crystal. The differing dependencies of $d_{p}^{SR}$ and $d_{p}^{ATR}$ on $\theta$, $n$ and $k$ result in different spectral profiles and sensitivities to vibrational modes between the two techniques \cite{AmmaJNCS2015}. Additionally, in SR-IR spectra, an oscillating background can appear, with period and characteristics dependent on the film thickness, due to interference from light reflections at the air-film and film-substrate interfaces. This interference is absent in ATR-IR spectra, where the reduced penetration depth precludes such effects, making ATR-IR a practical alternative to SR-IR spectroscopy for thin layers, although existing literature on its application in this area is surprisingly limited.

\section{\label{Results}Results and Discussion}

        \subsection{\label{ElaOpt}Elastic and optical properties}
        
                      \subsubsection{\label{asdep}As-deposited material}
        \begin{figure*}[h!]
        	\centering
        	\includegraphics[width=18 cm]{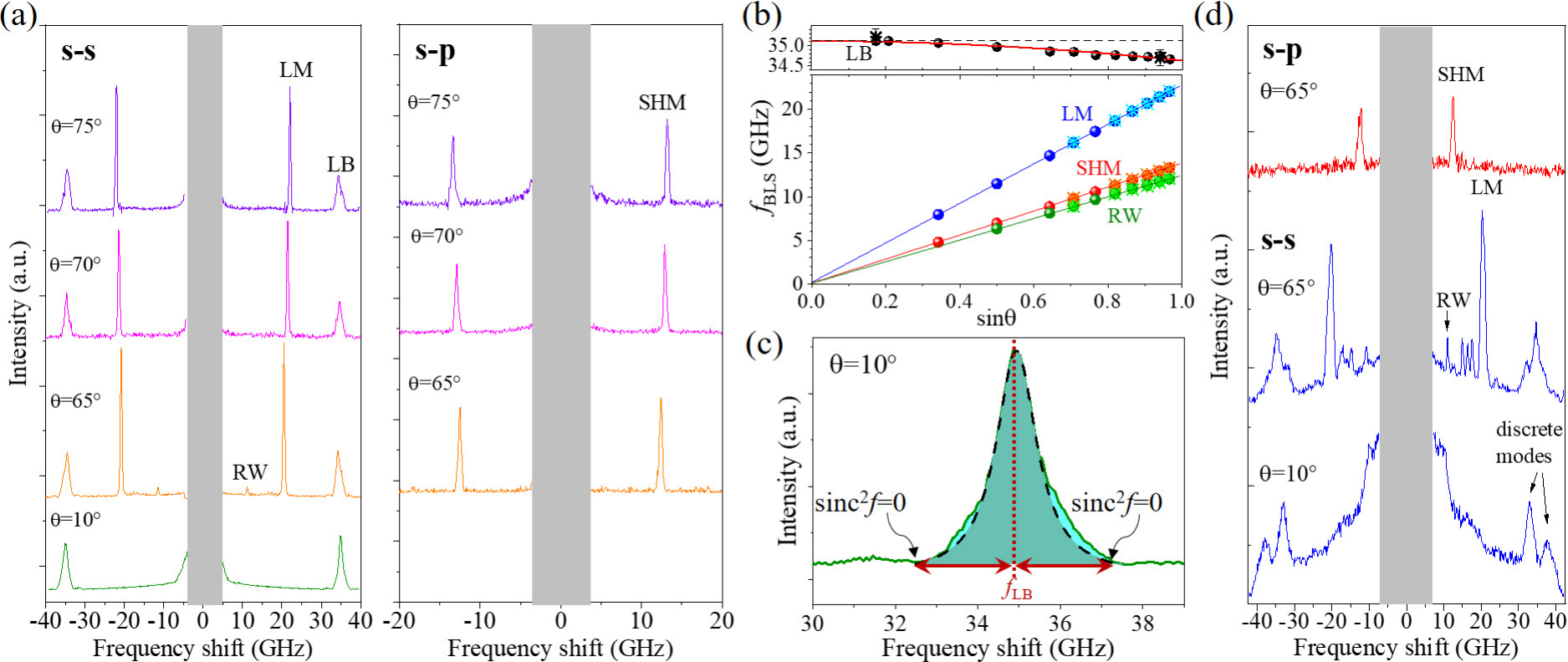}
        	\caption{\small (a) Polarized (s-s) and depolarized (s-p) BLS spectra of the as-deposited sample S21089 at selected incidence angles $\theta$. The Rayleigh, longitudinal (surface-like and bulk), and shear horizontal peaks are labeled as RW, LM, LB, and SHM, respectively. (b) Lower frame: Brillouin frequency $f_{BLS}$ of different surface-like acoustic waves as a function of the incidence angle. Solid spheres are data for the sample S21089, stars are data for the thinner samples S21088 and S19175. Solid lines represent linear fits through the origin. Error bars are smaller than the symbol size. Upper frame: Brillouin frequency $f_{LB}$ of bulk-like acoustic wave as a function of the incidence angle, from s-s spectra. Black solid spheres are data for the sample S21089, stars are data for the thinner samples S21088 and S19175. The horizontal dotted line highlights the difference between low and high-incidence $f_{LB}$ values. The red solid line is the numericaly simulated dispersion curve. (c) Enlargement of the BLS spectrum at quasi-normal incidence ($\theta$=10°) in the LB frequency region, where a $\textnormal{sinc}^{2}$ modulation is observed. The central frequency between the zeros of the $\textnormal{sinc}^{2}$ function, $f_{LB}$, is marked by a vertical red dotted line. The cyan areas in the figure represent deviations from a single-peak structure, as indicated by the black dashed line. (d) BLS spectra acquired at $\theta$=10° and 65° for the as-deposited thin sample S19175.}
        	\label{fig:BLSspectra}
        \end{figure*}
The Brillouin investigation begins with the $\sim2.5$ $\mu$m-thick sample S21089 in its as-deposited state. The BLS experiment reveals several peaks, selectively detected in polarized and depolarized spectra, that change differently their frequency position with the incidence angle, as shown in Figs.~\ref{fig:BLSspectra}(a) and \ref{fig:BLSspectra}(b). Three peaks whose frequencies scale with $\sin\theta$, in agreement with Eq.~(\ref{eq:vbvs}), are identified as surface-like acoustic waves: the LM mode ($f_{LM}$) in s-s spectra, the SHM mode ($f_{SHM}$) in s-p spectra, and the Rayleigh wave ($f_{RW}$) in s-s and p-p spectra. The Rayleigh signal, much weaker than the others, is observed only at a few angles in the s-s configuration, whereas in the p–p configuration it appears more clearly and over a broader angular range ($\theta\geq30$°), enabling systematic analysis. For each of these modes, the constant ratio $f_{BLS}/\sin\theta$ in Fig.~\ref{fig:BLSspectra}(b) demonstrates nondispersive propagation and, according to Eq.~(\ref{eq:vbvs}), yields a phase velocity independent of $\theta$ (Fig.~\ref{15} in Appendix~\ref{AppC}). This behavior is consistent with numerical expectations since all probed phonons satisfy $h/\Lambda\gtrsim 5$ --- exceeding the $h/\Lambda\sim 1$ threshold for substrate-independent acoustic propagation (Sec.~\ref{Acoustic}).
        
In addition, a strong broadened peak around 35 GHz in polarized spectra, nearly fixed with $\theta$, matches the expected behavior of bulk-like modes. Further analysis of this feature suggests a multipeak structure, consistent with predictions for films thin enough  to induce transversal phonon confinement and consequent discretization of the surface-normal phonon wavevector (Sec.~\ref{Acoustic}). Indeed, studies \cite{VacherPRB1997, BaldiJNCS2011} report a mean free path $\ell \sim 6$ $\mu$m for LB phonons in fused silica measured at ambient temperature via BLS. In Fig.~\ref{fig:BLSspectra}(c), the spectrum is clearly perceived as a triplet, with a central peak significantly more intense than the two lateral contributions, spaced approximately 1 GHz apart. This spacing agrees with the estimate $\Delta f=v_{LB}/2h \sim 1.2$ GHz, using $v_{LB}\sim$ 6000 m/s as in fused silica \cite{VacherPRB1997}. However, the multipeak structure is blurred due to the peak width being comparable to their separation. Varying the incidence angle further decreases the separation while changing the scattering efficiency of each mode, as evidenced by the differently asymmetric structures observed at higher incidence angles and after annealing at various temperatures (Fig.~\ref{16} in Appendix~ \ref{AppC}). As explained in Sec.~\ref{Acoustic}, $f_{LB}$ corresponds not to the frequency of maximum intensity but to the central frequency between the first zeros of the $\textnormal{sinc}^{2}$ envelope (see Fig.~\ref{fig:BLSspectra}(c)). 
        
The frequencies $f_{LM}$, $f_{SHM}$ and $f_{LB}$ allow direct access to three of the five independent elastic constants. By combining Eqs.~(\ref{eq:vbvs}) and (\ref{eq:cij}), $f_{LM}$ and $f_{SHM}$ are used to calculate $c_{11}$ and $c_{66}$, which are independent of $\theta$, while $f_{LB}(\theta=10^{\circ})\approx f_{LB}(\alpha=0^{\circ})$, together with the material refractive index, is used to determine $c_{33}$. For the remaining constants, $c_{13}$ and $c_{44}$, they are numerically estimated as those that best reproduce the experimentally measured velocities $v_{RW}$ and $v_{LB}(\theta)$, according to the procedure detailed in Sec.~\ref{B3} of the Appendix. This complete characterization, which would be difficult if not impossible with conventional mechanical testing, enables to capture a sizeable elastic anisotropic response of the material. A comparison of $c_{11}$ with $c_{33}$ --- which should be equal in an isotropic medium --- reveals $c_{33}>c_{11}$, meaning that the compressibility along the growth axis is lower than in directions parallel to the surface. The anisotropy ratio $c_{33}/c_{11}$ is $1.062\pm 0.004$, providing the first direct evidence of a $(6.2 \pm 0.4)\,\%$ compressive anisotropy in the as-deposited material. By contrast, $c_{44}$ and $c_{66}$ are identical within errors, revealing the absence of shear anisotropy ($c_{66}/c_{44}=1.01\pm 0.02$). Therefore, although the film is stiffer in compression along the growth direction, it exhibits similar resistance to shearing out of plane and in plane. Finally, $c_{13}$ is higher than $c_{12}=c_{11}-2c_{66}$ by $(10 \pm 1)\,\%$, still in contrast with the isotropic case (Table~\ref{T3} in Appendix~\ref{AppF}).   
        
A behavior consistent with transverse isotropy, where all directions parallel to the surface are equivalent, is confirmed by rotating the sample around its normal axis and observing that the velocity $v_{LM}$ of longitudinal waves remains invariant across different in-plane directions. Additionally, the velocity $v_{LB}$ of longitudinal bulk-like waves propagating at different angles within the film, rather than being independent of $\theta$ as in the isotropic case, exhibits a slight but significant variation with $\theta$ (as shown in Fig.~\ref{17} in the Appendix~\ref{AppC}), resulting $1\%$ lower at higher incidence ($6190 \pm 10$ m/s for $\theta = 75^\circ$) compared to quasi-normal incidence ($6260 \pm 10$ m/s). The dependence of $v_{LB}$ on $\theta$ directly reflects that of the frequency $f_{LB}$, shown in the upper panel of Fig.~\ref{fig:BLSspectra}(b).
        
The velocity $v_{LB}$, at any incidence, is calculated here from s-s spectra using $n_{532}$, the refractive index at the BLS wavelength obtained from the isotropic SE model. This warrants particular attention, because transverse elastic isotropy may be accompanied by uniaxial birefringence, with the film’s growth direction acting as the optic axis. In our experimental geometry (see Fig.~\ref{fig:Immagine1}), s-polarized incoming light is always polarized perpendicular to the optic axis and is thus governed by the ordinary refractive index $n_{o}$, which may differ from $n_{532}$ in case of birefringence. However, we find that $n_{532}$, within the uncertainty, matches $n_{o}$ at the same wavelength, as obtained by reproducing the SE data with the anisotropic model (Sec.~\ref{Char}), which indicates that the approximation $n_{532}\approx n_{o}$ used to calculate $v_{LB}$ is very accurate. It has to be noted that the anisotropic model describes the SE data with a difference between $n_{e}$ and $n_{o}$ of 0.001--0.002 in all samples investigated, suggesting that a small birefringence, if any, is beyond the experimental sensitivity. 
        
Comparing BLS and SE measurements further highlights the anisotropy of the material in terms of acoustic wave propagation. According to Eq.~(\ref{eq:vbvs}), the velocity ratio $v_{LB}/v_{LM}$ is provided at any incidence angle by the expression $(f_{LB}/f_{LM})\sin\theta/n_{532} \equiv n_{BLS}/n_{532}$, where $n_{BLS}$ denotes the quantity $(f_{LB}/f_{LM})\sin\theta$. This was evaluated for twelve spectra at $\theta\geq30^{\circ}$ in the as-deposited film. While largely independent of $\theta$ in the measured range, the velocity ratio is found to be $1.019\pm 0.004$, reflecting anisotropic acoustic wave propagation within the material. It should be noted that $n_{BLS}$ cannot be calculated at quasi-normal incidence, where $v_{LB}$ is higher than the values measured at $\theta\geq30^{\circ}$. This accounts for the $2\%$ difference from $n_{532}$, instead of the $3\%$ expected from the elasticity ratio $c_{33}/c_{11}=v_{LB}^{2}/v_{LM}^{2}$, where $v_{LB}$ is measured at $\theta=10^{\circ}$. Also, it is interesting to note that, if the film were isotropic in both elastic and optical properties, $n_{BLS}$ would coincide with the refractive index of the material at the BLS wavelength, $n_{532}$.
        
In conclusion, as a remarkable first achievement, we provide evidence that the elastic properties of the as-deposited sample S21089 differ substantially from those expected in an isotropic material. The possibility that the observed anisotropy is a measurement artifact is ruled out by results obtained from a reference isotropic sample (details are provided in Appendix~\ref{AppD}). 
        
For the thinner samples S21088 and S19175, BLS spectra allow the determination of the frequencies $f_{LM}$, $f_{SHM}$, and $f_{RW}$ of surface-like acoustic modes at $\theta \geq 45^{\circ}$ (stars in Fig.~\ref{fig:BLSspectra}(b)), which --- as expected for $h/\Lambda > 1.9$ --- are nondispersive and yield the same phase velocities as measured in sample S21089 (Fig.~\ref{15} in Appendix~\ref{AppC}). The determination of $f_{LB}$ at different incidence angles is subject to larger uncertainty due to the phonon discretization effect arising from the reduced sample thickness. Nevertheless, albeit with reduced precision, all elastic constants and the quantity $n_{BLS}$ can still be extracted by applying the same procedure used for sample S21089. As shown in Fig.~\ref{fig:BLSspectra}(d), the multipeak structure in the thinner samples is particularly pronounced at quasi-normal incidence, and the information provided will be further discussed in Sec.~\ref{Thickness}.

                 \subsubsection{\label{annealing}Post-deposition annealing effects}
\begin{figure*}[h!]
	\centering
	\includegraphics[width=14.5 cm]{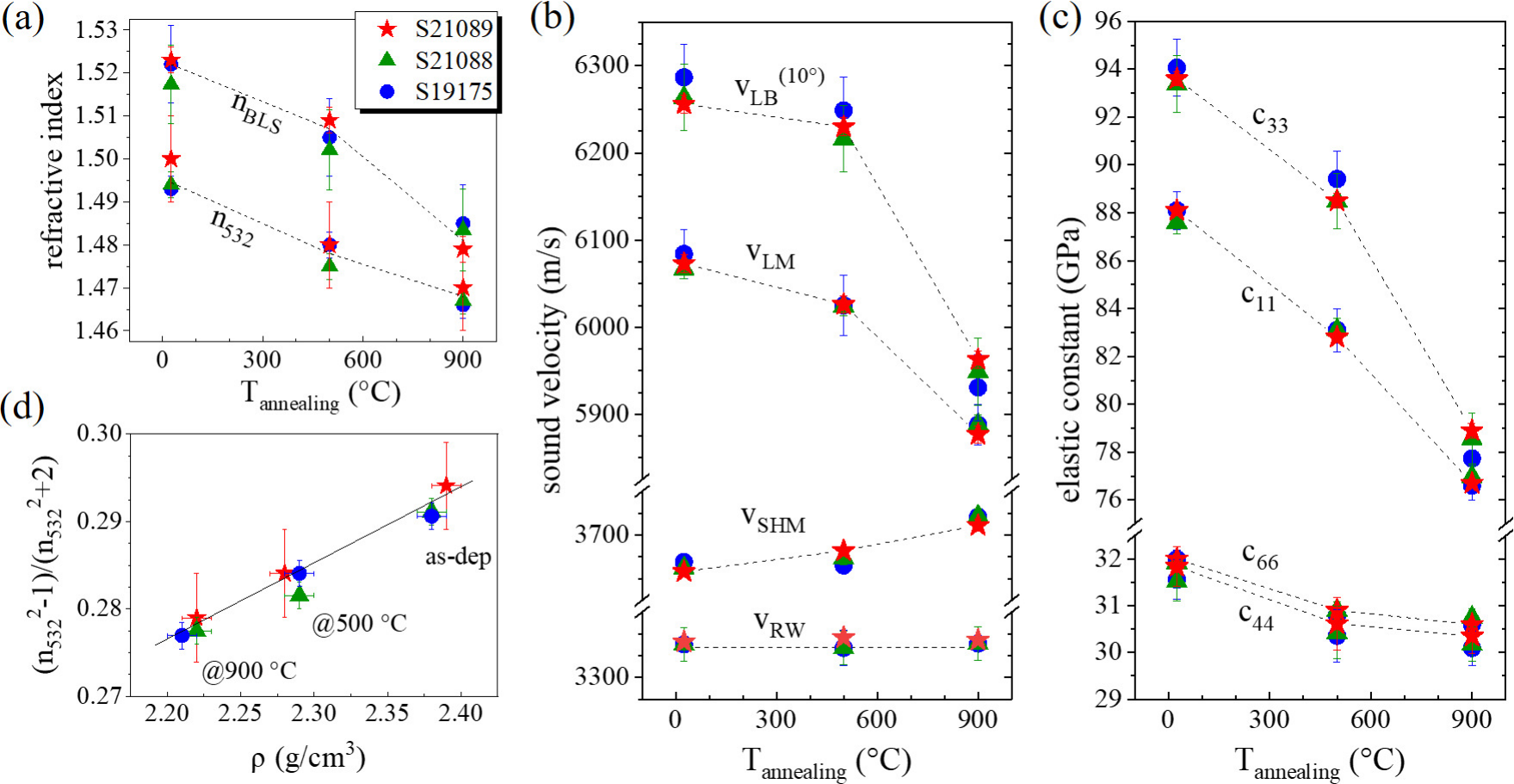}
	\caption{\small (a) Dependence on the annealing temperature of the refractive index $n_{532}$ obtained from the SE isotropic model and the BLS-apparent refractive index $n_{BLS}$. (b) Dependence of the sound velocities $v_{LM}$, $v_{SHM}$, $v_{RW}$, and $v_{LB}$ on the annealing temperature, with $v_{LB}$ measured at quasi-normal incidence. (c) Dependence on the annealing temperature of the elastic constants $c_{11}$, $c_{33}$, $c_{66}$, and  $c_{44}$. (d) Dependence of the quantity $(n_{532}^{2}-1)/(n_{532}^{2}+2)$ on the film density $\rho$, and best fit with the Lorentz-Lorenz equation (solid line). The legend is the same for all the panels. Dashed lines are just a visual guide.} 
	\label{fig:cij}
\end{figure*}
BLS spectra of samples annealed at 500 °C and 900 °C are qualitatively similar to those in the as-deposited state and allow for the calculation of the same quantities described in the previous section. An example is shown in the spectra of sample S21089 (Fig.~\ref{18} in Appendix~\ref{AppC}). The derived elastic constants $c_{11}$, $c_{33}$, $c_{66}$, $c_{44}$, and $c_{13}$, along with the density and refractive index used in the calculations, are listed in Table~\ref{T3} in the Appendix~\ref{AppF}. 
Figures~\ref{fig:cij}(a)--(c) display the refractive index, sound velocities, and elastic moduli of all investigated samples as a function of annealing temperature. The density is used as the x-axis in Fig.~\ref{fig:cij}(d). All the mentioned quantities can be measured in films with thickness of $\sim 720$ nm (S21088, S19175) and $\sim 2.5$ $\mu$m (S21089) and are identical within error, which demonstrates that macroscopic properties can be reliably probed regardless of film thickness, consistently with numerical expectations. 

We now examine the effects of post-deposition heat treatment in detail. Figures~\ref{fig:cij}(a) and \ref{fig:cij}(c) show that both the optical and elastic properties of IBS SiO$_{2}$ films are highly sensitive to annealing, with higher temperatures resulting in lower refractive index and stiffness coefficients (Table~\ref{T3} in Appendix~\ref{AppF}). From an elasticity perspective, annealing induces a general softening of the material: the decreases in $c_{33}$ and $c_{44}$ --- both reflecting out-of-plane properties --- indicate that the film becomes easier to compress and shear along the direction perpendicular to the surface, while the decreases in $c_{11}$ and $c_{66}$ --- both reflecting in-plane properties --- indicate diminished resistance to in-plane compressional and shear deformation. Fig.~\ref{fig:cij}(d) reveals that variations in refractive index correlate strongly with density changes, consistent with the Lorentz-Lorenz relation $(n_{532}^{2}-1)/(n_{532}^{2}+2)=\rho (N_{A} \alpha_{e})/(3M_{e})$, where $N_{A}$ is Avogadro’s number, $M_{e}$ the effective molecular weight, and $\alpha_{e}$ the effective polarizability of the material's structural units. Quantitatively, after annealing at 900 °C, $n_{532}$ drops by 2$\%$, and the density decreases by 7$\%$. While density reduction after 10--hour annealing at 500 °C has been reported for materials like SiO$_{2}$, Ta$_{2}$O$_{5}$, and TiO$_{2}$:Ta$_{2}$O$_{5}$ \cite{GranataPRM2018, PaoloneCOAT2022}, this study shows that for SiO$_{2}$, the effect persists even at much higher temperatures. 

Annealing reduces the elastic moduli to varying degrees [Fig.~\ref{fig:cij}(c)], and it is instructive to separate the purely acoustic and density-related contributions to these reductions. Heat treatment simultaneously affects sound velocity and atomic concentration, two factors that can have compensating effects on elasticity [see Eq.~(\ref{eq:cij})]. The data in Fig.~\ref{fig:cij}(b) show that the sound velocities of longitudinal and horizontal shear modes are affected in opposite ways, with longitudinal waves becoming slower and shear waves faster as the annealing temperature increases. The slowdown of longitudinal acoustic waves after annealing at 900 °C is considerably larger than that produced by annealing at 500 °C (see also Fig.~\ref{17} in Appendix~\ref{AppC}). By contrast, the Rayleigh-wave velocity, which mainly involves particle motion localized near the film surface, is largely insensitive not only to film thickness but also to heat treatment, as shown in Fig.~\ref{19} in the Appendix~\ref{AppC}. The density reduction decreases both $c_{11}$ and $c_{33}$ by up to 13\textendash16$\%$ at 900 °C and dominates the concurrent $4 \%$ decrease of $c_{44}$. By compensating the increase in $v_{SHM}$, it also leads to a net 4$\%$ reduction in $c_{66}$. As a result, compressional softening in both in-plane and out-of-plane directions is markedly more pronounced than shear softening, while the material preserves its shear isotropy.

An important point is how anisotropy is affected by the post-deposition treatment. Although all measured quantities decrease as the annealing temperature increases, we find that the differences $n_{BLS} - n_{532}$  and $c_{33} - c_{11}$ [Figs.~\ref{fig:cij}(a) and \ref{fig:cij}(c)] surprisingly remain unchanged within the experimental error after annealing at 500 °C. Instead, both differences are reduced by approximately a factor of 2.5 after annealing at 900 °C. Therefore, as a second remarkable finding, we demonstrate that the heat treatment so far adopted for mirrors in GW detectors (10 hours at 500 °C), although beneficial for reducing coating thermal noise, is unable to alter the compressive anisotropy that silica acquires during deposition. Notably, the reduced anisotropy observed after annealing at 900 °C correlates well with a higher degree of structural relaxation, as reported in IR \cite{HiroseJNCS2006} and Raman \cite{GranataPRM2018} studies. This suggests that recovering isotropy is correlated to mechanical loss reduction.

              \subsubsection{\label{GENS}Comparison with low-frequency elastic properties}

As anticipated, BLS spectroscopy probes the elastic constants of SiO$_{2}$ at frequencies of tens of GHz, which is far above the frequency range between about 10 Hz and a few hundred Hz where mirror coatings represent a limiting noise source for GW detectors. However, a recent study \cite{GranataCQG2020} reported the elastic characterization of the same material --- produced by LMA under identical conditions --- in the much lower frequency range of 1–30 kHz, which is more directly relevant for the detection of GW signals. Data are only available for the as-deposited material and after annealing at 500 °C \cite{GranataCQG2020}, obtained using a Gentle Nodal Suspension (GeNS) system. 

We now compare the results of the two techniques. First, one has to consider that GeNS measurements yield the real part of the Young’s modulus and Poisson’s ratio under the assumption of isotropic medium. A direct comparison with the present BLS results, which reveal elastic anisotropy, is therefore not straightforward. However, an approximate comparison can be made using the estimated values of $Y$ and $\nu$ that would reasonably be obtained by neglecting the anisotropy described in the previous section. To this end, we calculate $Y$ and $\nu$ according to Eq.~(\ref{eq:Ynu}), replacing $M$ with the average of the longitudinal moduli $c_{11}$ and $c_{33}$, and $G$ with the average of the shear moduli $c_{44}$ and $c_{66}$. The results are reported in Table~\ref{tab1}. Notably, as the annealing temperature increases, the BLS-estimated values of $Y$ and $\nu$ confirm material's softening.

Although a comparison up to 900 °C is not available, some observations deserve attention. As a first remark, the value of the Young’s modulus in the as-deposited materials is the same at GHz and kHz frequencies, meaning that the contribution of relaxation processes in the intermediate frequency range is negligible within the experimental sensitivity. The implication would be that the more accurate high-frequency value provided by BLS can be used in place of the low-frequency value provided by GeNS. Apparently, the same is not true in the materials annealed at 500 °C. This discrepancy may find an explanation in the density used in the two studies. The values for the as-deposited materials ($2.39\pm0.01$ and $2.38\pm0.01$ g/cm$^{3}$, respectively in Ref. \cite{GranataCQG2020} and in this work), despite determined by different methods coincide within the uncertainty. By contrast, those for the materials annealed at 500 °C ($2.28\pm0.01$ and $2.36\pm0.03$ g/cm$^{3}$, respectively) differ significantly. Notably, thanks to their smaller uncertainties, the BLS data clearly resolve the dependence of $Y$ on annealing temperature --- a trend that in the GeNS results is entirely masked by experimental uncertainty --- and reveal that, as the treatment temperature increases, $Y$ approaches the value of fused silica, measured in the present work and consistent with textbooks \cite{Auld1973} and technical datasheets \cite{Heraeus, Crystran, Corning}. While we are unable to provide an explanation of the higher density reported in Ref. \cite{GranataCQG2020} for the film annealed at 500 °C, we note that if the density measured in the present study were used, then the best-value of Young’s modulus determined by GeNS would be in perfect agreement with the BLS estimate. These considerations suggest that contributions from mechanical relaxations in the GHz to kHz range are within the experimental uncertainty.

\begin{table*}
	\caption{\small \label{tab1} Real part of the Young's modulus $Y$ and Poisson's ratio $\nu$ of IBS SiO$_{2}$ films, estimated under the assumption of isotropic material from BLS and GeNS. The values for fused silica at room temperature are also reported for comparison.} 
	\vspace{0.2 cm} 
	\centering{\scalebox{1}{\begin{tabular}{l | c c c | c c c} 
				\hline
				& \multicolumn{3}{c |}{$Y$ (GPa)}   & \multicolumn{3}{c}{$\nu$}  \\
				\hline 
				&   BLS\textsuperscript{(a)}  & GeNS\textsuperscript{(b)} & GeNS\textsuperscript{(c)}  &   BLS\textsuperscript{(a)}   &  GeNS\textsuperscript{(b)} &    GeNS\textsuperscript{(c)} \\
				\hline   
				\hline
				as-deposited & $78.5\pm0.4$ & $78\pm1$ & $79\pm1$ & $0.229\pm0.004$ & $0.14\pm0.01$ & $0.23\pm0.05$  \\
				10h@500 °C & $75.0\pm0.5$ & $78\pm1$ & $78\pm2$ & $0.220\pm0.005$    & $0.11\pm0.01$ & $0.20\pm0.05$  \\
				10h@900 °C & $71.8\pm0.3$ & &  & $0.178\pm0.004$ &   &    \\
				\hline
				fused SiO$_{2}$ &  & $\approx 72-73$ \textsuperscript{(d)} &  & 
				& $\approx 0.17$\textsuperscript{(d)} &   \\
				\hline
			\end{tabular}
	}}
	\vspace{0.1 cm}
	{\raggedright
		\footnotesize\\
		\textsuperscript{a}~This work; \\
		\textsuperscript{b}~Ref.~\cite{GranataCQG2020}, as published; \\
		\textsuperscript{c}~This work, corrected reanalysis; \\
		\textsuperscript{d}~This work \& Refs.~\cite{Auld1973, Heraeus, Crystran, Corning}.	\par}
\end{table*}

The experimental evidence is supported by theoretical expectation. The well-known model of thermally activated relaxations in amorphous solids by Gilroy and Phillips \cite{GilroyPM1981} predicts a loss angle with a power-law frequency dependence, $\phi=\phi_0\cdot(f/f_0)^{\alpha}$, valid for $2\pi f\ll1/\tau_0$, where $\tau_0\approx 10^{-13}$ s is a characteristic time, $f_0$ is a reference frequency, and $\phi_0\equiv \phi(f_0)$ and $\alpha$ are constants. From Eqs. (13) and (14) of ref. \cite{GilroyPM1981}, one derives the real part of the Young's modulus $Y$, relative to $Y_0 \equiv Y(f_0)$, as  $Y/Y_0=\left[(2 \pi f \tau_0)^{\alpha}-1\right]\cdot (2\phi_0)/[\pi \alpha (2\pi\tau_0\,f_0)^{\alpha}]$. Using this expression, the relative difference of the modulus at frequencies $f_1$ and $f_2$ is given by:
\begin{equation}
	\frac{Y(f_2)-Y(f_1)}{Y_0}=\frac{2 \phi_0}{\pi \alpha}\left[  \left(\frac{f_2}{f_0}\right)^{\alpha} - \left(\frac{f_1}{f_0}\right)^{\alpha}  \right]
\end{equation}
Reasonable parameters for a silica film are $f_0=10$ kHz, $\phi_0=10^{-4}$, and $\alpha\simeq 0.1$ \cite{GranataCQG2020}. With these values, the relative difference  between $f_1=1$ kHz and $f_2=30$ GHz is approximately 0.2\%, corresponding to about 0.2 GPa for a modulus of around 80 GPa. The result is that the Young's modulus measured with GeNS and BLS is expected to differ by an amount well within the experimental uncertainty. Therefore, the value provided by BLS can be used to constraint the analysis of GeNS measurements, and to reduce the effect of criticalities such as the influence by sample geometry and boundary conditions, the need of a model to simulate the experimental data, and prior knowledge of the substrate properties (limitations to which Brillouin spectroscopy is immune) \cite{CesariniRSI2009, GranataPRD2016, GranataCQG2020}. More importantly, BLS is capable of detecting anisotropy effects ---that correlate to the presence of residual dissipative mechanisms--- to which the GeNS analysis remains insensitive.

Finally, a remark on the Poisson’s ratio is in order.
All studies of silica for GW detectors assume isotropy in this parameter. However, unlike the Young’s modulus, the Poisson’s ratio estimated from BLS under the isotropic approximation is nearly twice the value previously reported from GeNS measurements \cite{GranataCQG2020} up to 500 °C of annealing temperature, and markedly reduces when the material is annealed well above 500 °C. This result is unexpected, and indicates a resistance to lateral deformation under axial stretching that is substantially lower than that adopted in earlier modeling and data analysis \cite{GranataPRD2016, GranataCQG2020, MalhaireJVSTA2023}. Prompted by the marked discrepancy identified in the present work, the GeNS measurements originally reported in Ref. \cite{GranataCQG2020} were carefully reanalyzed by Granata and Cagnoli, who are the holders of the original GeNS data and coauthors of both Ref. \cite{GranataCQG2020} and the present manuscript. The reanalysis revealed that the discrepancy originates from a transcription error in the numerical simulation code used to extract the elastic parameters in Ref. \cite{GranataCQG2020}. The discrepancy is therefore neither attributable to a frequency-dependent Poisson's ratio nor to geometric or boundary-condition artifacts of the GeNS technique. The corrected GeNS values are reported in Table~\ref{tab1}, together with the originally published estimates and the BLS results. The corrected values are fully consistent, within uncertainties, with the BLS estimates, similarly to what is observed for the Young’s modulus. Owing to their significantly smaller uncertainties (by about one order of magnitude), the BLS data clearly resolve the dependence on annealing temperature --- a trend that is only barely visible in the best-fit GeNS values and entirely masked by their experimental uncertainty --- revealing that both $Y$ and $\nu$ of IBS silica approach the values of fused silica as the annealing temperature increases \cite{Auld1973, Heraeus, Crystran, Corning}. 

This downshift toward fused-silica\textit{-like} properties, that emerges clearly from the BLS characterization, reasonably extends to the elastic dissipation behavior and has potential implications for the level of thermal noise in the material. In the absence of an analytic expression for the thermal noise in anisotropic media, an approximate estimate of its relative variation as a function of the annealing temperature is performed for sample S21088 by considering its isotropic-analogous properties. For these calculations, we use the elastic parameters reported in Table 1, the layer thickness derived in Sec.~\ref{Thickness}, and the loss angle $\Phi$ given in the Appendix~\ref{AppE}. Different analytic models \cite{HongPRD2013, YamPRD2015, TaitPRL2020, FejerLIGO2021, VajentePRL2021}, in their commonly used approximate expressions [Eqs.~(\ref{mono1}) and (\ref{mono2}) in the Appendix], indicate that the amplitude spectral density of thermal noise in the silica monolayer is reduced by about $60\%$ relative to the as-deposited state (corresponding to a factor of $2.5$) after annealing at 500 °C, and by a further $60\%$ relative to the 500 °C sample after annealing at 900 °C. The results are confirmed for substrates other than silicon, such as fused silica and sapphire. These findings suggest that suppression of residual anisotropy --- which concomitantly lowers both the real part of the elastic constants and their associated loss factors --- can reduce, albeit to different extents, the thermal-noise contribution of individual materials within high-reflectivity multilayer coatings. At present, extending such thermal treatments to full high-reflectivity stacks is constrained by the thermal stability of the high-index material currently employed, which limits the accessible annealing temperatures.  More refined estimates, both for monolayers and heterostructures, require models that allow for anisotropic tensors with independent loss factors for different elastic moduli. In this respect, numerical modeling ---although demanding--- appears to be the only rigorous approach.

        \subsection{\label{Local}Local structure}

\subsubsection{\label{par}Structure on planes parallel to the surface}
\begin{figure*}
	\centering
	\includegraphics[width=13 cm]{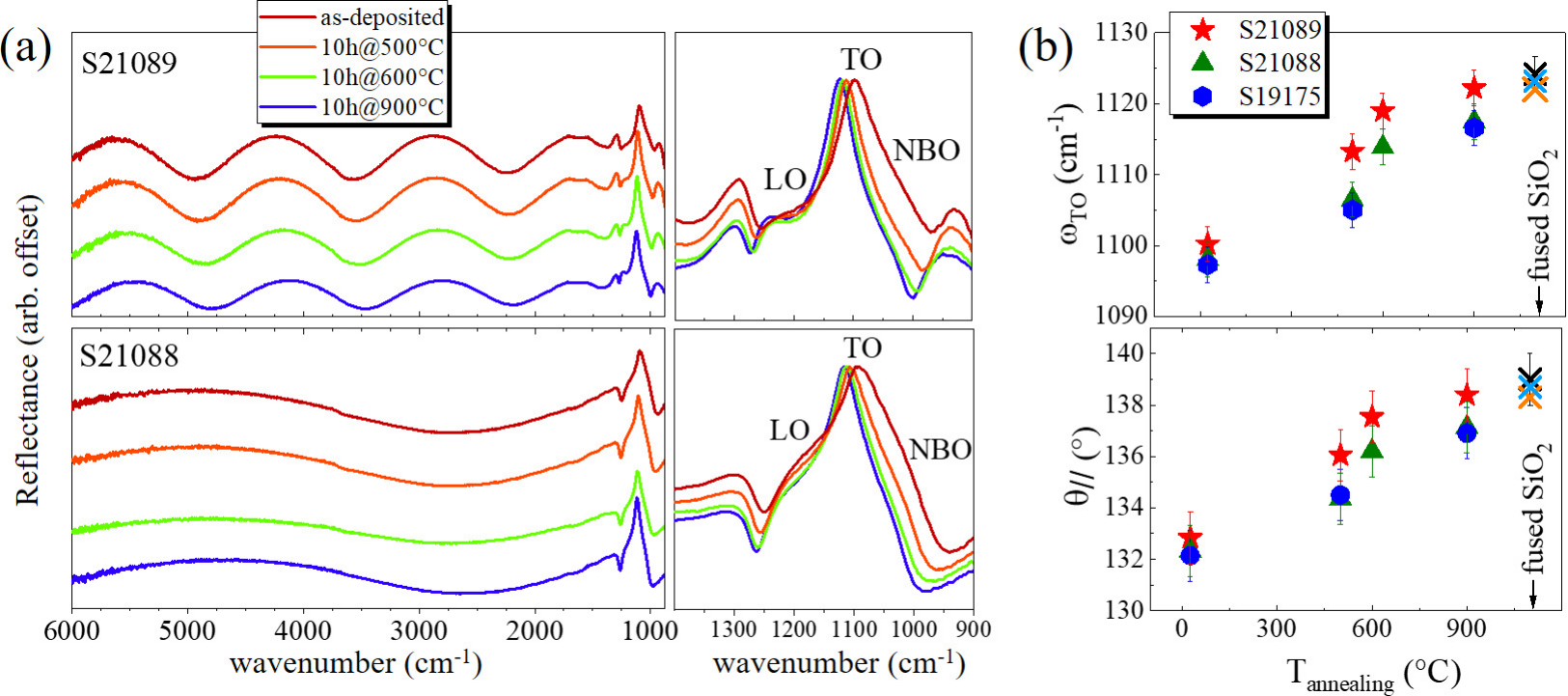}
	\caption{\small (a) SR-IR spectra of the samples S21089 and S21088 as-deposited and annealed at different temperatures, acquired at incidence angle $\theta$=30° in the wavenumber region of 6000-800 cm$^{-1}$, together with their enlargement in the absorption peaks region of 1400-900 cm$^{-1}$. (b) Frequency position of the TO mode, $\omega_{TO}$ (upper panel), and mean Si-O-Si bonding angle on the planes parallel to the surface, $\theta_{\parallel}$ (lower panel), as a function of the annealing temperature. The values of fused silica are reported for comparison with crosses (black: this work;  orange: Ref. \cite{TanOPE2005}; cyan: Ref. \cite{HiroseJNCS2006}).} \label{fig:spectraSR}
\end{figure*}

In the SR-IR spectra of samples S21089 and S21088 at different annealing temperatures [Fig.~\ref{fig:spectraSR}(a)] the TO component of the antisymmetric Si-O-Si stretching vibration stands out at $\sim$1100–1120 cm$^{-1}$ as the most intense signal, superimposed on an oscillating background. As recalled in Sec.~\ref{Optical}, this mode serves as a probe of the structure on surface-parallel planes, and acquisition at fixed incidence of $\theta$=30° ensures that the structure is probed in the two samples of different thicknesses up to an equal depth from the surface. According to the central force network model \cite{HosonoJAP1991, DeLosArcosVibSpect2021} the frequency position $\omega_{TO}$ of the TO mode depends on $\theta_{\parallel}$, the mean Si-O-Si bonding angle on planes parallel to the surface. Figure~\ref{fig:spectraSR}(b) indicates a blue shift of about 20 cm$^{-1}$ in $\omega_{TO}$, corresponding to an increase of about 5° in $\theta_{\parallel}$ on increasing the annealing temperature up to 900 °C. $\omega_{TO}$ could be determined with an uncertainty of about 2$\%$ because of the intense oscillating background present in the spectra. The result, confirmed in sample S19175, suggests that the structural properties and the changes induced by the post-deposition treatment are largely independent of the film thickness.  It is interesting to note that the values of $\omega_{TO}$ and $\theta_{\parallel}$ on increasing the annealing temperature align fairly well with the values of fused silica measured in the present work and consistent with literature data \cite{HiroseJNCS2006, TanOPE2005}.

\subsubsection{\label{perp}Structure in direction perpendicular to the surface}   

\begin{figure*}
	\centering
	\includegraphics[width=17 cm]{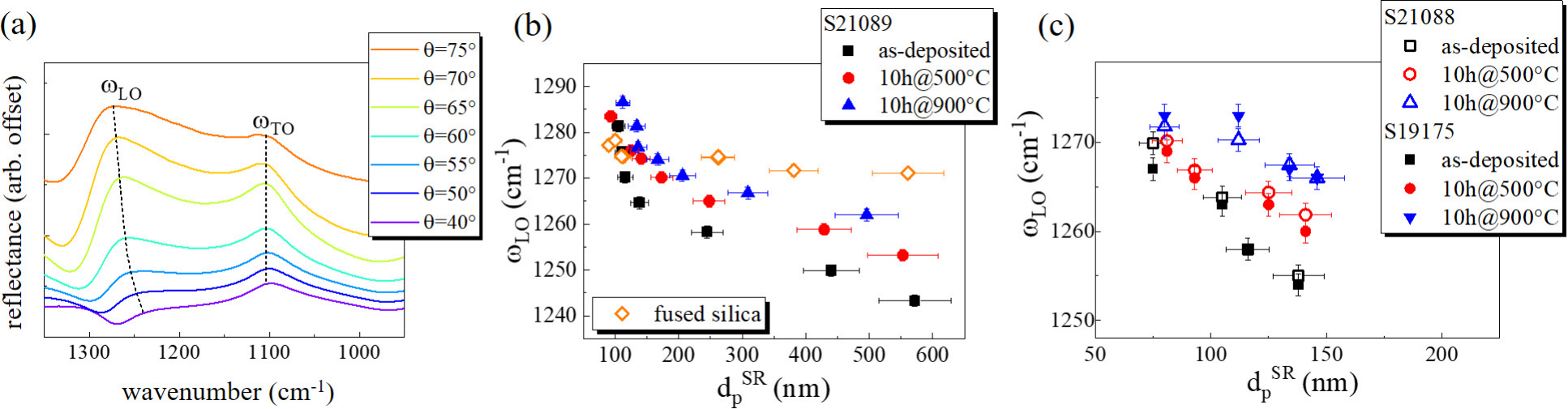}
	\caption{\small (a) SR-IR spectra of the as-deposited sample S21089, measured at different incidence angles $\theta$ using p-polarized radiation. (b) Peak position of the LO mode, $\omega_{LO}$, as a function of the radiation penetration depth, $d_{p}^{SR}$, in the sample S21089 and (c) the samples S21088 and S19175.} \label{fig:wLO}
\end{figure*}

\begin{table*}
	\caption{\small \label{tab2} Penetration depth $d_{p}^{SR}$ of the IR radiation of different wavenumber in SR-IR configuration, inside the as-deposited material. Estimated uncertainty is 10$\%$.} 
	\vspace{0.2 cm} 
	\centering{\scalebox{1}{\begin{tabular}{c | c c c c c c c c c} 
				& $d_{p}^{SR}$ (nm)& $d_{p}^{SR}$ (nm) & $d_{p}^{SR}$ (nm) & $d_{p}^{SR}$ (nm)& $d_{p}^{SR}$ (nm)& $d_{p}^{SR}$ (nm) & $d_{p}^{SR}$ (nm)& $d_{p}^{SR}$ (nm)& $d_{p}^{SR}$ (nm)\\ 
				cm$^{-1}$ & $\theta=30^{\circ}$ & $\theta=40^{\circ}$ & $\theta=50^{\circ}$ & $\theta=55^{\circ}$ & $\theta=60^{\circ}$ & $\theta=65^{\circ}$ & $\theta=70^{\circ}$ & $\theta=75^{\circ}$ & $\theta=80^{\circ}$\\ 
				\hline
				\hline
				1000 & 714 & 712 & 706 & 704 & 700 & 613 & 606 & 568 & 541 \\
				1100 & 356 & 352 & 344 & 340 & 334 & 333 & 330 & 327 & 321 \\
				1200 & 664 & 604 & 456 & 364 & 247 & 188 & 165 & 74 & 73 \\
				1250 & 676 & 572 & 245 & 138 & 116 & 111 & 106 & 75 & 75 \\
				\hline
				\hline
			\end{tabular}
	}}
\end{table*}

The polarization of IR radiation significantly influences the spectra in the region 1400--900 cm$^{-1}$. At fixed incidence, p-polarized radiation markedly enhances the relative intensity of the TO and LO bands compared to s-polarized radiation, while having a negligible effect on their frequencies (Fig.~\ref{20} in Appendix~\ref{AppC}). In constrast, varying the angle of incidence leaves $\omega_{TO}$ unchanged but causes a significant shift in the LO mode [Fig.~\ref{fig:wLO}(a)] \cite{WangJCP2003}.

Different incidence angles result in different values of penetration depth, depending also on $n$ and $k$ [Eq.~(\ref{eq:dpSR})]. Optical quantities in the absorption region were obtained from Kramers-Kr\"{o}nig (KK) transformation of SR-IR spectra of SiO$_{2}$ films on SiO$_{2}$ substrates, with perfectly flat background, deposited by LMA under identical conditions. The values of $n$ and $k$ obtained this way for the untreated and annealed material can be found in Table~\ref{T4} in Appendix~\ref{AppF}. They exhibit a strong frequency dispersion, and a good agreement with the value of $n$ measured by SE at higher frequency. Notice that this behavior reflects into a great frequency dependence of $d_{p}^{SR}$. As shown in Tab.~\ref{tab2} for the as-deposited material, at 1000--1100 cm$^{-1}$, characteristic range of the TO band, $d_{p}^{SR}$ is approximately independent of $\theta$ from 30° to 80°. Conversely, there is a significant dependence on $\theta$ at 1200--1250 cm$^{-1}$, characteristic range of the LO band. The values of $d_{p}^{SR}$ indicate that, by changing the radiation incidence, the LO band is capable of probing layers of film material with varying thicknesses from the surface. 

The frequency position $\omega_{LO}$ of the LO mode is associated with $\theta_{\perp}$, the average angle between tetrahedral silicate units along the direction perpendicular to the surface \cite{TanOPE2005}. 
$\omega_{LO}$ for the sample S21089 is reported in Fig.~\ref{fig:wLO}(b) as a function of $d_{p}^{SR}$, in the range of $\theta$ from 40° to 75°. Its value decreases over the penetration range of 70--600 nm, more markedly from 70 to 150 nm, suggesting that the film exhibits a structural gradient along the surface-normal direction, likely driven by a variation in the distribution of angles at bridging oxygens along the growth axis, formed during the deposition process. The data indicate that this gradient is particularly pronounced in the as-deposited film, and progressively reduces on increasing the annealing temperature up to 900°C, tending to the case of fused silica where the gradient is absent. 
Similar behavior is found in thinner films (samples S21088 and S19175) where reliable values of $\omega_{LO}$ have been obtained in the limited range 65° $< \theta <$ 80° [Fig.~\ref{fig:wLO}(c)]. Nonetheless, it is evident a decrease in the 70--150 nm penetration range, less pronounced with increasing the annealing temperature: $\sim$15 cm$^{-1}$ in the untreated material, and $\sim$5 cm$^{-1}$ after annealing at 900°C. Therefore, in all samples, within the first hundred nanometers from the surface the distribution of Si-O-Si angles is consistent with the picture of a material grown with compressive stress and a lower surface density. The post-deposition heat treatment reduces the heterogeneity in the growth direction by increasing the average Si-O-Si bonding angle, thus leading to an average density decrease.

\subsubsection{\label{chem}Chemical defects}

\begin{figure*}
	\centering
	\includegraphics[width=15 cm]{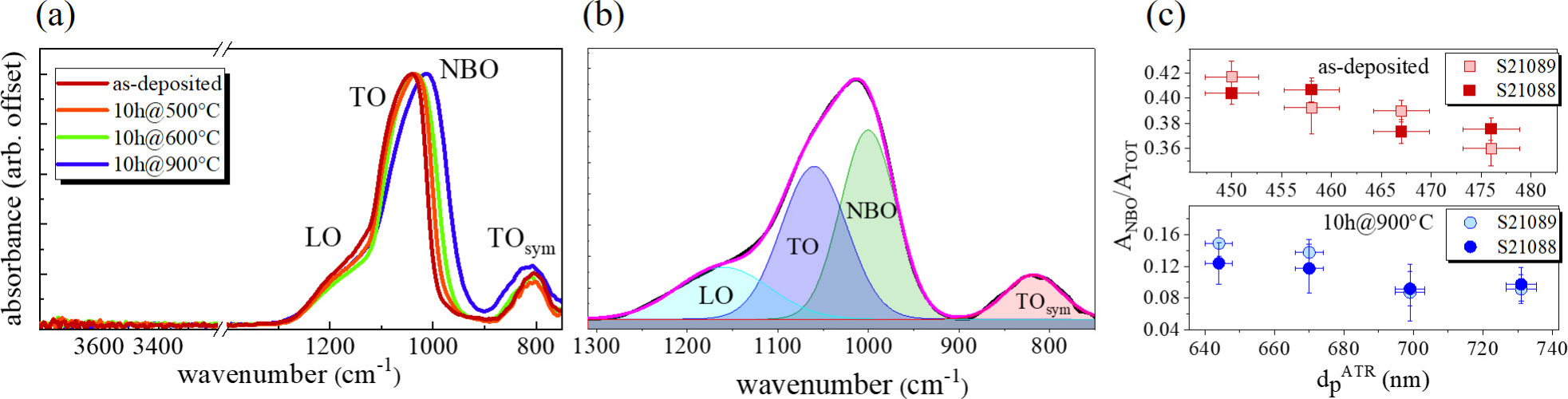}
	\caption{\small (a) ATR-IR spectra of the sample S21089, as-deposited and annealed at different temperatures, acquired in the range 3800--700 cm$^{-1}$. (b) ATR-IR profile in the as-deposited material fitted by four Gaussian components. (c) Relative area of the NBO band as a function of the radiation penetration depth, $d_{p}^{ATR}$, in the samples as-deposited (upper panel) and annealed at 900 °C (lower panel).} \label{fig:spectraATR}
\end{figure*} 

The reduced penetration depth of the IR radiation in ATR configuration accounts for the absence of interference fringes in the spectra, shown in Fig.~\ref{fig:spectraATR}(a). In the absorption peaks region they exhibit a maximum intensity at frequencies where the stretching mode of NBO-containing groups resonates ($\sim$1000 cm$^{-1}$). Thus, ATR-IR spectra enable a convenient evaluation of NBO-related chemical defects in the silica structure. These defects are here attributed to the presence of Si-O$^{-}$ groups inside the material rather than to the presence of Si-OH groups from moisture capture, since no OH-stretching signal around 3600 cm$^{-1}$ is detected.

A quantitative estimate of the NBO structures is obtained by fitting the ATR-IR profile in the range 1310--750 cm$^{-1}$ with four Gaussian components, reproducing the LO, TO, and NBO bands, plus the symmetric stretching mode of the Si-O-Si unit [Fig.~\ref{fig:spectraATR}(b)]. The relative area of the NBO component in the as-deposited material and that annealed at 900 °C is reported as a function of $d_{p}^{ATR}$ in Fig.~\ref{fig:spectraATR}(c). Along with a significant reduction in NBO structures after annealing, the data show a slight increase in these defects at smaller radiation penetration depths, suggesting that a higher concentration of NBOs is localized near the surface. This indicates that an additional source of inhomogeneity may be related to the material's ability to capture cations, likely Ar$^{+}$ involved in the deposition process. These cations are intuitively expected to induce the formation of Si-O$^{-}$ groups, particularly near the surface. Therefore, anisotropy is also associated to the presence of chemical defects that persist in the film even after high temperature treatment.

\subsection{\label{Thickness}Film thickness}

In thin films such as sample S21088, both BLS and SR-IR spectra exhibit features resulting from wave interference, useful for determining the film thickness. In SR-IR spectra the interfering waves are IR electromagnetic radiation, whereas in BLS spectra they are acoustic waves. Despite the differing nature of these waves, the dependence of the interference phenomenon on $h$ can be exploited as explained hereafter.

\begin{figure*}
	\centering
	\includegraphics[width=13 cm]{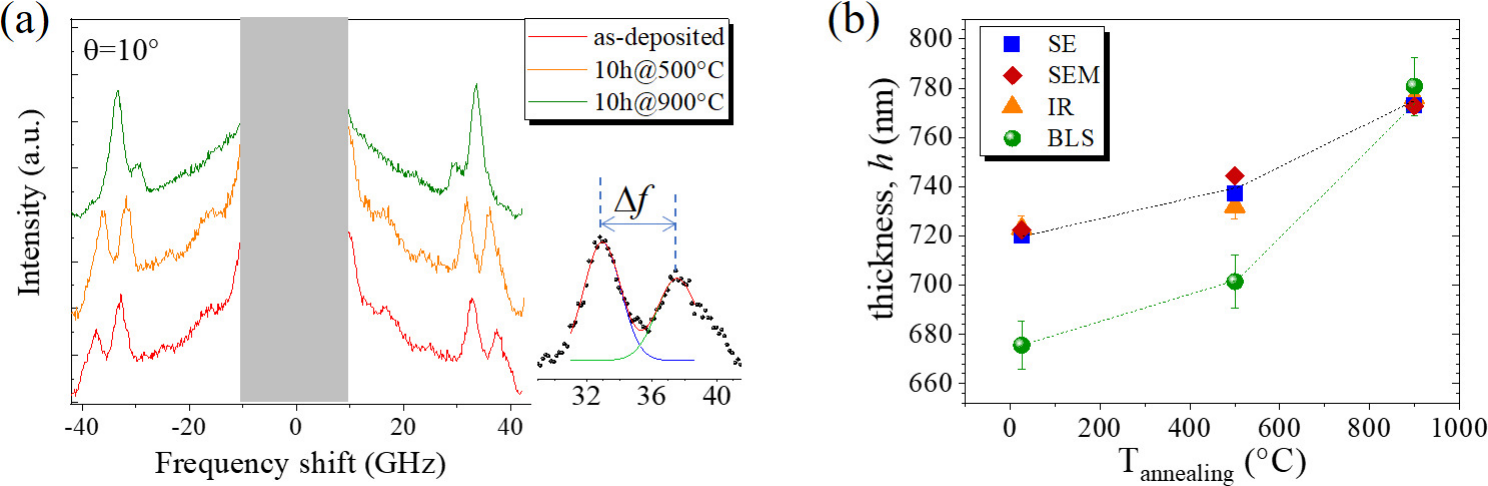}
	\caption{\small (a) Polarized BLS spectra at quasi-normal incidence of the sample S21088, as-deposited and annealed at 500 °C and 900 °C. The frequency separation $\Delta f$ of the discrete peaks in the as-deposited film are shown on the right. Solid lines are a fit with Gaussian functions of equal width. (b) Thickness $h$ of the sample S21088 as a function of the annealing temperature, estimated by different techniques as indicated in the legend. Dotted lines are a guide for the eyes.}
	\label{fig:thick}
\end{figure*}

On one hand, the interference pattern in SR-IR spectra arises from the difference in optical distance traveled by light of each wavelength $\lambda$ reflected at the two film interfaces. Taking into account the dependence on $\lambda$ of the material's refractive index, the film thickness can be determined from regions of the spectrum where individual interference fringes can be resolved, according to $h_{IR}=N/2[n(\lambda_{1})/\lambda_{1} -n(\lambda_{2})/\lambda_{2}]$, where $N$ is the number of fringes observed from wavelength $\lambda_{1}$ to $\lambda_{2}>\lambda_{1}$, and $n(\lambda_{1,2})$ is the refractive index at these wavelengths \cite{HuibersLANG1997}. For sample S21088, $h_{IR}$ is calculated from Fig.~\ref{fig:spectraSR}(a) with $\lambda_{1}$=1667 nm and $\lambda_{2}$=2500 nm by using $n(\lambda)$ measured by SE (Fig.~\ref{fig:n} in Appendix~\ref{AppA}). On the other hand, the multiple peaks in BLS spectra at quasi-normal incidence [Fig.~\ref{fig:thick}(a)] arise from the scattering of light by discrete acoustic modes normal to the film surface, that exist as standing waves because they satisfy the condition for constructive interference of propagating acoustic waves reflected at the two film interfaces. The film thickness can be determined from the frequency separation $\Delta f$ of these peaks, according to $h_{BLS}=v_{LB}/2\Delta f$. $\Delta f$ is obtained by a simple data analysis, as illustrated in Fig.~\ref{fig:thick}(a). The velocity $v_{LB}$ of longitudinal acoustic waves traveling in the surface-normal direction is taken from sample S21089, as it has been demonstrated that the acoustic properties of the material are independent of film thickness. 

In Fig.~\ref{fig:thick}(b), $h_{IR}$ and $h_{BLS}$ are reported as a function of the annealing temperature, and compared with the values provided with nanometer-level accuracy by SEM and SE. An excellent agreement of $h_{IR}$ is observed at all temperatures. The data, while confirming the previously established result \cite{GranataCQG2020} that annealing at 500 °C causes a thickness increase, reveal that the effect becomes nearly three times larger at 900 °C (+7.4 $\%$ relative to the as-deposited film) and correlates with a decrease in material's density. Notably, assuming that the mass and surface area remain unchanged upon annealing, this thickening fully accounts for the observed density reduction. Unexpectedly, the agreement of $h_{BLS}$ is limited to the highest temperature. Specifically, in the untreated material and that annealed at 500 °C, $h_{BLS}$ underestimates $h$ by $\sim$5--6$\%$, while the disagreement vanishes in the material annealed at 900 °C, where elastic anisotropy is drastically reduced. The result is confirmed in sample S19175 (Fig.~\ref{21} in Appendix~\ref{AppC}), ruling out any measurement artifact. It should be noted that only optical properties of the material are required to calculate $h_{IR}$. Conversely, the calculation of $h_{BLS}$ relies on acoustic parameters ($v_{LB}$, $\Delta f$), which are linked through a simplified model assuming a homogeneous film where specific boundary conditions at the interfaces must be fulfilled by acoustic waves (Sec.~\ref{Acoustic}). One plausible reason for the observed mismatch between $h_{BLS}$ and the actual film thickness in both the as-deposited and 500 °C-annealed samples is the simplicity of the adopted model, which does not account for the structural heterogeneity along the surface-normal direction revealed by the IR analysis. The discrepancy can therefore be interpreted not merely as a limitation of the model, but as a signature of inhomogeneity, directly linked to elastic anisotropy in our samples. Consistently, it disappears in the 900 °C-annealed sample, where both the structural gradient and elastic anisotropy are strongly suppressed.

\section{\label{Conclusions}Conclusions}  

The elastic and structural properties of IBS thin films have so far been assumed isotropic without direct verification, despite their central role in governing coating thermal noise in GW detectors. Any unaccounted anisotropy in these properties may contribute to the persistent gap between predicted and measured noise \cite{GranataCQG2020, GranataPRD2016} and limit future sensitivity improvements. Silica remains a cornerstone material for mirror coatings, making its study \textit{per se} crucial and, at the same time, a gateway to uncovering a fundamental and previously overlooked link between elastic and structural anisotropy and mechanical dissipation.  Establishing this link can feed new predictive models, with direct impact on both coating design and processing strategies.

Addressing this issue requires an experimental approach that goes beyond current characterization practices. This study combines macroscopic information from BLS measurements with structural insight from IR spectroscopy. BLS provides a complete characterization of the elastic response --- difficult, if not impossible, to achieve with conventional mechanical testing --- revealing that ion-beam sputtered SiO$_{2}$ exhibits cylindrical symmetry and a sizeable out-of-plane compressive anisotropy ($+ 6\%$) in the as-deposited state. At the atomic scale, IR reflectivity measurements, including depth profiling of bridging and non-bridging oxygens, confirm in-plane structural isotropy while uncovering a gradient along the growth axis, consistent with compressive stress and lower surface density. Annealing leads to a softening of all elastic constants. Critically, however, the anisotropy remains unchanged after the post-deposition heat treatment currently used in ground-based detectors (10 hours at 500 °C), and is almost completely removed only after annealing at the highest temperature tested (900 °C). The structural gradient --- not directly resolvable by BLS --- is likewise strongly reduced at high temperature, consistent with the trend toward restoring macroscopic isotropy revealed by BLS, and with both the elastic response and the structural profile converging toward those of fused silica.

Beyond their independent diagnostic power, both BLS and IR yield sub-micrometric thickness estimates that can be benchmarked against SEM and SE. IR, SE and SEM consistently show that annealing drives a directional recovery, with pronounced film expansion ($>7\%$) exclusively along the growth axis, while the in-plane dimensions remain essentially unchanged. By contrast, the BLS-based thickness agrees with IR, SE, and SEM estimates only at the highest annealing temperature, where elastic isotropy is nearly restored. The discrepancy at lower temperatures thus emerges as a direct marker of structural inhomogeneity, which correlates with elastic anisotropy and vanishes as the material approaches isotropic behavior. 

In addition, our study has direct implications for how the elastic properties of mirror-coating materials are determined. BLS probes the real part of the elastic constants at GHz frequencies, clearly resolving their dependence on annealing temperature. By contrast, the standard technique in the field, GeNS, operates at kHz frequencies --- closer to the GW-relevant band --- but with lower accuracy and no sensitivity to elastic anisotropy. The comparison shows that relaxation processes between kHz and GHz contribute negligibly to the isotropic-effective Young's modulus, validating the high-frequency BLS values as reliable proxies for the low-frequency response. Crucially, this study enables a robust reassessment of the widely used GeNS Poisson’s ratio values for silica, removing a systematic bias in current inputs to coating-thermal-noise estimates. Using the more realistic values determined here will directly improve future simulations.

Our study challenges the long-standing assumption of elastic and structural isotropy in amorphous materials for GW mirror coatings, providing quantitative evidence of anisotropy in ion-beam sputtered SiO$_{2}$ and linking it to mechanical dissipation. A coherent physical picture emerges in which the downshift toward fused-silica\textit{-like} isotropic properties translates, within isotropic-equivalent modeling, into a substantial reduction of thermal noise, by about a factor of 2.5. Crucially, our results indicate that restoring isotropy --- and thus reducing thermal noise --- requires annealing temperatures significantly above the current standard.

The implications extend well beyond silica, offering a concrete pathway to unlock sensitivity gains in future GW detectors. Low-loss materials are essential not only for upgrading current facilities but also for designing next-generation detectors, including the Einstein Telescope \cite{ET}, where mirror coatings may depart from the conventional high/low-index doublet paradigm and may involve other materials than currently used. In this broader context, anisotropy-aware characterization is indispensable for assessing candidate materials, guiding the evolution of deposition and post-processing strategies, and refining coating design. Achieving quantitatively reliable thermal-noise predictions will also require sustained efforts toward numerical modeling frameworks capable of incorporating transverse isotropy tensors with independent loss factors and, when relevant, thickness variations induced by anisotropy suppression.

Beyond gravitational-wave detection, the thin-film characterization approach presented in this work may prove valuable in fields such as information and communication technologies and optoelectronics, where precise control of elastic and structural properties is likewise critical.

\begin{acknowledgments}
	G. Cagnoli, M. G., and D. H. gratefully acknowledge the support of the French Agence Nationale de la Recherche (ANR) through grant n. ANR-18-CE08-0023 to the ViSIONs project. 
	G. Carlotti, S. Corezzi, and P. S., acknowledge support from the European Union -NextGenerationEU, Mission 4, Component 2, under the Italian Ministry
	of University and Research (MUR) National Innovation Ecosystem Grant
	ECS00000041 - VITALITY - CUP J97G22000170005. G. Carlotti acknowledges support from Fondazione Cassa di Risparmio Terni e Narni (Project n. FCTR24UNIPG).
\end{acknowledgments}

\section*{Author Contributions}
S. Corezzi and P. S. conceived and supervised the work. D. H. and M. G. provided the samples. B. B., S. Corezzi, and G. Carlotti designed and performed the BLS study. B. B. and P. S. designed and performed the IR study. M. M. and S. Colace performed and analyzed the SE measurements. G. F. performed and analyzed the XRR measurements. A. D. M. performed and analyzed the SEM measurements. M. B. and M. C. provided rosources. B. B. and S. Corezzi wrote the manuscript draft. B. B., S. Corezzi, and P. S. wrote the paper with contributions from G. Carlotti, G. Cagnoli, M. M., M. G., and L. S.. All authors read and reviewed the data and manuscript.


\appendix

\section{\label{AppA}Characterization data}
\begin{figure}[H]
		\centering
		\includegraphics[width=6.5 cm]{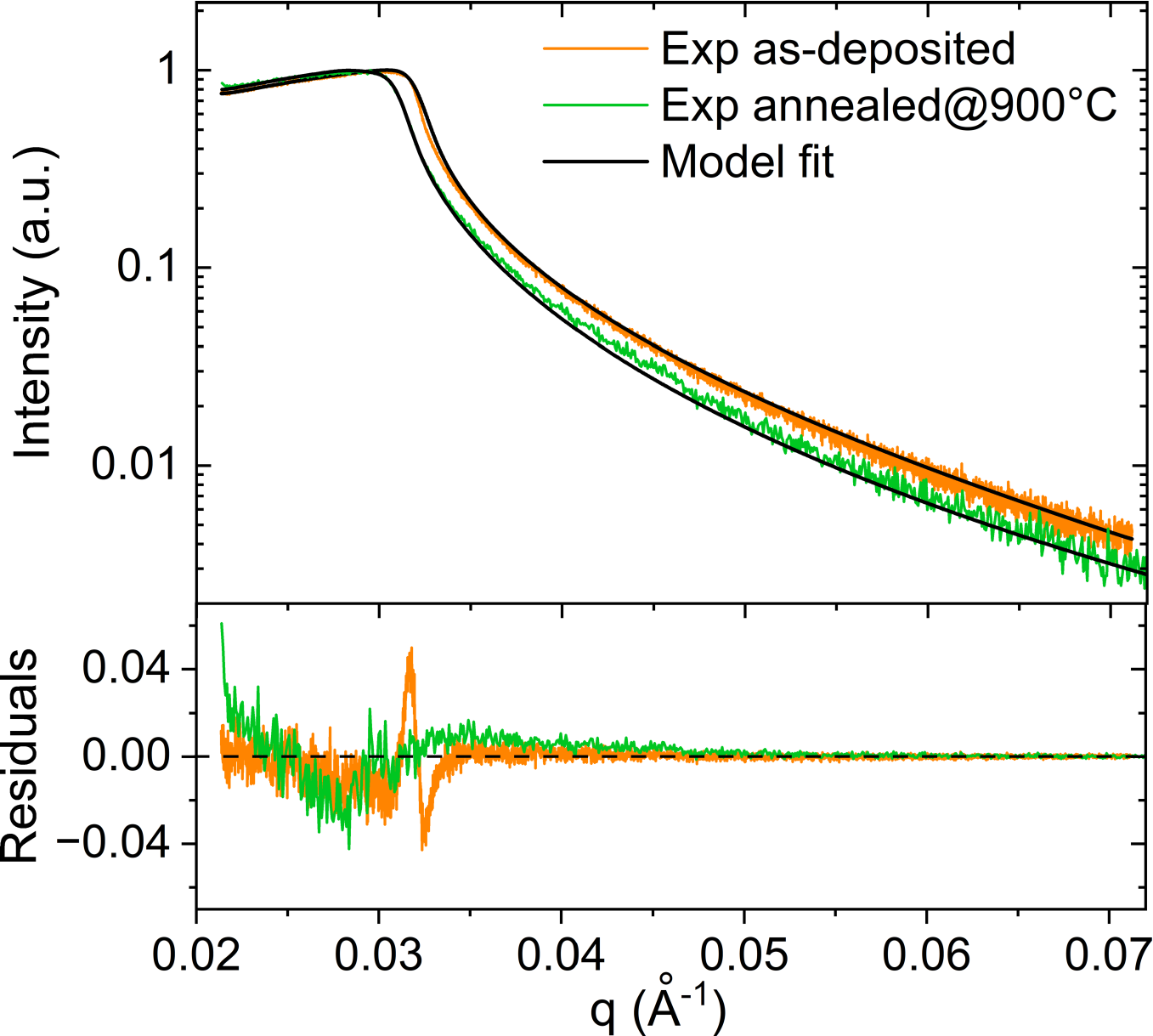}
		\caption{\small Upper panel: XRR spectra of the sample S21088 as-deposited and annealed at 900 °C (colored lines), together with their best-fit curve (black lines). Bottom panel: Residuals between the experimental data and the fit curve.}
		\label{fig:XRR}
\end{figure}

\begin{figure}[H]
		\centering
		\includegraphics[width=6.5 cm]{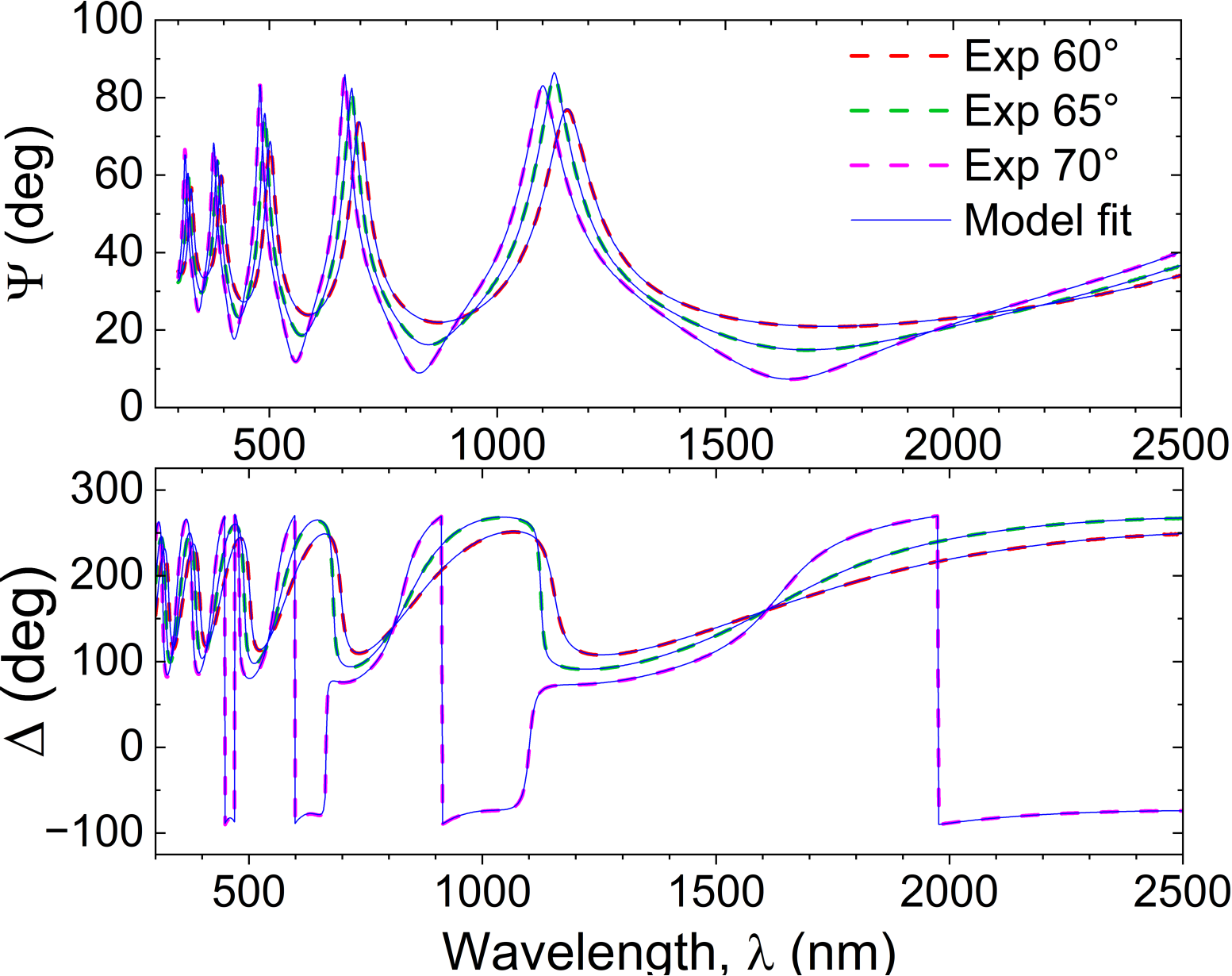}
		\caption{\small Spectroscopic ellipsometry (SE). Wavelength dependence of the angles $\Psi$ and $\Delta$ in the as-deposited sample S21088, at three values of the incidence angle θ. The fit curve is generated by the isotropic model using a Cauchy function supplemented with a Pole oscillator.
		\label{fig:SE}}
\end{figure}

\begin{figure}[H]
		\centering
		\includegraphics[width=6 cm]{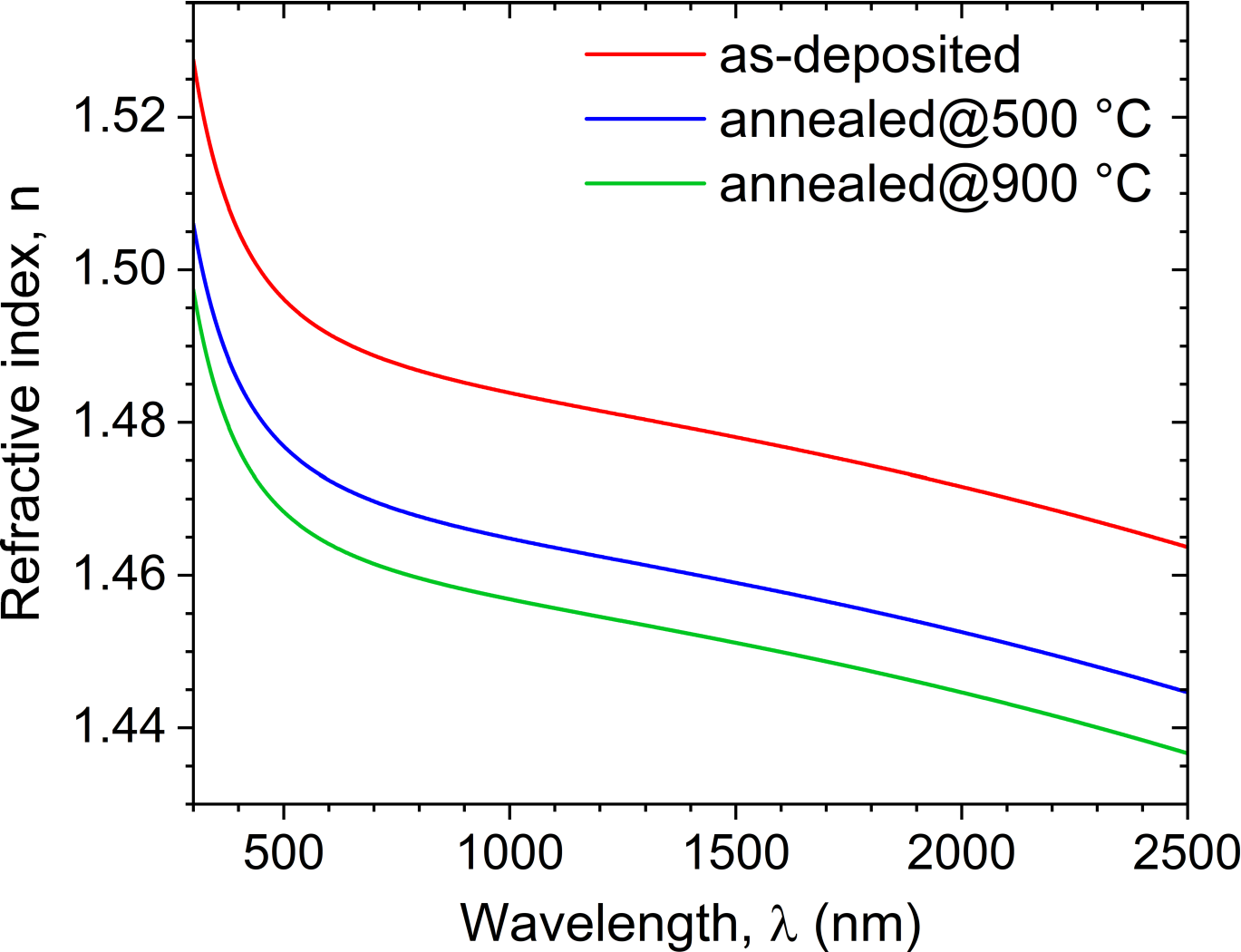}
		\caption{\small Wavelength dependence of the refractive index, $n$, obtained by reproducing the SE experimental data ($\Psi, \Delta$) with the isotropic model, for the sample S21088 in the as-deposited state and after annealing at 500 °C and 900 °C.
		\label{fig:n}}
\end{figure}

\begin{figure} [H]
	\centering
	\includegraphics[width=8 cm]{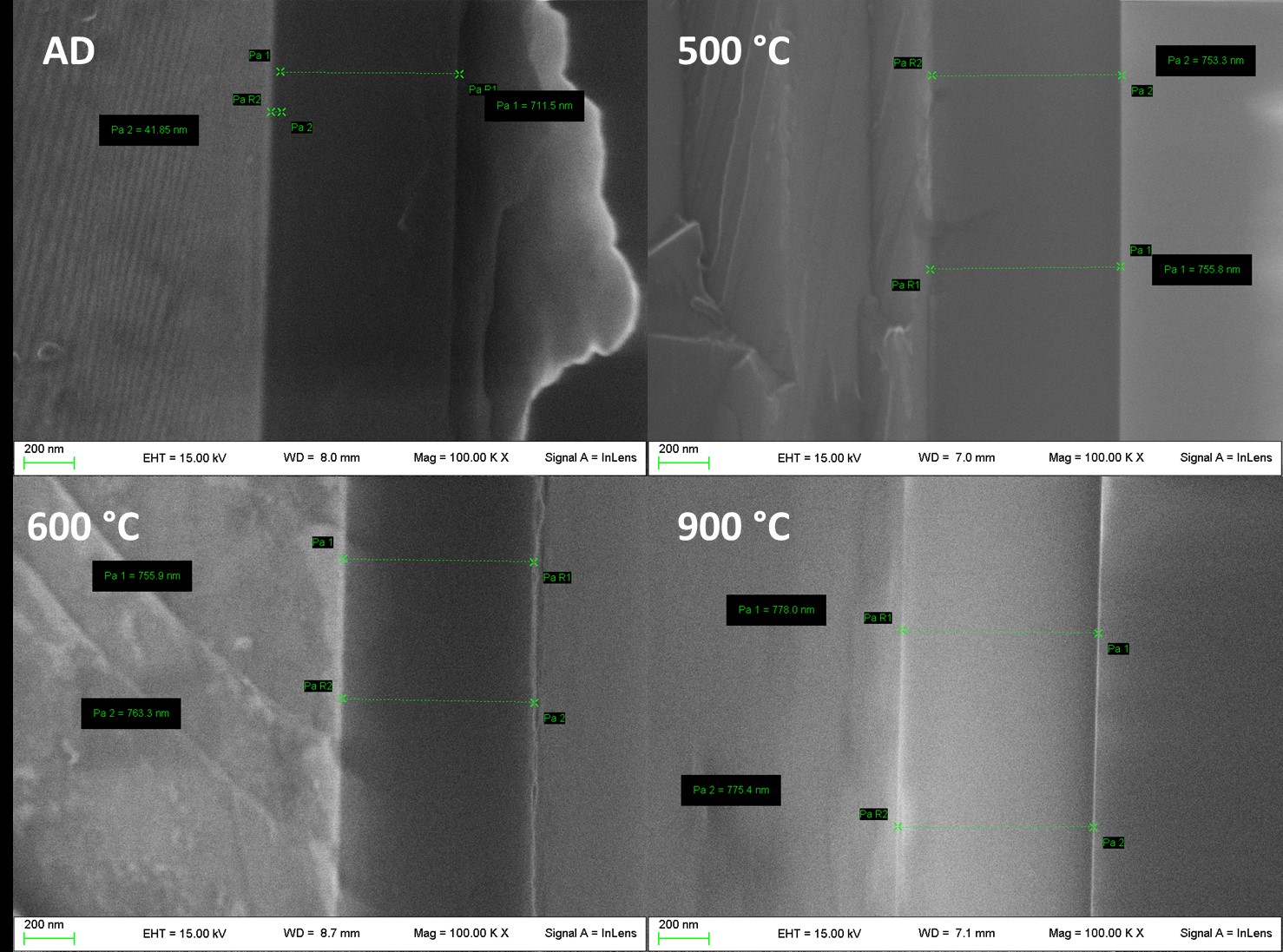}
	\caption{\small SEM images of the sample S21088 in the as-deposited state and after 10-h annealing at 500 °C, 600 °C and 900 °C. The thin layer of SiO$_{2}$ deposited on the thick silicon substrate is visible. Scale bar is 200 nm.}
	\label{fig:SEM}
\end{figure}

This appendix collects characterization data for sample S21088, including measurements obtained by XRR (Fig.~\ref{fig:XRR}), SE (Figs.~\ref{fig:SE} and \ref{fig:n}), and SEM (Fig.~\ref{fig:SEM}). Detailed descriptions of the experimental methods and discussion of the results are reported in Sec. II B of the main text.


\section{\label{AppB}Elasticity in a transversely isotropic semi-infinite medium}

        \subsection{\label{B1}Elasticity tensor}

The elastic behavior of a transversely isotropic medium (cylindrical symmetry; space group: $D_{\infty h}$)  is described by five independent elastic constants. With axis $x_{3}$ perpendicular to the film surface, and axes $x_{1}$ and $x_{2}$ in plane, these constants are $c_{11}$, $c_{33}$, $c_{44}$, $c_{66}$, $c_{13}$. The fourth-order elasticity tensor $c_{ijkl}$, symmetric under index exchange $i \leftrightarrow j$, $k \leftrightarrow l$, and $(ij) \leftrightarrow (kl)$, can be represented in the compact Voigt notation as a symmetric (6 x 6) matrix given by: 
\begin{equation}
	c_{ij} =
	\begin{pmatrix}
		c_{11} & c_{12} & c_{13} & 0 & 0 & 0 \\
		c_{12} & c_{11} & c_{13} & 0 & 0 & 0 \\
		c_{13} & c_{13} & c_{33} & 0 & 0 & 0 \\
		0 & 0 & 0 & c_{44} & 0 & 0 \\
		0 & 0 & 0 & 0 & c_{44} & 0 \\
		0 & 0 & 0 & 0 &0 & c_{66}
	\end{pmatrix}
	\label{matrix}
\end{equation}
with the symmetry constraint $c_{12}=c_{11}-2c_{66}$. Physical meaning: the coefficients $c_{11}$ and $c_{33}$ (longitudinal stiffness in the isotropic plane and along the symmetry axis) describe the resistance to horizontal and vertical compression; $c_{44}$ and $c_{66}$ (vertical and horizontal shear modulus) describe the resistance to shearing along planes containing the symmetry axis and along the isotropic plane; $c_{13}$ (coupling constant between axial and in-plane strains) describes how vertical compression induces lateral deformation and vice versa; $c_{12}$ (coupling constant between deformations in the isotropic plane) is not independent, and can be computed from $c_{11}$ and $c_{66}$.

         \subsection{\label{B2}Relation to bulk and surface wave velocities}

\textit{(i)} The phase velocities $v$ of bulk acoustic waves propagating in an arbitrary direction within a transversely isotropic half-space is governed by the Christoffel equation
\begin{equation}
	\det\!\left[\Gamma_{ik} - \rho v^2 \delta_{ik}\right] = 0
	\label{bulk}
\end{equation}
where $\Gamma_{ik}(\mathbf{n}) = c_{ijkl} n_j n_l$ is the Christoffel matrix build with the elastic constants in Eq.~(\ref{matrix}), $\mathbf{n} = (n_1, n_2, n_3)$ the unit propagation vector, and $\rho$ the mass density. By solving Eq.~(\ref{bulk}) yields three bulk wave modes, i.e. longitudinal, shear vertical, and shear horizontal (SH), whose velocities depend on the propagation angle ($\alpha$) relative to the symmetry axis. Special cases of propagation include those in Eq. 2 of the main text, namely:
\begin{eqnarray}
	\mathrm{along\ symmetry\ axis\ (x_{3})} & \rightarrow & v_{LB}(\alpha=0^{\circ})=\sqrt{\frac{c_{33}}{\rho}}\nonumber \\
	\mathrm{in\ isotropic\ plane\ (e.g.,\ x_{1})} & \rightarrow &v_{LM}=\sqrt{\frac{c_{11}}{\rho}}~; \nonumber \\ 	 & & v_{SHM}=\sqrt{\frac{c_{66}}{\rho}}~. 
	\label{v}
\end{eqnarray}
\noindent
Although the constants $c_{44}$ and $c_{13}$ do not appear in the above special cases, according to Eq.~(\ref{bulk}) the velocity of longitudinal bulk waves for oblique directions, $v_{LB}(\alpha \neq 0^{\circ})$, depends explicitly on $c_{11}$, $c_{33}$, $c_{13}$, $c_{44}$, and $\rho$.

\textit{(ii)} The phase velocity $v_R$ of the Rayleigh-wave, propagating along a surface direction (e.g., $x_1$), follows from solutions to the Christoffel equation that are exponentially decaying with depth ($x_3 \rightarrow \infty$), and by imposing the traction-free boundary condition at the surface. This leads to a secular equation depending on $c_{11}, c_{33}, c_{13}, c_{44}$, and $\rho$, whose smallest positive root (less than the slowest bulk shear velocity) defines the Rayleigh-wave velocity $v_R$.
The analytic relation between $v_{RW}$ and elastic constants, derived by Dobrzynski and Maradudin \cite{MaradudinPRB1976}, is the following: 
\begin{equation}
	c_{33}(v_{RW}^{2} - \dfrac{c_{44}}{\rho}) (v_{RW}^{2} - \dfrac{c_{11}}{\rho}  + \dfrac{c_{13}^2}{c_{33}\rho})^2=c_{44}v_{RW}^{4}(v_{RW}^{2} - \dfrac{c_{11}}{\rho})~.
	\label{eq:Maradudin}
\end{equation} 

        \subsection{\label{B3}Procedure for complete determination of elastic moduli}

According to Eqs.~(\ref{v}), with the material density known, the elastic constants $c_{11}$ and $c_{66}$ are directly determined by measuring the velocities $v_{LM}$ and $v_{SHM}$ of longitudinal and shear-horizontal waves propagating parallel to the surface, while $c_{33}$ is determined by measuring the velocity $v_{LB}(\alpha=0^{\circ})$ of longitudinal waves propagating along the normal-to-surface direction. The remaining independent constants, $c_{13}$ and $c_{44}$, can be evaluated from the Rayleigh surface-wave velocity $v_{RW}$ and the longitudinal bulk waves velocity $v_{LB}(\alpha \neq 0^{\circ})$ for different oblique directions. In fact, both $v_{RW}$ and $v_{LB}$ for each propagation angle depend on $c_{11}, c_{33}, c_{13}, c_{44}$, and $\rho$.
Thus, once $c_{11}$, $c_{33}$ and $\rho$ are measured directly, only $c_{13}$ and $c_{44}$ remain unknown. Their values can be obtained by jointly fitting the theoretical predictions of the Christoffel [Eq.~(\ref{bulk})] and Maradudin [Eq.~(\ref{eq:Maradudin})] equations to the experimental wave velocities. This procedure uniquely determines the elastic constants $c_{44}$ and $c_{13}$, leading to complete determination of the material's elastic behavior.


\section{\label{AppC}Additional BLS and IR data}      
\begin{figure} [H]
	\centering
	\includegraphics[width=8 cm]{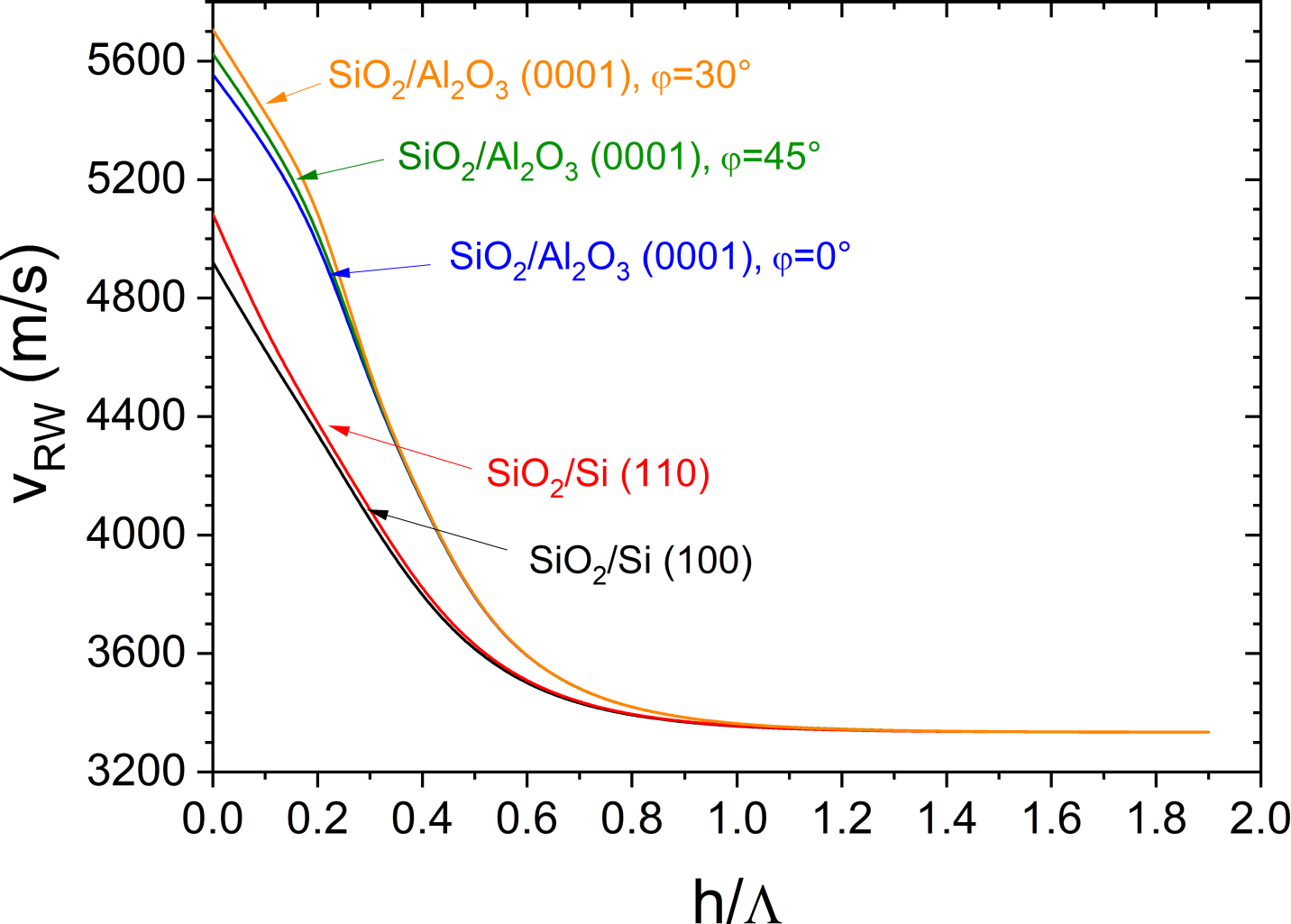}
	\caption{\small Numerical simulation of the Rayleigh-wave phase velocity, $v_{RW}$, for a SiO$_{2}$ thin film deposited on (100)-Si or (0001)-Al$_{2}$O$_{3}$ (C-plane sapphire) substrate, as a function of the ratio $h/\Lambda$ between film thickness, $h$, and phonon wavelength, $\Lambda$. The simulation is presented for two(three) propagation directions, labeled by the in-plane angle $\phi$ relative to the $x$($a_{1}$)-axis of silicon(sapphire). It can be seen that the influence of the substrate is negligible when the condition $h/\Lambda \gtrsim 1$ is fulfilled.}
	\label{14}
\end{figure}

\begin{figure*}
	\centering
	\includegraphics[width=1\linewidth]{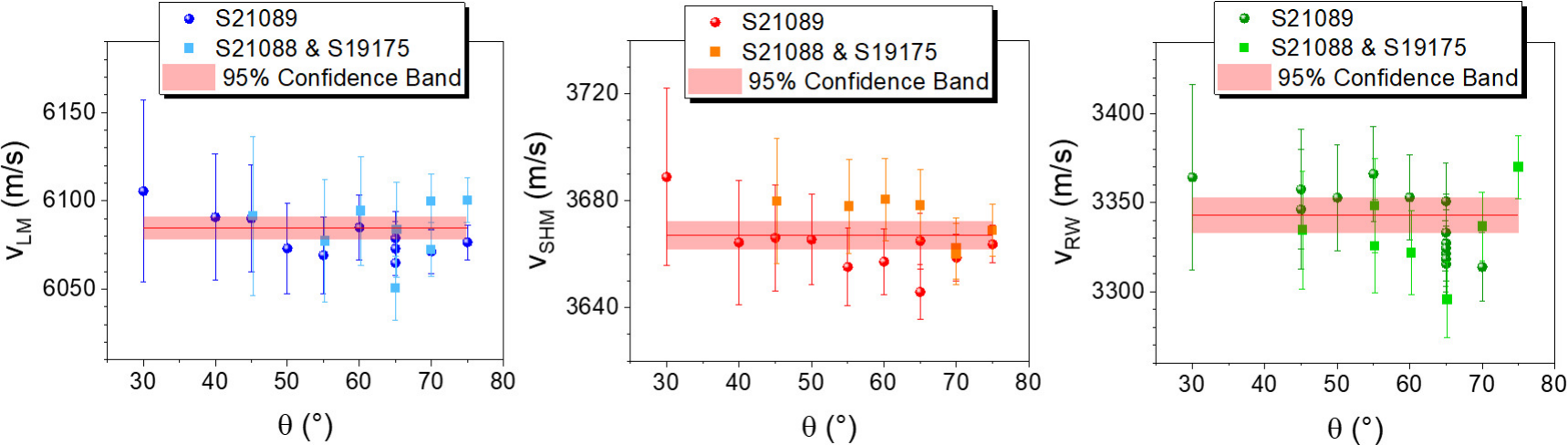}
	\caption{\small Phase velocity of surface-like acoustic waves ---longitudinal mode ($v_{LM}$), shear horizontal mode ($v_{SHM}$), and Rayleigh surface wave ($v_{RW}$) --- as a function of the incidence angle $\theta$, in films of thickness 2.5 $\mu$m (S21089) and 720 nm (S21088, S19175) in the as-deposited state. Solid lines are the best-constant fit of the data. All three modes are nondispersive, and the velocities measured in films of different thickness are identical within errors.}
	\label{15}
\end{figure*}

\begin{figure*}
	\centering
	\begin{minipage}{17.5 cm}
		\centering
		\includegraphics[width=17 cm]{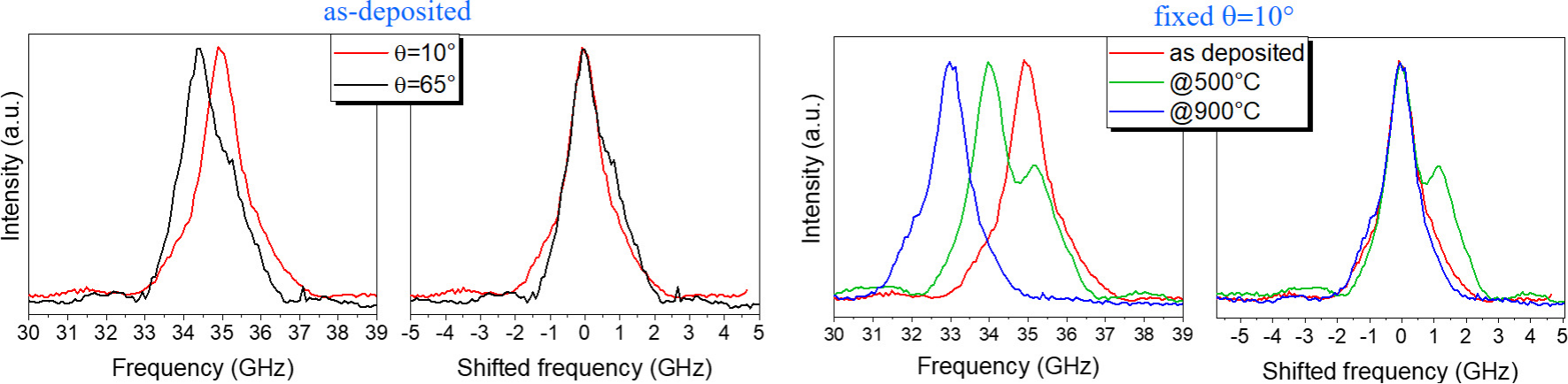}
	\end{minipage}
	\caption{\small BLS spectra in the region of longitudinal bulk-like waves. (Left) Comparison between polarized BLS spectra at $\theta$=10° and 65° of the as-seposited sample S21089. (Right) Comparison between polarized BLS spectra at $\theta$=10° of sample S21089, in the as-deposited state and after annealing at 500 °C and 900 °C. To highlight differences between the spectra, in both cases the comparison is also shown by shifting each spectrum by its frequency of maximum intensity.
		\label{16}}
\end{figure*}

\begin{figure}[H]
	\centering
	\includegraphics[width=7 cm]{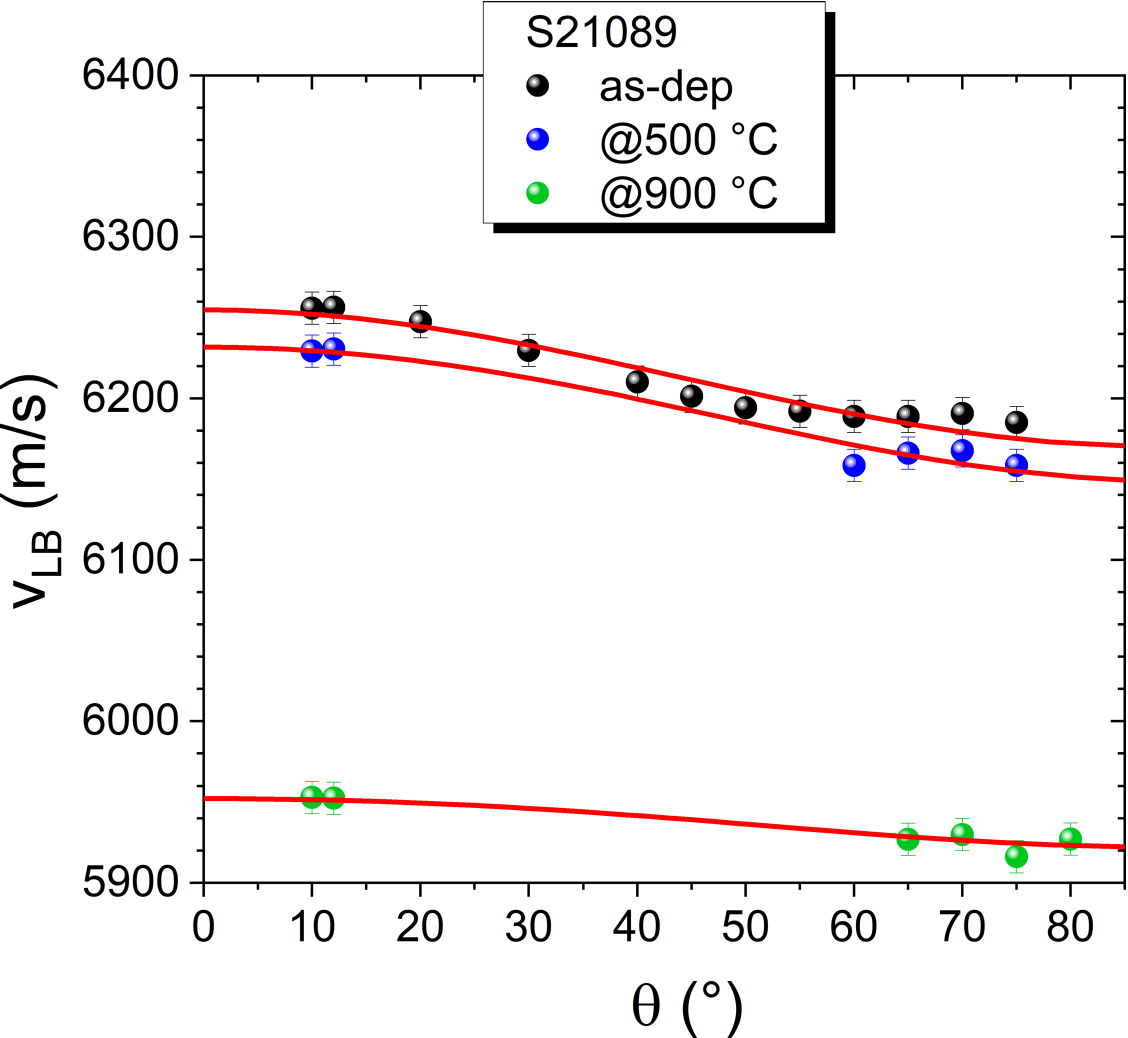}
	\caption{\small Phase velocity $v_{LB}$ of the longitudinal bulk-like wave, as a function of the incidence angle $\theta$, in the 2.5 $\mu$m-thick film in different states. Solid lines are numerical simulation curves calculated from the measured values of $c_{11}$, $c_{33}$, $\rho$ and the best-fit values of $c_{44}$, $c_{13}$.}
	\label{17}
\end{figure}

This appendix collects additional BLS (Figs.~\ref{16} and \ref{17}) and SR-IR spectra (Fig.~\ref{20}), along with data on acoustic velocities obtained from numerical simulations (Fig.~\ref{14}) and BLS experiments (Figs.~\ref{15},\ref{17}, and \ref{19}), and data of film thickness (Fig.~\ref{21}) derived from BLS measurements. The figures included here complement the results presented in the main text, providing a more complete representation of the experimental data, and illustrating trends and features discussed in Sec. IV of the article.   

\FloatBarrier

\begin{figure} [H]
	\centering
	\includegraphics[width=8 cm]{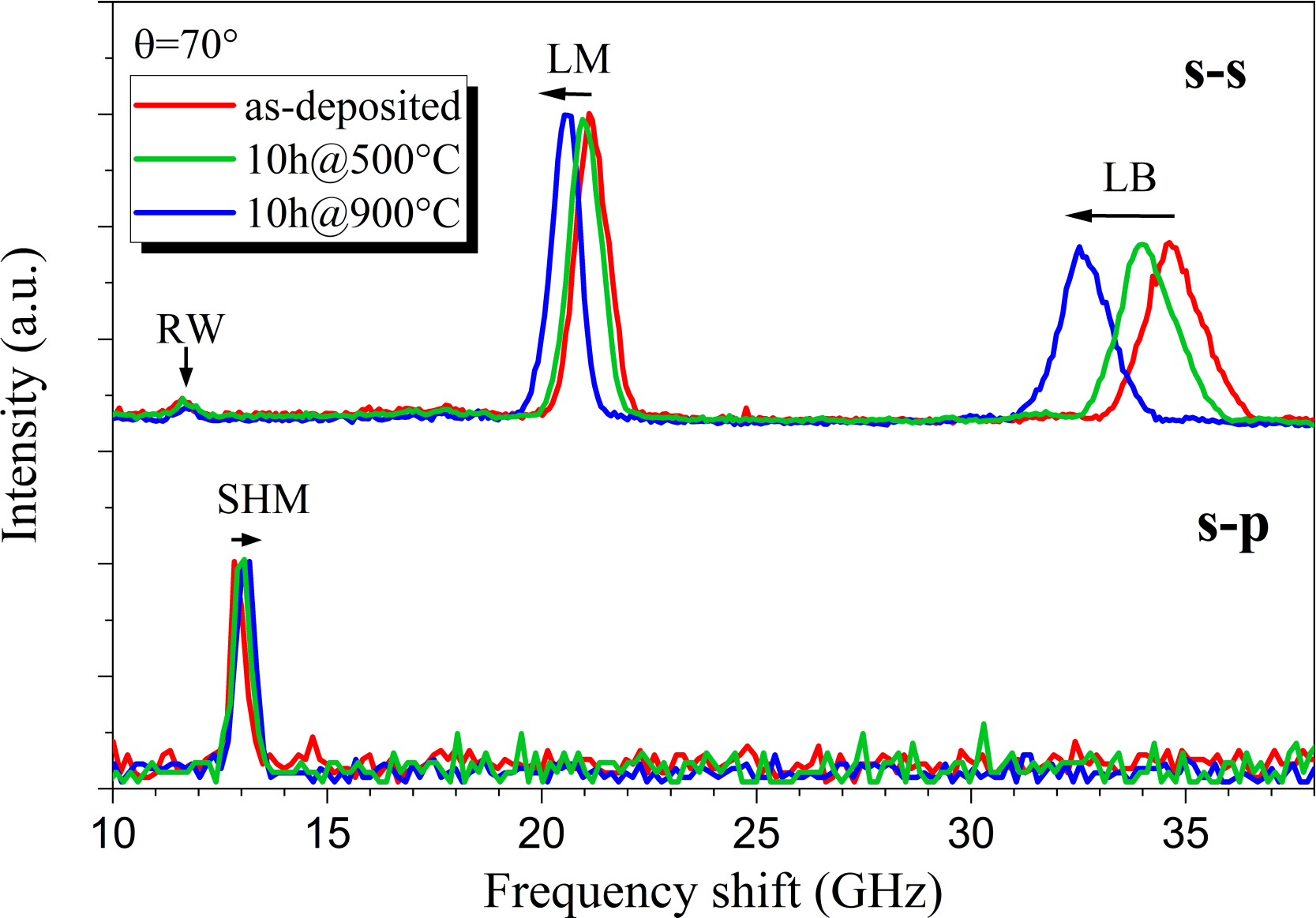}
	\caption{\small BLS spectra vs. annealing. Polarized (s-s) and depolarized (s-p) BLS spectra of sample S21089, as-deposited and annealed at 500 °C and 900 °C, collected at a fixed incidence angle of $\theta$=70° (only the Stokes side is shown for clarity). Arrows indicate the variation in the frequency of longitudinal (LB, LM) and shear (SHM) acoustic waves and the frequency invariance of the Rayleigh wave (RW).}
	\label{18}
\end{figure}

\begin{figure}[H]
	\centering
	\includegraphics[width=7 cm]{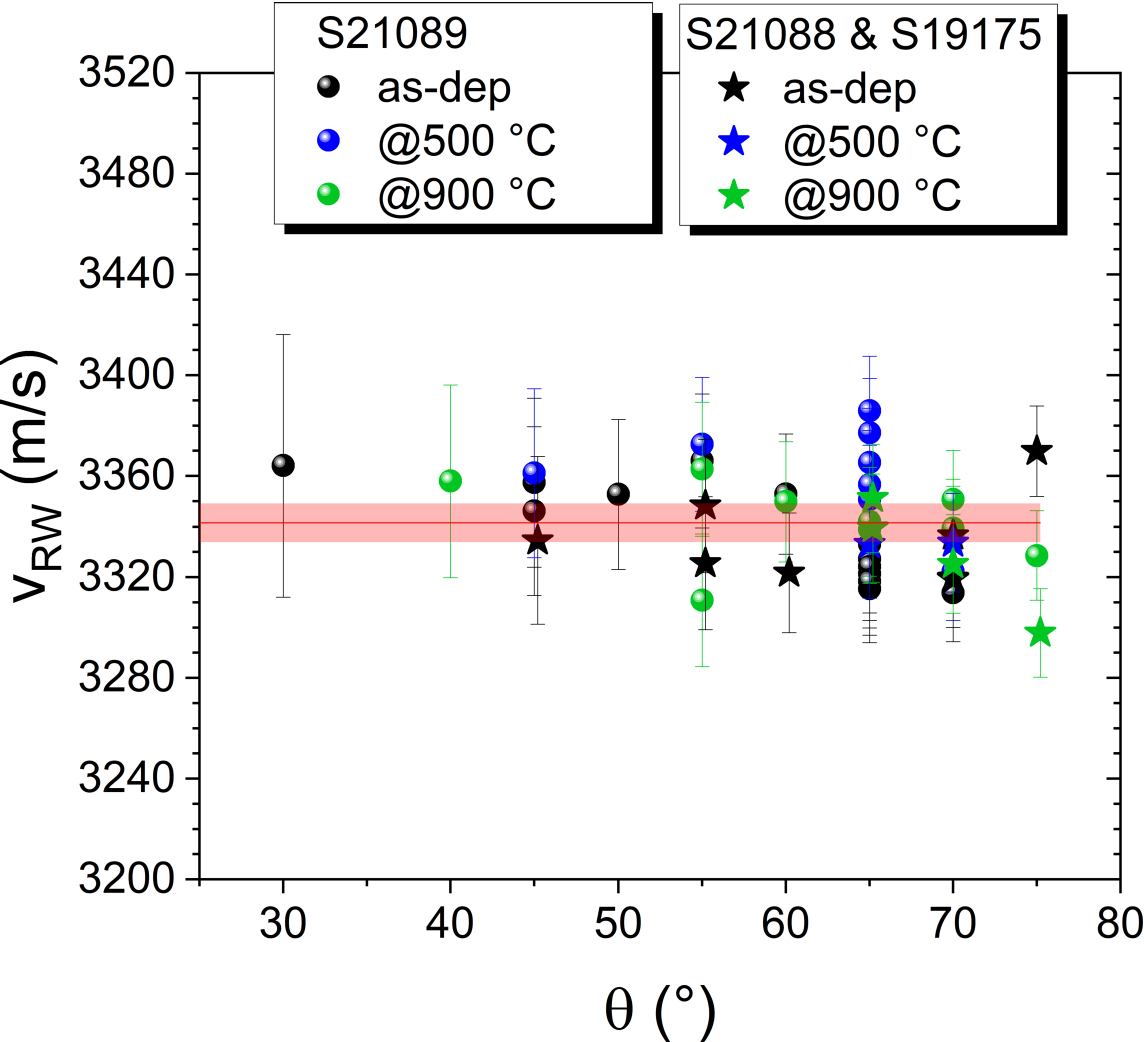}
	\caption{\small  Phase velocity $v_{RW}$ of the Rayleigh surface wave, as a function of the incidence angle $\theta$, in films of thickness 2.5 $\mu$m (S21089) and 720 nm (S21088, S19175) in different states. The solid line is the best-constant fit of all data, with the $95\%$ confidence band highlighted.}
	\label{19}
\end{figure}          

\begin{figure}[H]
		\centering
		\includegraphics[width=7.5 cm]{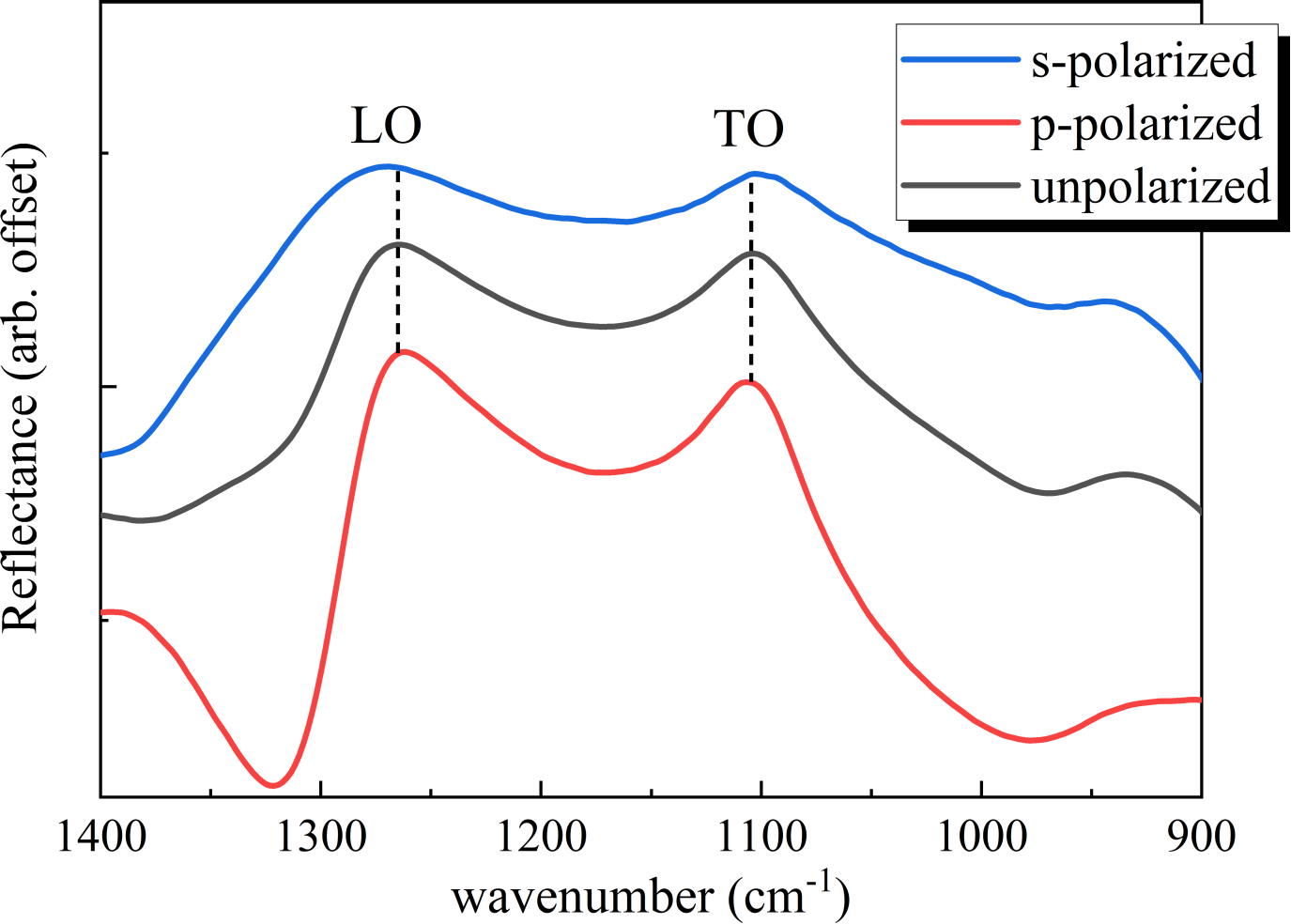}
		\caption{\small Longitudinal (LO) and transverse (TO) optical modes. Specular reflection infrared (SR-IR) spectra of sample S21089 in the as-deposited state, acquired at a fixed incidence angle of $\theta$ = 67° with different IR radiation polarizations. This angle was selected to reduce overlap with interference fringes and to amplify the intensity of the LO band.}
		\label{20}
\end{figure}

\begin{figure}[H]
		\centering
		\includegraphics[width=7.5 cm]{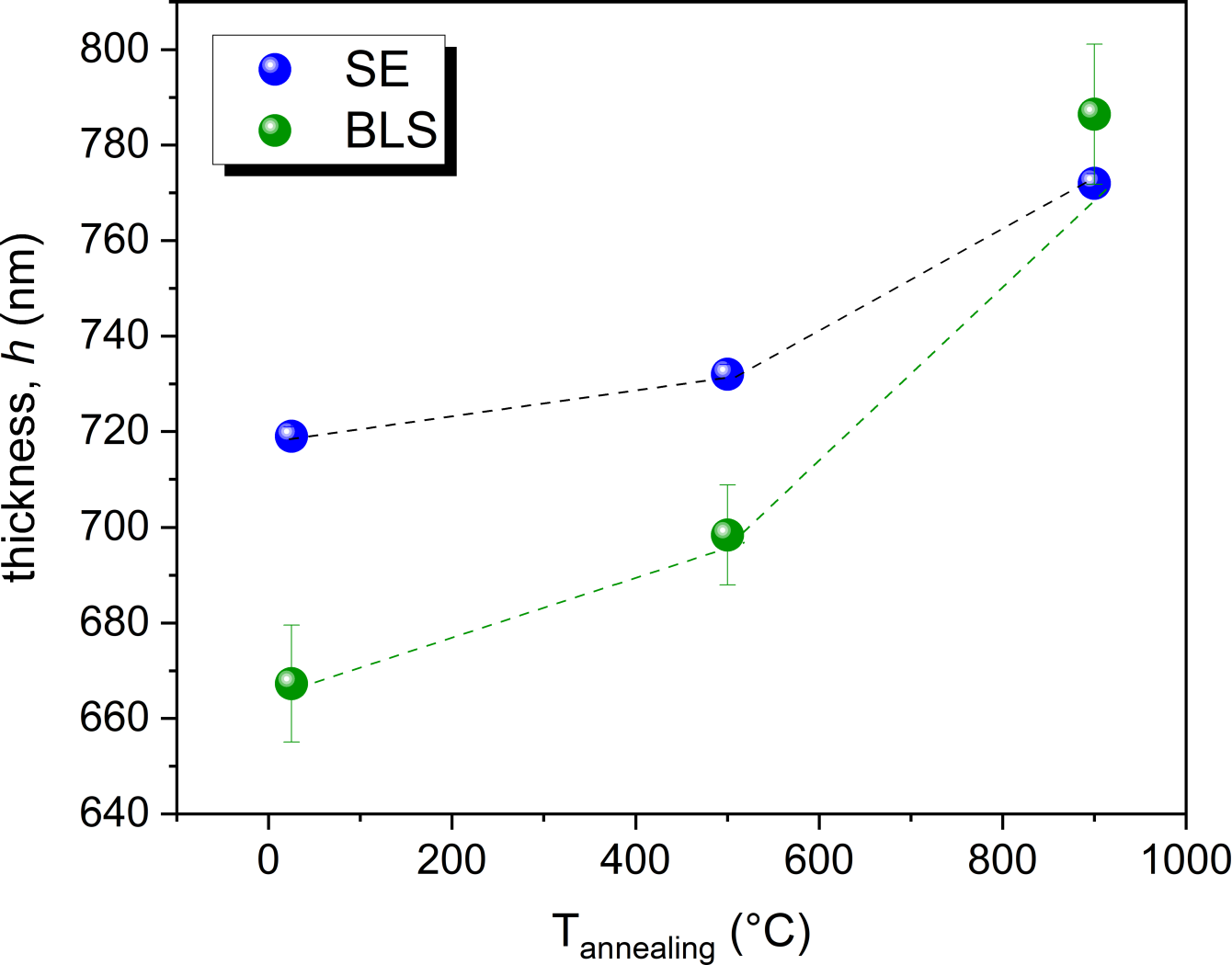}
		\caption{\small Variation of film thickness, \textit{h}, with annealing temperature for sample S19175, as determined by SE and BLS spectroscopy.}
		\label{21}
\end{figure}

\FloatBarrier

\section{\label{AppD}BLS results for D 263\textsuperscript{\textregistered} M cover glass}

\begin{figure} [H]
	\centering
	\includegraphics[width=7.5 cm]{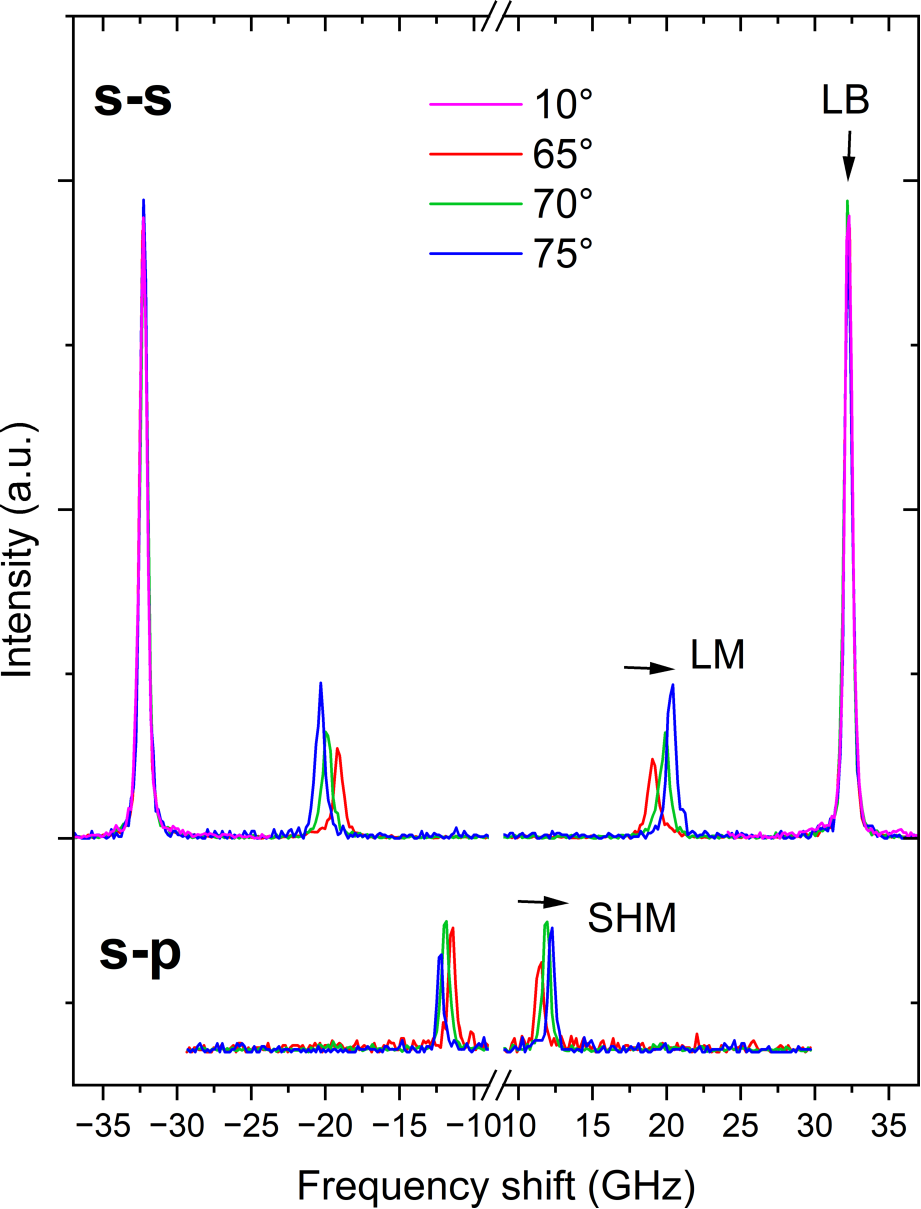}
	\caption{\small Polarized (s-s) and depolarized (s-p) BLS spectra of D 263\textsuperscript{\textregistered} M cover glass at varying incidence angles. Arrows indicate that the frequencies $f_{LM}$ and $f_{SHM}$ of LM and SHM acoustic waves increase with incidence angle, whereas the frequency $f_{LB}$ of LB acoustic wave remains unchanged.}
	\label{22}
\end{figure}

The borosilicate cover glass D 263\textsuperscript{\textregistered} M by SCHOTT (\url{https://www.schott.com/en-gr/products}), $145\pm 15$ $\mu$m-thick, served as a reference isotropic material. A reflective substrate was created by sputter-coating several hundred nanometers of chromium onto one side, replicating the scattering geometry in Fig. 1 of the manuscript. BLS spectra were collected using the same setup and measurement conditions as those used for the IBS SiO$_{2}$ films produced by LMA. Polarized (s-s) and depolarized (s-p) spectra were acquired at incidence angles $\theta$ = 10°, 65°, 70°, 75° (Fig.~\ref{22}). Consistent with thick films of glassy materials, the high-frequency spectrum is unimodal and well-fitted by a Lorentzian line shape.\\ 
As expected for elastically and optically isotropic materials, the calculated quantity $n_{BLS}=(f_{LB}/f_{LM})sin\theta=1.528\pm0.003$ is found to be independent of $\theta$ and agrees, within the experimental error, with the refractive index at the BLS wavelength, $n_{532}=1.5265\pm 0.0015$, derived from the certified value of standard refractive index at 546.1 nm, $n_{e}=1.5255\pm 0.0015$. \\
Additionally, the Poisson's ratio $\nu$ is determined from BLS frequencies using the relation:\\ $\nu= (2G-M)/[2(G-M)]=(2 f_{SHM}^2 - f_{LM}^2)/[2 (f_{SHM}^2 - f_{LM}^2)]$. The calculated value, 0.220$\pm$0.005, is in excellent agreement with the reported value of 0.21.

 \FloatBarrier

\section{\label{AppE}Thermal noise estimate}

We report here commonly used analytic models of coating thermal noise, explicitly assuming isotropic materials. For each model, we first present the general expression for a multilayer stack, followed by the simplified form for a single layer. All expressions are given in terms of the amplitude spectral density of thermal noise, defined as 
\begin{equation*}
	x(f) = \sqrt{S_x(f)}~,
\end{equation*}
where $S_x(f)$ is the power spectral density of thermal fluctuations of the coating surface.

(i) According to Hong et al. \cite{HongPRD2013}, assuming equal mechanical losses for bulk and shear motions, the amplitude spectral density for a multilayer stack is:
\begin{equation*}
	x(f) = \sqrt{\dfrac{2 k_B T}{\pi^2 f w^2} \dfrac{1-\nu_S-2\nu_S^2}{Y_S} \sum_j b_j h_j \Phi_j}
\end{equation*}
where $k_{B}$ is the Boltzmann constant, $T$ the mirror temperature, $f$ the frequency, $w$ the radius of the laser beam on the coating, $Y_{S}$ and $\nu_{S}$ the Young's modulus and Poisson's ratio of the substrate, and $h_{j}$ and $\Phi_{j}$ are the thickness and loss angle of each layer forming the coating. $b_{j}$ is a weighting factor whose expression, after the correction to Yam et al. \cite{YamPRD2015} reported by Tait et al. \cite{TaitPRL2020}, is given by:
\begin{eqnarray*}
	b_{j}&=&\dfrac{(1-2\nu_{j})(1+\nu_{j})}{(1-2\nu_{S})(1+\nu_{S})} \dfrac{1}{1-\nu_{j}} \\
	& &\times \left[ \left(1-n_j \dfrac{\partial\theta_c}{\partial\theta_j} \right)^2 \dfrac{Y_{S}}{Y_{j}}  +  \dfrac{(1-\nu_{S}-2\nu_{S}^{2})^2}{(1+\nu_{j})^2 (1-2\nu_{j})} \dfrac{Y_{j}}{Y_{S}}      \right]
\end{eqnarray*}
where $\nu_{j}$ and $\nu_{S}$ are the Poisson's ratios of each coating layer and the substrate. $n_j$ is the refractive index of the $j$th layer and $\frac{\partial\theta_c}{\partial\theta_j}$  describes the field penetration in each layer. Ignoring the field penetration ($\frac{\partial\theta_c}{\partial\theta_j} \rightarrow 0$), for a monolayer, the amplitude spectral density simplifies to:
\begin{eqnarray}
	x(f)&=&\sqrt{\dfrac{2 k_B T}{\pi^2 f w^2} \dfrac{1-\nu_S-2\nu_S^2}{Y_S} \, h \, \Phi \, \dfrac{(1-2\nu)(1+\nu)}{(1-2\nu_S)(1+\nu_S)} \dfrac{1}{1-\nu}} \nonumber \\
	& &\times \sqrt{\left[ \dfrac{Y_S}{Y} + \dfrac{(1-\nu_S-2\nu_S^2)^2}{(1+\nu)^2(1-2\nu)} \dfrac{Y}{Y_S} \right] }.
	\label{mono1}
\end{eqnarray}

(ii) According to the approach of Fejer \cite{FejerLIGO2021}, then used in Vajente et al. \cite{VajentePRL2021}, the amplitude spectral density of thermal noise, in the limit of equal bulk and shear loss angles, is given by:
\begin{widetext}
	\begin{equation*}
	x(f) = \sqrt{
		\frac{2 k_B T h}{\pi^2 f w^2}
		\Bigg[
		\left\langle \frac{Y}{1-\nu^2}\Phi \right\rangle
		\frac{(1+\nu_S)^2 (1-2\nu_S)^2}{Y_S^2} +
		\left\langle \frac{(1+\nu)(1-2\nu)}{(1-\nu)Y}\Phi \right\rangle
		\Bigg] }
	\end{equation*}
where $h$ is the total thickness of the coating, while the angular bracket expression $\langle x \rangle$ indicates the \textit{effective medium} average of the material property $x$ through the multilayer stack, weighted by the physical thickness of the layers. For a monolayer, this reduces to:
\begin{equation}
	x(f) = \sqrt{
		\frac{2 k_B T}{\pi^2 f w^2} \, h \, \Phi \,
		\Bigg[
		\frac{Y}{1-\nu^2}
		\frac{(1+\nu_S)^2 (1-2\nu_S)^2}{Y_S^2} +
		\frac{(1+\nu)(1-2\nu)}{(1-\nu) Y}
		\Bigg]. }
	\label{mono2}
\end{equation}
\end{widetext}

The relative variation upon annealing of the amplitude spectral density of thermal noise in the range 1--10 kHz for sample S21088 was calculated using both Eq.~\ref{mono1} and Eq.~\ref{mono2}, by adopting the following parameter sets ($\nu_{S}$, $Y_{S}$) for the substrate: (0.28, 130 GPa) for Si(100), (0.17, 73.2 GPa) for fused silica, (0.29, 400 GPa) for sapphire, and by using the following parameters for the film material:
\begin{table} [H]
	\centering{\scalebox{1}{\begin{tabular}{l  c c c c } 
				\hline
				& $h$ (nm) & $\Phi$ (rad) & $\nu$ & $Y$ (GPa) \\ 
				\hline 
				\hline
				as-deposited & 720 & $4.5 \cdot 10^{-4}$ \textsuperscript{[a]} & 0.229 & 78.5  \\
				10h@500 °C &  737 & $7 \cdot 10^{-5}$ \textsuperscript{[a]} & 0.220 & 75.0  \\
				10h@900 °C &  773 & $1 \cdot 10^{-5}$ \textsuperscript{[b]} & 0.178 & 71.8  \\
				\hline
			\end{tabular}
	}}
	\vspace{0.1 cm}
	{\raggedright
		\footnotesize\\
		\textsuperscript{[a]}~from Ref.~\cite{GranataCQG2020} \\
		\textsuperscript{[b]}~from Ref.~\cite{McGheePRL2023}\\
		\par}
\end{table}

\FloatBarrier

\onecolumngrid

\section{\label{AppF}Additional Tables}
This appendix reports additional tabulated data supporting the results discussed in the article. Table III summarizes the elastic and optical properties of sample S21089, including some of the parameters presented in Figs. 5(a), 5(c), and 5(d), along with additional values not shown in graphical form in the main text. Table IV collects optical properties in the infrared spectral region, which are discussed in Sec. IV B 2 of the main text.

\begin{table*}
	\caption{Elastic and optical properties of sample S21089. Refractive index at $\lambda$=532.3 nm, $n_{532}$, obtained from the SE isotropic model. BLS-apparent refractive index $n_{BLS}\equiv (f_{LB}/f_{LM})\sin\theta$. Mass density $\rho$, determined by XRR. Elastic constants $c_{ij}$, obtained by BLS. The values of $n_{BLS}$, $c_{11}$, $c_{33}$, and $c_{66}$ are reported as average $\pm$SD of several independent measurements; the values of $c_{44}$ and $c_{13}$ are obtained as best-fit results of a numerical simulation model; $c_{12}$ is calculated as $c_{11}-2c_{66}$, for symmetry constraint. The values of $n_{532}$ refer to the sample S21088, since they are identical to those of sample S21089 but affected by lower uncertainty.} 
	\centering{\scalebox{1}{\begin{tabular}{l c c c c c c c c c } 
				\hline
				& $n_{532}$ & $n_{BLS}$ & $\rho$ (g/cm$^{3}$) & $c_{11}$ (GPa) & $c_{33}$ (GPa)& $c_{66}$ (GPa) & $c_{44}$ (GPa) & $c_{13}$ (GPa) & $c_{12}$ (GPa)  \\
				\hline 
				\hline
				as-deposited & $1.494\pm0.003$ & $1.523\pm0.003$ & $2.39\pm0.01$ & $88.1\pm0.2$ & $93.6\pm0.3$ & $32.0\pm0.1$ & $31.8\pm0.4$ & $26.8\pm0.8$ & $24.1\pm0.3$ \\
				10h@500 °C & $1.475\pm0.003$ & $1.509\pm0.003$ & $2.28\pm0.01$ & $82.8\pm0.2$ & $88.5\pm0.3$ & $30.9\pm0.1$ & $30.6\pm0.6$ & $24.9\pm1.1$ & $21.0\pm0.3$  \\
				10h@900 °C & $1.467\pm0.003$ & $1.479\pm0.003$ & $2.22\pm0.01$ & $76.7\pm0.2$ & $78.9\pm0.3$ & $30.6\pm0.1$ & $30.4\pm0.4$ & $17.6\pm0.7$ & $15.5\pm0.3$ \\
				\hline
			\end{tabular}
	}}
\label{T3}

\vspace{20mm}

\caption{Real ($n$) and imaginary part ($k$) of the complex refractive index obtained from the Kramers-Kr\"{o}nig (KK) transformation of SR-IR spectra of SiO$_{2}$ films on SiO$_{2}$ substrate. The upper part of the table reports the values at selected wavevectors in the range 1000--1250 cm$^{-1}$. The lower part of the table compares the value of $n$ estimated by the KK transformation at 4000 cm$^{-1}$ with that measured by SE for sample S21088.}
\centering{\scalebox{1}{\begin{tabular}{ccccccccc}
			\hline
			& by IR \(\pm\)0.03 & &  &  &  &  & \\
			& \( n_{1000cm^{-1}}\) & \( k_{1000cm^{-1}}\) & \( n_{1100cm^{-1}}\) & \( k_{1100cm^{-1}}\) & \( n_{1200cm^{-1}}\) & \( k_{1200cm^{-1}}\) & \( n_{1250cm^{-1}}\) & \( k_{1250cm^{-1}}\) \\
			as deposited    & 3.30 & 1.02 & 0.54 & 1.99 & 0.48 & 0.78 & 0.36 &  0.43\\
			\hline
			10h@500°C    & 3.33 & 0.64 & 0.44 & 2.25 & 0.44 & 0.81 & 0.32 & 0.43 \\
			\hline
			10h@900°C    & 2.86 & 0.40 & 0.45 & 2.32 & 0.37 & 0.79 & 0.28 & 0.40 \\
			\hline
			\hline
			
		\end{tabular}
}}

\centering{\scalebox{1}{\begin{tabular}{c c || c c}
			
			& by SE \(\pm\) 0.005 & by IR \(\pm\)0.03 & \\
			& \( n_{4000cm^{-1}}\) & \( n_{4000cm^{-1}}\) & \( k_{4000cm^{-1}}\)\\
			\hline
			as deposited & 1.464 & 1.43 & 0.08 \\
			\hline
			10h@500°C    & 1.445 & 1.45 & 0.05 \\
			\hline
			10h@900°C    & 1.437 & 1.44 & 0.03 \\
			\hline
		\end{tabular}
}}
\label{T4}

\end{table*}

\FloatBarrier
\twocolumngrid


\begin{thebibliography}{56}%
	\makeatletter
	\providecommand \@ifxundefined [1]{%
		\@ifx{#1\undefined}
	}%
	\providecommand \@ifnum [1]{%
		\ifnum #1\expandafter \@firstoftwo
		\else \expandafter \@secondoftwo
		\fi
	}%
	\providecommand \@ifx [1]{%
		\ifx #1\expandafter \@firstoftwo
		\else \expandafter \@secondoftwo
		\fi
	}%
	\providecommand \natexlab [1]{#1}%
	\providecommand \enquote  [1]{``#1''}%
	\providecommand \bibnamefont  [1]{#1}%
	\providecommand \bibfnamefont [1]{#1}%
	\providecommand \citenamefont [1]{#1}%
	\providecommand \href@noop [0]{\@secondoftwo}%
	\providecommand \href [0]{\begingroup \@sanitize@url \@href}%
	\providecommand \@href[1]{\@@startlink{#1}\@@href}%
	\providecommand \@@href[1]{\endgroup#1\@@endlink}%
	\providecommand \@sanitize@url [0]{\catcode `\\12\catcode `\$12\catcode
		`\&12\catcode `\#12\catcode `\^12\catcode `\_12\catcode `\%12\relax}%
	\providecommand \@@startlink[1]{}%
	\providecommand \@@endlink[0]{}%
	\providecommand \url  [0]{\begingroup\@sanitize@url \@url }%
	\providecommand \@url [1]{\endgroup\@href {#1}{\urlprefix }}%
	\providecommand \urlprefix  [0]{URL }%
	\providecommand \Eprint [0]{\href }%
	\providecommand \doibase [0]{https://doi.org/}%
	\providecommand \selectlanguage [0]{\@gobble}%
	\providecommand \bibinfo  [0]{\@secondoftwo}%
	\providecommand \bibfield  [0]{\@secondoftwo}%
	\providecommand \translation [1]{[#1]}%
	\providecommand \BibitemOpen [0]{}%
	\providecommand \bibitemStop [0]{}%
	\providecommand \bibitemNoStop [0]{.\EOS\space}%
	\providecommand \EOS [0]{\spacefactor3000\relax}%
	\providecommand \BibitemShut  [1]{\csname bibitem#1\endcsname}%
	\let\auto@bib@innerbib\@empty
	\bibitem [{\citenamefont {{The LIGO Scientific Collaboration et
				al.}}(2015)}]{Aasi_CQG2015}%
	\BibitemOpen
	\bibfield  {author} {\bibinfo {author} {\bibnamefont {{The LIGO Scientific
					Collaboration et al.}}},\ }\bibfield  {title} {\bibinfo {title} {Advanced
			{LIGO}},\ }\href {https://doi.org/10.1088/0264-9381/32/7/074001} {\bibfield
		{journal} {\bibinfo  {journal} {Classical and Quantum Gravity}\ }\textbf
		{\bibinfo {volume} {32}},\ \bibinfo {pages} {074001} (\bibinfo {year}
		{2015})}\BibitemShut {NoStop}%
	\bibitem [{\citenamefont {{F. Acernese et al.}}(2015)}]{Acernese_CQG2015}%
	\BibitemOpen
	\bibfield  {author} {\bibinfo {author} {\bibnamefont {{F. Acernese et
					al.}}},\ }\bibfield  {title} {\bibinfo {title} {Advanced {Virgo}: a
			second-generation interferometric gravitational wave detector},\ }\href
	{https://doi.org/10.1088/0264-9381/32/2/024001} {\bibfield  {journal}
		{\bibinfo  {journal} {Classical and Quantum Gravity}\ }\textbf {\bibinfo
			{volume} {32}},\ \bibinfo {pages} {024001} (\bibinfo {year}
		{2015})}\BibitemShut {NoStop}%
	\bibitem [{\citenamefont {{T. Akutsu et al.}}(2020)}]{Akutsu_PTEP2020}%
	\BibitemOpen
	\bibfield  {author} {\bibinfo {author} {\bibnamefont {{T. Akutsu et al.}}},\
	}\bibfield  {title} {\bibinfo {title} {Overview of {KAGRA}: Detector design
			and construction history},\ }\href {https://doi.org/10.1093/ptep/ptaa125}
	{\bibfield  {journal} {\bibinfo  {journal} {Progress of Theoretical and
				Experimental Physics}\ }\textbf {\bibinfo {volume} {2021}},\ \bibinfo {pages}
		{05A101} (\bibinfo {year} {2020})}\BibitemShut {NoStop}%
	\bibitem [{LMA()}]{LMA}%
	\BibitemOpen
	\href {https://lma.in2p3.fr} {\bibinfo {title} {Laboratoire des mat\'eriaux
			avanc\'es; https://lma.in2p3.fr}}\BibitemShut {NoStop}%
	\bibitem [{\citenamefont {Pinard}\ \emph {et~al.}(2017)\citenamefont {Pinard},
		\citenamefont {Michel}, \citenamefont {Sassolas}, \citenamefont {Balzarini},
		\citenamefont {Degallaix}, \citenamefont {Dolique}, \citenamefont {Flaminio},
		\citenamefont {Forest}, \citenamefont {Granata}, \citenamefont {Lagrange},
		\citenamefont {Straniero}, \citenamefont {Teillon},\ and\ \citenamefont
		{Cagnoli}}]{PinardAO2017}%
	\BibitemOpen
	\bibfield  {author} {\bibinfo {author} {\bibfnamefont {L.}~\bibnamefont
			{Pinard}}, \bibinfo {author} {\bibfnamefont {C.}~\bibnamefont {Michel}},
		\bibinfo {author} {\bibfnamefont {B.}~\bibnamefont {Sassolas}}, \bibinfo
		{author} {\bibfnamefont {L.}~\bibnamefont {Balzarini}}, \bibinfo {author}
		{\bibfnamefont {J.}~\bibnamefont {Degallaix}}, \bibinfo {author}
		{\bibfnamefont {V.}~\bibnamefont {Dolique}}, \bibinfo {author} {\bibfnamefont
			{R.}~\bibnamefont {Flaminio}}, \bibinfo {author} {\bibfnamefont
			{D.}~\bibnamefont {Forest}}, \bibinfo {author} {\bibfnamefont
			{M.}~\bibnamefont {Granata}}, \bibinfo {author} {\bibfnamefont
			{B.}~\bibnamefont {Lagrange}}, \bibinfo {author} {\bibfnamefont
			{N.}~\bibnamefont {Straniero}}, \bibinfo {author} {\bibfnamefont
			{J.}~\bibnamefont {Teillon}},\ and\ \bibinfo {author} {\bibfnamefont
			{G.}~\bibnamefont {Cagnoli}},\ }\bibfield  {title} {\bibinfo {title} {Mirrors
			used in the ligo interferometers for first detection of gravitational
			waves},\ }\href {https://doi.org/10.1364/AO.56.000C11} {\bibfield  {journal}
		{\bibinfo  {journal} {Appl. Opt.}\ }\textbf {\bibinfo {volume} {56}},\
		\bibinfo {pages} {C11} (\bibinfo {year} {2017})}\BibitemShut {NoStop}%
	\bibitem [{\citenamefont {Granata}\ \emph
		{et~al.}(2020{\natexlab{a}})\citenamefont {Granata}, \citenamefont {Amato},
		\citenamefont {Cagnoli}, \citenamefont {Coulon}, \citenamefont {Degallaix},
		\citenamefont {Forest}, \citenamefont {Mereni}, \citenamefont {Michel},
		\citenamefont {Pinard}, \citenamefont {Sassolas},\ and\ \citenamefont
		{Teillon}}]{GranataAO2020}%
	\BibitemOpen
	\bibfield  {author} {\bibinfo {author} {\bibfnamefont {M.}~\bibnamefont
			{Granata}}, \bibinfo {author} {\bibfnamefont {A.}~\bibnamefont {Amato}},
		\bibinfo {author} {\bibfnamefont {G.}~\bibnamefont {Cagnoli}}, \bibinfo
		{author} {\bibfnamefont {M.}~\bibnamefont {Coulon}}, \bibinfo {author}
		{\bibfnamefont {J.}~\bibnamefont {Degallaix}}, \bibinfo {author}
		{\bibfnamefont {D.}~\bibnamefont {Forest}}, \bibinfo {author} {\bibfnamefont
			{L.}~\bibnamefont {Mereni}}, \bibinfo {author} {\bibfnamefont
			{C.}~\bibnamefont {Michel}}, \bibinfo {author} {\bibfnamefont
			{L.}~\bibnamefont {Pinard}}, \bibinfo {author} {\bibfnamefont
			{B.}~\bibnamefont {Sassolas}},\ and\ \bibinfo {author} {\bibfnamefont
			{J.}~\bibnamefont {Teillon}},\ }\bibfield  {title} {\bibinfo {title}
		{Progress in the measurement and reduction of thermal noise in optical
			coatings for gravitational-wave detectors},\ }\href
	{https://doi.org/10.1364/AO.377293} {\bibfield  {journal} {\bibinfo
			{journal} {Appl. Opt.}\ }\textbf {\bibinfo {volume} {59}},\ \bibinfo {pages}
		{A229} (\bibinfo {year} {2020}{\natexlab{a}})}\BibitemShut {NoStop}%
	\bibitem [{\citenamefont {Durante}\ \emph {et~al.}(2024)\citenamefont
		{Durante}, \citenamefont {Magnozzi}, \citenamefont {Fiumara}, \citenamefont
		{Neilson}, \citenamefont {Canepa}, \citenamefont {Avallone}, \citenamefont
		{Bobba}, \citenamefont {Carapella}, \citenamefont {Chiadini}, \citenamefont
		{{De Salvo}}, \citenamefont {{De Simone}}, \citenamefont {{Di Giorgio}},
		\citenamefont {Fittipaldi}, \citenamefont {Micco}, \citenamefont {Pinto},
		\citenamefont {Vecchione}, \citenamefont {Pierro},\ and\ \citenamefont
		{Granata}}]{DuranteOptMat2024}%
	\BibitemOpen
	\bibfield  {author} {\bibinfo {author} {\bibfnamefont {O.}~\bibnamefont
			{Durante}}, \bibinfo {author} {\bibfnamefont {M.}~\bibnamefont {Magnozzi}},
		\bibinfo {author} {\bibfnamefont {V.}~\bibnamefont {Fiumara}}, \bibinfo
		{author} {\bibfnamefont {J.}~\bibnamefont {Neilson}}, \bibinfo {author}
		{\bibfnamefont {M.}~\bibnamefont {Canepa}}, \bibinfo {author} {\bibfnamefont
			{G.}~\bibnamefont {Avallone}}, \bibinfo {author} {\bibfnamefont
			{F.}~\bibnamefont {Bobba}}, \bibinfo {author} {\bibfnamefont
			{G.}~\bibnamefont {Carapella}}, \bibinfo {author} {\bibfnamefont
			{F.}~\bibnamefont {Chiadini}}, \bibinfo {author} {\bibfnamefont
			{R.}~\bibnamefont {{De Salvo}}}, \bibinfo {author} {\bibfnamefont
			{R.}~\bibnamefont {{De Simone}}}, \bibinfo {author} {\bibfnamefont
			{C.}~\bibnamefont {{Di Giorgio}}}, \bibinfo {author} {\bibfnamefont
			{R.}~\bibnamefont {Fittipaldi}}, \bibinfo {author} {\bibfnamefont
			{A.}~\bibnamefont {Micco}}, \bibinfo {author} {\bibfnamefont {I.~M.}\
			\bibnamefont {Pinto}}, \bibinfo {author} {\bibfnamefont {A.}~\bibnamefont
			{Vecchione}}, \bibinfo {author} {\bibfnamefont {V.}~\bibnamefont {Pierro}},\
		and\ \bibinfo {author} {\bibfnamefont {V.}~\bibnamefont {Granata}},\
	}\bibfield  {title} {\bibinfo {title} {Toward the optimization of {S}i{O}$_2$
			and {T}i{O}$_2$-based metamaterials: Morphological, structural, and optical
			characterization},\ }\href
	{https://doi.org/https://doi.org/10.1016/j.optmat.2024.116038} {\bibfield
		{journal} {\bibinfo  {journal} {Optical Materials}\ }\textbf {\bibinfo
			{volume} {157}},\ \bibinfo {pages} {116038} (\bibinfo {year}
		{2024})}\BibitemShut {NoStop}%
	\bibitem [{\citenamefont {Yang}\ \emph {et~al.}(2020)\citenamefont {Yang},
		\citenamefont {Fazio}, \citenamefont {Vajente}, \citenamefont {Ananyeva},
		\citenamefont {Billingsley}, \citenamefont {Markosyan}, \citenamefont
		{Bassiri}, \citenamefont {Fejer},\ and\ \citenamefont
		{Menoni}}]{YangACSANM2020}%
	\BibitemOpen
	\bibfield  {author} {\bibinfo {author} {\bibfnamefont {L.}~\bibnamefont
			{Yang}}, \bibinfo {author} {\bibfnamefont {M.}~\bibnamefont {Fazio}},
		\bibinfo {author} {\bibfnamefont {G.}~\bibnamefont {Vajente}}, \bibinfo
		{author} {\bibfnamefont {A.}~\bibnamefont {Ananyeva}}, \bibinfo {author}
		{\bibfnamefont {G.}~\bibnamefont {Billingsley}}, \bibinfo {author}
		{\bibfnamefont {A.}~\bibnamefont {Markosyan}}, \bibinfo {author}
		{\bibfnamefont {R.}~\bibnamefont {Bassiri}}, \bibinfo {author} {\bibfnamefont
			{M.~M.}\ \bibnamefont {Fejer}},\ and\ \bibinfo {author} {\bibfnamefont
			{C.~S.}\ \bibnamefont {Menoni}},\ }\bibfield  {title} {\bibinfo {title}
		{Structural evolution that affects the room-temperature internal friction of
			binary oxid: Implications for ultrastable optical cavities},\ }\href
	{https://doi.org/10.1021/acsanm.0c02798} {\bibfield  {journal} {\bibinfo
			{journal} {ACS Applied Nano Materials}\ }\textbf {\bibinfo {volume} {3}},\
		\bibinfo {pages} {12308} (\bibinfo {year} {2020})}\BibitemShut {NoStop}%
	\bibitem [{\citenamefont {Granata}\ \emph {et~al.}(2018)\citenamefont
		{Granata}, \citenamefont {Coillet}, \citenamefont {Martinez}, \citenamefont
		{Dolique}, \citenamefont {Amato}, \citenamefont {Canepa}, \citenamefont
		{Margueritat}, \citenamefont {Martinet}, \citenamefont {Mermet},
		\citenamefont {Michel}, \citenamefont {Pinard}, \citenamefont {Sassolas},\
		and\ \citenamefont {Cagnoli}}]{GranataPRM2018}%
	\BibitemOpen
	\bibfield  {author} {\bibinfo {author} {\bibfnamefont {M.}~\bibnamefont
			{Granata}}, \bibinfo {author} {\bibfnamefont {E.}~\bibnamefont {Coillet}},
		\bibinfo {author} {\bibfnamefont {V.}~\bibnamefont {Martinez}}, \bibinfo
		{author} {\bibfnamefont {V.}~\bibnamefont {Dolique}}, \bibinfo {author}
		{\bibfnamefont {A.}~\bibnamefont {Amato}}, \bibinfo {author} {\bibfnamefont
			{M.}~\bibnamefont {Canepa}}, \bibinfo {author} {\bibfnamefont
			{J.}~\bibnamefont {Margueritat}}, \bibinfo {author} {\bibfnamefont
			{C.}~\bibnamefont {Martinet}}, \bibinfo {author} {\bibfnamefont
			{A.}~\bibnamefont {Mermet}}, \bibinfo {author} {\bibfnamefont
			{C.}~\bibnamefont {Michel}}, \bibinfo {author} {\bibfnamefont
			{L.}~\bibnamefont {Pinard}}, \bibinfo {author} {\bibfnamefont
			{B.}~\bibnamefont {Sassolas}},\ and\ \bibinfo {author} {\bibfnamefont
			{G.}~\bibnamefont {Cagnoli}},\ }\bibfield  {title} {\bibinfo {title}
		{Correlated evolution of structure and mechanical loss of a sputtered silica
			film},\ }\href {https://doi.org/10.1103/PhysRevMaterials.2.053607} {\bibfield
		{journal} {\bibinfo  {journal} {Phys. Rev. Mater.}\ }\textbf {\bibinfo
			{volume} {2}},\ \bibinfo {pages} {053607} (\bibinfo {year}
		{2018})}\BibitemShut {NoStop}%
	\bibitem [{\citenamefont {Amato}\ \emph {et~al.}(2019)\citenamefont {Amato},
		\citenamefont {Terreni}, \citenamefont {Granata}, \citenamefont {Michel},
		\citenamefont {Pinard}, \citenamefont {Gemme}, \citenamefont {Canepa},\ and\
		\citenamefont {Cagnoli}}]{AmatoJVSTB2019}%
	\BibitemOpen
	\bibfield  {author} {\bibinfo {author} {\bibfnamefont {A.}~\bibnamefont
			{Amato}}, \bibinfo {author} {\bibfnamefont {S.}~\bibnamefont {Terreni}},
		\bibinfo {author} {\bibfnamefont {M.}~\bibnamefont {Granata}}, \bibinfo
		{author} {\bibfnamefont {C.}~\bibnamefont {Michel}}, \bibinfo {author}
		{\bibfnamefont {L.}~\bibnamefont {Pinard}}, \bibinfo {author} {\bibfnamefont
			{G.}~\bibnamefont {Gemme}}, \bibinfo {author} {\bibfnamefont
			{M.}~\bibnamefont {Canepa}},\ and\ \bibinfo {author} {\bibfnamefont
			{G.}~\bibnamefont {Cagnoli}},\ }\bibfield  {title} {\bibinfo {title} {Effect
			of heating treatment and mixture on optical properties of coating materials
			used in gravitational-wave detectors},\ }\href
	{https://doi.org/10.1116/1.5122661} {\bibfield  {journal} {\bibinfo
			{journal} {Journal of Vacuum Science \& Technology B}\ }\textbf {\bibinfo
			{volume} {37}},\ \bibinfo {pages} {062913} (\bibinfo {year}
		{2019})}\BibitemShut {NoStop}%
	\bibitem [{\citenamefont {Colace}\ \emph {et~al.}(2024)\citenamefont {Colace},
		\citenamefont {Samandari}, \citenamefont {Granata}, \citenamefont {Amato},
		\citenamefont {Caminale}, \citenamefont {Michel}, \citenamefont {Gemme},
		\citenamefont {Pinard}, \citenamefont {Canepa},\ and\ \citenamefont
		{Magnozzi}}]{ColaceCQG2024}%
	\BibitemOpen
	\bibfield  {author} {\bibinfo {author} {\bibfnamefont {S.}~\bibnamefont
			{Colace}}, \bibinfo {author} {\bibfnamefont {S.}~\bibnamefont {Samandari}},
		\bibinfo {author} {\bibfnamefont {M.}~\bibnamefont {Granata}}, \bibinfo
		{author} {\bibfnamefont {A.}~\bibnamefont {Amato}}, \bibinfo {author}
		{\bibfnamefont {M.}~\bibnamefont {Caminale}}, \bibinfo {author}
		{\bibfnamefont {C.}~\bibnamefont {Michel}}, \bibinfo {author} {\bibfnamefont
			{G.}~\bibnamefont {Gemme}}, \bibinfo {author} {\bibfnamefont
			{L.}~\bibnamefont {Pinard}}, \bibinfo {author} {\bibfnamefont
			{M.}~\bibnamefont {Canepa}},\ and\ \bibinfo {author} {\bibfnamefont
			{M.}~\bibnamefont {Magnozzi}},\ }\bibfield  {title} {\bibinfo {title}
		{Monitoring the evolution of optical coatings during thermal annealing with
			real-time, in-situ spectroscopic ellipsometry},\ }\href
	{https://doi.org/10.1088/1361-6382/ad672c} {\bibfield  {journal} {\bibinfo
			{journal} {Classical and Quantum Gravity}\ }\textbf {\bibinfo {volume}
			{41}},\ \bibinfo {pages} {175016} (\bibinfo {year} {2024})}\BibitemShut
	{NoStop}%
	\bibitem [{\citenamefont {Amato}\ \emph {et~al.}(2025)\citenamefont {Amato},
		\citenamefont {Bazzan}, \citenamefont {Cagnoli}, \citenamefont {Canepa},
		\citenamefont {Coulon}, \citenamefont {Degallaix}, \citenamefont {Demos},
		\citenamefont {Di~Michele}, \citenamefont {Evans}, \citenamefont {Fabrizi},
		\citenamefont {Favaro}, \citenamefont {Forest}, \citenamefont {Gras},
		\citenamefont {Hofman}, \citenamefont {Lema\^{\i}tre}, \citenamefont
		{Maggioni}, \citenamefont {Magnozzi}, \citenamefont {Martinez}, \citenamefont
		{Mereni}, \citenamefont {Michel}, \citenamefont {Milotti}, \citenamefont
		{Montani}, \citenamefont {Paolone}, \citenamefont {Pereira}, \citenamefont
		{Piergiovanni}, \citenamefont {Pierro}, \citenamefont {Pinard}, \citenamefont
		{Pinto}, \citenamefont {Placidi}, \citenamefont {Samandari}, \citenamefont
		{Sassolas}, \citenamefont {Shcheblanov}, \citenamefont {Teillon},
		\citenamefont {Vickridge},\ and\ \citenamefont {Granata}}]{AmatoPRD2025}%
	\BibitemOpen
	\bibfield  {author} {\bibinfo {author} {\bibfnamefont {A.}~\bibnamefont
			{Amato}}, \bibinfo {author} {\bibfnamefont {M.}~\bibnamefont {Bazzan}},
		\bibinfo {author} {\bibfnamefont {G.}~\bibnamefont {Cagnoli}}, \bibinfo
		{author} {\bibfnamefont {M.}~\bibnamefont {Canepa}}, \bibinfo {author}
		{\bibfnamefont {M.}~\bibnamefont {Coulon}}, \bibinfo {author} {\bibfnamefont
			{J.}~\bibnamefont {Degallaix}}, \bibinfo {author} {\bibfnamefont
			{N.}~\bibnamefont {Demos}}, \bibinfo {author} {\bibfnamefont
			{A.}~\bibnamefont {Di~Michele}}, \bibinfo {author} {\bibfnamefont
			{M.}~\bibnamefont {Evans}}, \bibinfo {author} {\bibfnamefont
			{F.}~\bibnamefont {Fabrizi}}, \bibinfo {author} {\bibfnamefont
			{G.}~\bibnamefont {Favaro}}, \bibinfo {author} {\bibfnamefont
			{D.}~\bibnamefont {Forest}}, \bibinfo {author} {\bibfnamefont
			{S.}~\bibnamefont {Gras}}, \bibinfo {author} {\bibfnamefont {D.}~\bibnamefont
			{Hofman}}, \bibinfo {author} {\bibfnamefont {A.}~\bibnamefont
			{Lema\^{\i}tre}}, \bibinfo {author} {\bibfnamefont {G.}~\bibnamefont
			{Maggioni}}, \bibinfo {author} {\bibfnamefont {M.}~\bibnamefont {Magnozzi}},
		\bibinfo {author} {\bibfnamefont {V.}~\bibnamefont {Martinez}}, \bibinfo
		{author} {\bibfnamefont {L.}~\bibnamefont {Mereni}}, \bibinfo {author}
		{\bibfnamefont {C.}~\bibnamefont {Michel}}, \bibinfo {author} {\bibfnamefont
			{V.}~\bibnamefont {Milotti}}, \bibinfo {author} {\bibfnamefont
			{M.}~\bibnamefont {Montani}}, \bibinfo {author} {\bibfnamefont
			{A.}~\bibnamefont {Paolone}}, \bibinfo {author} {\bibfnamefont
			{A.}~\bibnamefont {Pereira}}, \bibinfo {author} {\bibfnamefont
			{F.}~\bibnamefont {Piergiovanni}}, \bibinfo {author} {\bibfnamefont
			{V.}~\bibnamefont {Pierro}}, \bibinfo {author} {\bibfnamefont
			{L.}~\bibnamefont {Pinard}}, \bibinfo {author} {\bibfnamefont {I.~M.}\
			\bibnamefont {Pinto}}, \bibinfo {author} {\bibfnamefont {E.}~\bibnamefont
			{Placidi}}, \bibinfo {author} {\bibfnamefont {S.}~\bibnamefont {Samandari}},
		\bibinfo {author} {\bibfnamefont {B.}~\bibnamefont {Sassolas}}, \bibinfo
		{author} {\bibfnamefont {N.}~\bibnamefont {Shcheblanov}}, \bibinfo {author}
		{\bibfnamefont {J.}~\bibnamefont {Teillon}}, \bibinfo {author} {\bibfnamefont
			{I.}~\bibnamefont {Vickridge}},\ and\ \bibinfo {author} {\bibfnamefont
			{M.}~\bibnamefont {Granata}},\ }\bibfield  {title} {\bibinfo {title}
		{Development of ion-beam sputtered silicon nitride thin films for low-noise
			mirror coatings of gravitational-wave detectors},\ }\href
	{https://doi.org/10.1103/PhysRevD.111.042003} {\bibfield  {journal} {\bibinfo
			{journal} {Phys. Rev. D}\ }\textbf {\bibinfo {volume} {111}},\ \bibinfo
		{pages} {042003} (\bibinfo {year} {2025})}\BibitemShut {NoStop}%
	\bibitem [{\citenamefont {Vajente}\ \emph {et~al.}(2021)\citenamefont
		{Vajente}, \citenamefont {Yang}, \citenamefont {Davenport}, \citenamefont
		{Fazio}, \citenamefont {Ananyeva}, \citenamefont {Zhang}, \citenamefont
		{Billingsley}, \citenamefont {Prasai}, \citenamefont {Markosyan},
		\citenamefont {Bassiri}, \citenamefont {Fejer}, \citenamefont {Chicoine},
		\citenamefont {Schiettekatte},\ and\ \citenamefont
		{Menoni}}]{VajentePRL2021}%
	\BibitemOpen
	\bibfield  {author} {\bibinfo {author} {\bibfnamefont {G.}~\bibnamefont
			{Vajente}}, \bibinfo {author} {\bibfnamefont {L.}~\bibnamefont {Yang}},
		\bibinfo {author} {\bibfnamefont {A.}~\bibnamefont {Davenport}}, \bibinfo
		{author} {\bibfnamefont {M.}~\bibnamefont {Fazio}}, \bibinfo {author}
		{\bibfnamefont {A.}~\bibnamefont {Ananyeva}}, \bibinfo {author}
		{\bibfnamefont {L.}~\bibnamefont {Zhang}}, \bibinfo {author} {\bibfnamefont
			{G.}~\bibnamefont {Billingsley}}, \bibinfo {author} {\bibfnamefont
			{K.}~\bibnamefont {Prasai}}, \bibinfo {author} {\bibfnamefont
			{A.}~\bibnamefont {Markosyan}}, \bibinfo {author} {\bibfnamefont
			{R.}~\bibnamefont {Bassiri}}, \bibinfo {author} {\bibfnamefont {M.~M.}\
			\bibnamefont {Fejer}}, \bibinfo {author} {\bibfnamefont {M.}~\bibnamefont
			{Chicoine}}, \bibinfo {author} {\bibfnamefont {F.~m.~c.}\ \bibnamefont
			{Schiettekatte}},\ and\ \bibinfo {author} {\bibfnamefont {C.~S.}\
			\bibnamefont {Menoni}},\ }\bibfield  {title} {\bibinfo {title} {Low
			mechanical loss {T}i{O}$_2$:{G}e{O}$_2$ coatings for reduced thermal noise in
			gravitational wave interferometers},\ }\href
	{https://doi.org/10.1103/PhysRevLett.127.071101} {\bibfield  {journal}
		{\bibinfo  {journal} {Phys. Rev. Lett.}\ }\textbf {\bibinfo {volume} {127}},\
		\bibinfo {pages} {071101} (\bibinfo {year} {2021})}\BibitemShut {NoStop}%
	\bibitem [{\citenamefont {McGhee}\ \emph {et~al.}(2023)\citenamefont {McGhee},
		\citenamefont {Spagnuolo}, \citenamefont {Demos}, \citenamefont {Tait},
		\citenamefont {Murray}, \citenamefont {Chicoine}, \citenamefont {Dabadie},
		\citenamefont {Gras}, \citenamefont {Hough}, \citenamefont {Iandolo},
		\citenamefont {Johnston}, \citenamefont {Martinez}, \citenamefont {Patane},
		\citenamefont {Rowan}, \citenamefont {Schiettekatte}, \citenamefont {Smith},
		\citenamefont {Terkowski}, \citenamefont {Zhang}, \citenamefont {Evans},
		\citenamefont {Martin},\ and\ \citenamefont {Steinlechner}}]{McGheePRL2023}%
	\BibitemOpen
	\bibfield  {author} {\bibinfo {author} {\bibfnamefont {G.~I.}\ \bibnamefont
			{McGhee}}, \bibinfo {author} {\bibfnamefont {V.}~\bibnamefont {Spagnuolo}},
		\bibinfo {author} {\bibfnamefont {N.}~\bibnamefont {Demos}}, \bibinfo
		{author} {\bibfnamefont {S.~C.}\ \bibnamefont {Tait}}, \bibinfo {author}
		{\bibfnamefont {P.~G.}\ \bibnamefont {Murray}}, \bibinfo {author}
		{\bibfnamefont {M.}~\bibnamefont {Chicoine}}, \bibinfo {author}
		{\bibfnamefont {P.}~\bibnamefont {Dabadie}}, \bibinfo {author} {\bibfnamefont
			{S.}~\bibnamefont {Gras}}, \bibinfo {author} {\bibfnamefont {J.}~\bibnamefont
			{Hough}}, \bibinfo {author} {\bibfnamefont {G.~A.}\ \bibnamefont {Iandolo}},
		\bibinfo {author} {\bibfnamefont {R.}~\bibnamefont {Johnston}}, \bibinfo
		{author} {\bibfnamefont {V.}~\bibnamefont {Martinez}}, \bibinfo {author}
		{\bibfnamefont {O.}~\bibnamefont {Patane}}, \bibinfo {author} {\bibfnamefont
			{S.}~\bibnamefont {Rowan}}, \bibinfo {author} {\bibfnamefont {F.~m.~c.}\
			\bibnamefont {Schiettekatte}}, \bibinfo {author} {\bibfnamefont {J.~R.}\
			\bibnamefont {Smith}}, \bibinfo {author} {\bibfnamefont {L.}~\bibnamefont
			{Terkowski}}, \bibinfo {author} {\bibfnamefont {L.}~\bibnamefont {Zhang}},
		\bibinfo {author} {\bibfnamefont {M.}~\bibnamefont {Evans}}, \bibinfo
		{author} {\bibfnamefont {I.~W.}\ \bibnamefont {Martin}},\ and\ \bibinfo
		{author} {\bibfnamefont {J.}~\bibnamefont {Steinlechner}},\ }\bibfield
	{title} {\bibinfo {title} {Titania mixed with silica: A low thermal-noise
			coating material for gravitational-wave detectors},\ }\href
	{https://doi.org/10.1103/PhysRevLett.131.171401} {\bibfield  {journal}
		{\bibinfo  {journal} {Phys. Rev. Lett.}\ }\textbf {\bibinfo {volume} {131}},\
		\bibinfo {pages} {171401} (\bibinfo {year} {2023})}\BibitemShut {NoStop}%
	\bibitem [{\citenamefont {Granata}\ \emph
		{et~al.}(2020{\natexlab{b}})\citenamefont {Granata}, \citenamefont {Amato},
		\citenamefont {Balzarini}, \citenamefont {Canepa}, \citenamefont {Degallaix},
		\citenamefont {Forest}, \citenamefont {Dolique}, \citenamefont {Mereni},
		\citenamefont {Michel}, \citenamefont {Pinard}, \citenamefont {Sassolas},
		\citenamefont {Teillon},\ and\ \citenamefont {Cagnoli}}]{GranataCQG2020}%
	\BibitemOpen
	\bibfield  {author} {\bibinfo {author} {\bibfnamefont {M.}~\bibnamefont
			{Granata}}, \bibinfo {author} {\bibfnamefont {A.}~\bibnamefont {Amato}},
		\bibinfo {author} {\bibfnamefont {L.}~\bibnamefont {Balzarini}}, \bibinfo
		{author} {\bibfnamefont {M.}~\bibnamefont {Canepa}}, \bibinfo {author}
		{\bibfnamefont {J.}~\bibnamefont {Degallaix}}, \bibinfo {author}
		{\bibfnamefont {D.}~\bibnamefont {Forest}}, \bibinfo {author} {\bibfnamefont
			{V.}~\bibnamefont {Dolique}}, \bibinfo {author} {\bibfnamefont
			{L.}~\bibnamefont {Mereni}}, \bibinfo {author} {\bibfnamefont
			{C.}~\bibnamefont {Michel}}, \bibinfo {author} {\bibfnamefont
			{L.}~\bibnamefont {Pinard}}, \bibinfo {author} {\bibfnamefont
			{B.}~\bibnamefont {Sassolas}}, \bibinfo {author} {\bibfnamefont
			{J.}~\bibnamefont {Teillon}},\ and\ \bibinfo {author} {\bibfnamefont
			{G.}~\bibnamefont {Cagnoli}},\ }\bibfield  {title} {\bibinfo {title}
		{Amorphous optical coatings of present gravitational-wave interferometers*},\
	}\href {https://doi.org/10.1088/1361-6382/ab77e9} {\bibfield  {journal}
		{\bibinfo  {journal} {Classical and Quantum Gravity}\ }\textbf {\bibinfo
			{volume} {37}},\ \bibinfo {pages} {095004} (\bibinfo {year}
		{2020}{\natexlab{b}})}\BibitemShut {NoStop}%
	\bibitem [{\citenamefont {Granata}\ \emph {et~al.}(2016)\citenamefont
		{Granata}, \citenamefont {Saracco}, \citenamefont {Morgado}, \citenamefont
		{Cajgfinger}, \citenamefont {Cagnoli}, \citenamefont {Degallaix},
		\citenamefont {Dolique}, \citenamefont {Forest}, \citenamefont {Franc},
		\citenamefont {Michel}, \citenamefont {Pinard},\ and\ \citenamefont
		{Flaminio}}]{GranataPRD2016}%
	\BibitemOpen
	\bibfield  {author} {\bibinfo {author} {\bibfnamefont {M.}~\bibnamefont
			{Granata}}, \bibinfo {author} {\bibfnamefont {E.}~\bibnamefont {Saracco}},
		\bibinfo {author} {\bibfnamefont {N.}~\bibnamefont {Morgado}}, \bibinfo
		{author} {\bibfnamefont {A.}~\bibnamefont {Cajgfinger}}, \bibinfo {author}
		{\bibfnamefont {G.}~\bibnamefont {Cagnoli}}, \bibinfo {author} {\bibfnamefont
			{J.}~\bibnamefont {Degallaix}}, \bibinfo {author} {\bibfnamefont
			{V.}~\bibnamefont {Dolique}}, \bibinfo {author} {\bibfnamefont
			{D.}~\bibnamefont {Forest}}, \bibinfo {author} {\bibfnamefont
			{J.}~\bibnamefont {Franc}}, \bibinfo {author} {\bibfnamefont
			{C.}~\bibnamefont {Michel}}, \bibinfo {author} {\bibfnamefont
			{L.}~\bibnamefont {Pinard}},\ and\ \bibinfo {author} {\bibfnamefont
			{R.}~\bibnamefont {Flaminio}},\ }\bibfield  {title} {\bibinfo {title}
		{Mechanical loss in state-of-the-art amorphous optical coatings},\ }\href
	{https://doi.org/10.1103/PhysRevD.93.012007} {\bibfield  {journal} {\bibinfo
			{journal} {Phys. Rev. D}\ }\textbf {\bibinfo {volume} {93}},\ \bibinfo
		{pages} {012007} (\bibinfo {year} {2016})}\BibitemShut {NoStop}%
	\bibitem [{\citenamefont {Malhaire}\ \emph {et~al.}(2023)\citenamefont
		{Malhaire}, \citenamefont {Granata}, \citenamefont {Hofman}, \citenamefont
		{Amato}, \citenamefont {Martinez}, \citenamefont {Cagnoli}, \citenamefont
		{Lemaitre},\ and\ \citenamefont {Shcheblanov}}]{MalhaireJVSTA2023}%
	\BibitemOpen
	\bibfield  {author} {\bibinfo {author} {\bibfnamefont {C.}~\bibnamefont
			{Malhaire}}, \bibinfo {author} {\bibfnamefont {M.}~\bibnamefont {Granata}},
		\bibinfo {author} {\bibfnamefont {D.}~\bibnamefont {Hofman}}, \bibinfo
		{author} {\bibfnamefont {A.}~\bibnamefont {Amato}}, \bibinfo {author}
		{\bibfnamefont {V.}~\bibnamefont {Martinez}}, \bibinfo {author}
		{\bibfnamefont {G.}~\bibnamefont {Cagnoli}}, \bibinfo {author} {\bibfnamefont
			{A.}~\bibnamefont {Lemaitre}},\ and\ \bibinfo {author} {\bibfnamefont
			{N.}~\bibnamefont {Shcheblanov}},\ }\bibfield  {title} {\bibinfo {title}
		{Determination of stress in thin films using micro-machined buckled
			membranes},\ }\href {https://doi.org/10.1116/6.0002590} {\bibfield  {journal}
		{\bibinfo  {journal} {Journal of Vacuum Science \& Technology A}\ }\textbf
		{\bibinfo {volume} {41}},\ \bibinfo {pages} {043401} (\bibinfo {year}
		{2023})}\BibitemShut {NoStop}%
	\bibitem [{\citenamefont {Hong}\ \emph {et~al.}(2013)\citenamefont {Hong},
		\citenamefont {Yang}, \citenamefont {Gustafson}, \citenamefont {Adhikari},\
		and\ \citenamefont {Chen}}]{HongPRD2013}%
	\BibitemOpen
	\bibfield  {author} {\bibinfo {author} {\bibfnamefont {T.}~\bibnamefont
			{Hong}}, \bibinfo {author} {\bibfnamefont {H.}~\bibnamefont {Yang}}, \bibinfo
		{author} {\bibfnamefont {E.~K.}\ \bibnamefont {Gustafson}}, \bibinfo {author}
		{\bibfnamefont {R.~X.}\ \bibnamefont {Adhikari}},\ and\ \bibinfo {author}
		{\bibfnamefont {Y.}~\bibnamefont {Chen}},\ }\bibfield  {title} {\bibinfo
		{title} {Brownian thermal noise in multilayer coated mirrors},\ }\href
	{https://doi.org/10.1103/PhysRevD.87.082001} {\bibfield  {journal} {\bibinfo
			{journal} {Phys. Rev. D}\ }\textbf {\bibinfo {volume} {87}},\ \bibinfo
		{pages} {082001} (\bibinfo {year} {2013})}\BibitemShut {NoStop}%
	\bibitem [{\citenamefont {Yam}\ \emph {et~al.}(2015)\citenamefont {Yam},
		\citenamefont {Gras},\ and\ \citenamefont {Evans}}]{YamPRD2015}%
	\BibitemOpen
	\bibfield  {author} {\bibinfo {author} {\bibfnamefont {W.}~\bibnamefont
			{Yam}}, \bibinfo {author} {\bibfnamefont {S.}~\bibnamefont {Gras}},\ and\
		\bibinfo {author} {\bibfnamefont {M.}~\bibnamefont {Evans}},\ }\bibfield
	{title} {\bibinfo {title} {Multimaterial coatings with reduced thermal
			noise},\ }\href {https://doi.org/10.1103/PhysRevD.91.042002} {\bibfield
		{journal} {\bibinfo  {journal} {Phys. Rev. D}\ }\textbf {\bibinfo {volume}
			{91}},\ \bibinfo {pages} {042002} (\bibinfo {year} {2015})}\BibitemShut
	{NoStop}%
	\bibitem [{\citenamefont {Tait}\ \emph {et~al.}(2020)\citenamefont {Tait},
		\citenamefont {Steinlechner}, \citenamefont {Kinley-Hanlon}, \citenamefont
		{Murray}, \citenamefont {Hough}, \citenamefont {McGhee}, \citenamefont
		{Pein}, \citenamefont {Rowan}, \citenamefont {Schnabel}, \citenamefont
		{Smith}, \citenamefont {Terkowski},\ and\ \citenamefont
		{Martin}}]{TaitPRL2020}%
	\BibitemOpen
	\bibfield  {author} {\bibinfo {author} {\bibfnamefont {S.~C.}\ \bibnamefont
			{Tait}}, \bibinfo {author} {\bibfnamefont {J.}~\bibnamefont {Steinlechner}},
		\bibinfo {author} {\bibfnamefont {M.~M.}\ \bibnamefont {Kinley-Hanlon}},
		\bibinfo {author} {\bibfnamefont {P.~G.}\ \bibnamefont {Murray}}, \bibinfo
		{author} {\bibfnamefont {J.}~\bibnamefont {Hough}}, \bibinfo {author}
		{\bibfnamefont {G.}~\bibnamefont {McGhee}}, \bibinfo {author} {\bibfnamefont
			{F.}~\bibnamefont {Pein}}, \bibinfo {author} {\bibfnamefont {S.}~\bibnamefont
			{Rowan}}, \bibinfo {author} {\bibfnamefont {R.}~\bibnamefont {Schnabel}},
		\bibinfo {author} {\bibfnamefont {C.}~\bibnamefont {Smith}}, \bibinfo
		{author} {\bibfnamefont {L.}~\bibnamefont {Terkowski}},\ and\ \bibinfo
		{author} {\bibfnamefont {I.~W.}\ \bibnamefont {Martin}},\ }\bibfield  {title}
	{\bibinfo {title} {Demonstration of the multimaterial coating concept to
			reduce thermal noise in gravitational-wave detectors},\ }\href
	{https://doi.org/10.1103/PhysRevLett.125.011102} {\bibfield  {journal}
		{\bibinfo  {journal} {Phys. Rev. Lett.}\ }\textbf {\bibinfo {volume} {125}},\
		\bibinfo {pages} {011102} (\bibinfo {year} {2020})}\BibitemShut {NoStop}%
	\bibitem [{\citenamefont {Fejer}(2021)}]{FejerLIGO2021}%
	\BibitemOpen
	\bibfield  {author} {\bibinfo {author} {\bibfnamefont {M.~M.}\ \bibnamefont
			{Fejer}},\ }\href@noop {} {\bibinfo {title} {Effective medium description of
			multilayer coatings}},\ \bibinfo {howpublished} {LIGO Document T2100186}
	(\bibinfo {year} {2021}),\ \bibinfo {note}
	{https://dcc.ligo.org/LIGO-T2100186/public}\BibitemShut {NoStop}%
	\bibitem [{\citenamefont {Lovelace}\ \emph {et~al.}(2017)\citenamefont
		{Lovelace}, \citenamefont {Demos},\ and\ \citenamefont
		{Khan}}]{LovelaceCQG2018}%
	\BibitemOpen
	\bibfield  {author} {\bibinfo {author} {\bibfnamefont {G.}~\bibnamefont
			{Lovelace}}, \bibinfo {author} {\bibfnamefont {N.}~\bibnamefont {Demos}},\
		and\ \bibinfo {author} {\bibfnamefont {H.}~\bibnamefont {Khan}},\ }\bibfield
	{title} {\bibinfo {title} {Numerically modeling brownian thermal noise in
			amorphous and crystalline thin coatings},\ }\href
	{https://doi.org/10.1088/1361-6382/aa9ccc} {\bibfield  {journal} {\bibinfo
			{journal} {Classical and Quantum Gravity}\ }\textbf {\bibinfo {volume}
			{35}},\ \bibinfo {pages} {025017} (\bibinfo {year} {2017})}\BibitemShut
	{NoStop}%
	\bibitem [{\citenamefont {Vignaud}\ and\ \citenamefont
		{Gibaud}(2019)}]{VignaudJApplCry2019}%
	\BibitemOpen
	\bibfield  {author} {\bibinfo {author} {\bibfnamefont {G.}~\bibnamefont
			{Vignaud}}\ and\ \bibinfo {author} {\bibfnamefont {A.}~\bibnamefont
			{Gibaud}},\ }\bibfield  {title} {\bibinfo {title} {{{\it REFLEX}: a program
				for the analysis of specular X-ray and neutron reflectivity data}},\ }\href
	{https://doi.org/10.1107/S1600576718018186} {\bibfield  {journal} {\bibinfo
			{journal} {Journal of Applied Crystallography}\ }\textbf {\bibinfo {volume}
			{52}},\ \bibinfo {pages} {201} (\bibinfo {year} {2019})}\BibitemShut
	{NoStop}%
	\bibitem [{\citenamefont {Woollam}\ \emph {et~al.}(1999)\citenamefont
		{Woollam}, \citenamefont {Johs}, \citenamefont {Herzinger}, \citenamefont
		{Hilfiker}, \citenamefont {Synowicki},\ and\ \citenamefont
		{Bungay}}]{WoollamSPIE1999}%
	\BibitemOpen
	\bibfield  {author} {\bibinfo {author} {\bibfnamefont {J.~A.}\ \bibnamefont
			{Woollam}}, \bibinfo {author} {\bibfnamefont {B.~D.}\ \bibnamefont {Johs}},
		\bibinfo {author} {\bibfnamefont {C.~M.}\ \bibnamefont {Herzinger}}, \bibinfo
		{author} {\bibfnamefont {J.~N.}\ \bibnamefont {Hilfiker}}, \bibinfo {author}
		{\bibfnamefont {R.~A.}\ \bibnamefont {Synowicki}},\ and\ \bibinfo {author}
		{\bibfnamefont {C.~L.}\ \bibnamefont {Bungay}},\ }\bibfield  {title}
	{\bibinfo {title} {{Overview of variable-angle spectroscopic ellipsometry
				({VASE}): I. Basic theory and typical applications}},\ }in\ \href
	{https://doi.org/10.1117/12.351660} {\emph {\bibinfo {booktitle} {Optical
				Metrology: A Critical Review}}},\ Vol.\ \bibinfo {volume} {10294},\ \bibinfo
	{editor} {edited by\ \bibinfo {editor} {\bibfnamefont {G.~A.}\ \bibnamefont
			{Al-Jumaily}}},\ \bibinfo {organization} {International Society for Optics
		and Photonics}\ (\bibinfo  {publisher} {SPIE},\ \bibinfo {year} {1999})\ p.\
	\bibinfo {pages} {1029402}\BibitemShut {NoStop}%
	\bibitem [{\citenamefont {Magnozzi}\ \emph {et~al.}(2018)\citenamefont
		{Magnozzi}, \citenamefont {Terreni}, \citenamefont {Anghinolfi},
		\citenamefont {Uttiya}, \citenamefont {Carnasciali}, \citenamefont {Gemme},
		\citenamefont {Neri}, \citenamefont {Principe}, \citenamefont {Pinto},
		\citenamefont {Kuo}, \citenamefont {Chao},\ and\ \citenamefont
		{Canepa}}]{MagnozziOptMat2018}%
	\BibitemOpen
	\bibfield  {author} {\bibinfo {author} {\bibfnamefont {M.}~\bibnamefont
			{Magnozzi}}, \bibinfo {author} {\bibfnamefont {S.}~\bibnamefont {Terreni}},
		\bibinfo {author} {\bibfnamefont {L.}~\bibnamefont {Anghinolfi}}, \bibinfo
		{author} {\bibfnamefont {S.}~\bibnamefont {Uttiya}}, \bibinfo {author}
		{\bibfnamefont {M.}~\bibnamefont {Carnasciali}}, \bibinfo {author}
		{\bibfnamefont {G.}~\bibnamefont {Gemme}}, \bibinfo {author} {\bibfnamefont
			{M.}~\bibnamefont {Neri}}, \bibinfo {author} {\bibfnamefont {M.}~\bibnamefont
			{Principe}}, \bibinfo {author} {\bibfnamefont {I.}~\bibnamefont {Pinto}},
		\bibinfo {author} {\bibfnamefont {L.-C.}\ \bibnamefont {Kuo}}, \bibinfo
		{author} {\bibfnamefont {S.}~\bibnamefont {Chao}},\ and\ \bibinfo {author}
		{\bibfnamefont {M.}~\bibnamefont {Canepa}},\ }\bibfield  {title} {\bibinfo
		{title} {Optical properties of amorphous {S}i{O}$_2$-{T}i{O}$_2$
			multi-nanolayered coatings for 1064-nm mirror technology},\ }\href
	{https://doi.org/https://doi.org/10.1016/j.optmat.2017.09.043} {\bibfield
		{journal} {\bibinfo  {journal} {Opt. Mater.}\ }\textbf {\bibinfo {volume}
			{75}},\ \bibinfo {pages} {94} (\bibinfo {year} {2018})}\BibitemShut {NoStop}%
	\bibitem [{\citenamefont {Amato}\ \emph {et~al.}(2023)\citenamefont {Amato},
		\citenamefont {Magnozzi}, \citenamefont {Shcheblanov}, \citenamefont
		{Lemaître}, \citenamefont {Cagnoli}, \citenamefont {Granata}, \citenamefont
		{Michel}, \citenamefont {Gemme}, \citenamefont {Pinard},\ and\ \citenamefont
		{Canepa}}]{AmatoACSAOM2023}%
	\BibitemOpen
	\bibfield  {author} {\bibinfo {author} {\bibfnamefont {A.}~\bibnamefont
			{Amato}}, \bibinfo {author} {\bibfnamefont {M.}~\bibnamefont {Magnozzi}},
		\bibinfo {author} {\bibfnamefont {N.}~\bibnamefont {Shcheblanov}}, \bibinfo
		{author} {\bibfnamefont {A.}~\bibnamefont {Lemaître}}, \bibinfo {author}
		{\bibfnamefont {G.}~\bibnamefont {Cagnoli}}, \bibinfo {author} {\bibfnamefont
			{M.}~\bibnamefont {Granata}}, \bibinfo {author} {\bibfnamefont
			{C.}~\bibnamefont {Michel}}, \bibinfo {author} {\bibfnamefont
			{G.}~\bibnamefont {Gemme}}, \bibinfo {author} {\bibfnamefont
			{L.}~\bibnamefont {Pinard}},\ and\ \bibinfo {author} {\bibfnamefont
			{M.}~\bibnamefont {Canepa}},\ }\bibfield  {title} {\bibinfo {title}
		{Enhancing {Titania}-{Tantala} {Amorphous} {Materials} as {High}-{Index}
			{Layers} in {Bragg} {Reflectors} of {Gravitational}-{Wave} {Detectors}},\
	}\href {https://doi.org/10.1021/acsaom.2c00077} {\bibfield  {journal}
		{\bibinfo  {journal} {ACS Applied Optical Materials}\ }\textbf {\bibinfo
			{volume} {1}},\ \bibinfo {pages} {395} (\bibinfo {year} {2023})}\BibitemShut
	{NoStop}%
	\bibitem [{\citenamefont {{Brillouin, Léon}}(1922)}]{Brillouin1922}%
	\BibitemOpen
	\bibfield  {author} {\bibinfo {author} {\bibnamefont {{Brillouin, Léon}}},\
	}\bibfield  {title} {\bibinfo {title} {Diffusion de la lumière et des rayons
			x par un corps transparent homogène - influence de l'agitation thermique},\
	}\href {https://doi.org/10.1051/anphys/192209170088} {\bibfield  {journal}
		{\bibinfo  {journal} {Ann. Phys.}\ }\textbf {\bibinfo {volume} {9}},\
		\bibinfo {pages} {88} (\bibinfo {year} {1922})}\BibitemShut {NoStop}%
	\bibitem [{\citenamefont {Merklein}\ \emph {et~al.}(2022)\citenamefont
		{Merklein}, \citenamefont {Kabakova}, \citenamefont {Zarifi},\ and\
		\citenamefont {Eggleton}}]{BLSreview1}%
	\BibitemOpen
	\bibfield  {author} {\bibinfo {author} {\bibfnamefont {M.}~\bibnamefont
			{Merklein}}, \bibinfo {author} {\bibfnamefont {I.~V.}\ \bibnamefont
			{Kabakova}}, \bibinfo {author} {\bibfnamefont {A.}~\bibnamefont {Zarifi}},\
		and\ \bibinfo {author} {\bibfnamefont {B.~J.}\ \bibnamefont {Eggleton}},\
	}\bibfield  {title} {\bibinfo {title} {100 years of {B}rillouin scattering:
			{H}istorical and future perspectives},\ }\href
	{https://doi.org/10.1063/5.0095488} {\bibfield  {journal} {\bibinfo
			{journal} {Applied Physics Reviews}\ }\textbf {\bibinfo {volume} {9}},\
		\bibinfo {pages} {041306} (\bibinfo {year} {2022})}\BibitemShut {NoStop}%
	\bibitem [{\citenamefont {Kojima}(2022)}]{BLSreview2}%
	\BibitemOpen
	\bibfield  {author} {\bibinfo {author} {\bibfnamefont {S.}~\bibnamefont
			{Kojima}},\ }\bibfield  {title} {\bibinfo {title} {100th {A}nniversary of
			{B}rillouin scattering: {I}mpact on materials science},\ }\href
	{https://doi.org/10.3390/ma15103518} {\bibfield  {journal} {\bibinfo
			{journal} {Materials}\ }\textbf {\bibinfo {volume} {15}},\ \bibinfo {pages}
		{3518} (\bibinfo {year} {2022})}\BibitemShut {NoStop}%
	\bibitem [{\citenamefont {Zhang}\ \emph {et~al.}(2001)\citenamefont {Zhang},
		\citenamefont {Sooryakumar}, \citenamefont {Every},\ and\ \citenamefont
		{Manghnani}}]{ZhangPRB2001}%
	\BibitemOpen
	\bibfield  {author} {\bibinfo {author} {\bibfnamefont {X.}~\bibnamefont
			{Zhang}}, \bibinfo {author} {\bibfnamefont {R.}~\bibnamefont {Sooryakumar}},
		\bibinfo {author} {\bibfnamefont {A.~G.}\ \bibnamefont {Every}},\ and\
		\bibinfo {author} {\bibfnamefont {M.~H.}\ \bibnamefont {Manghnani}},\
	}\bibfield  {title} {\bibinfo {title} {Observation of organ-pipe acoustic
			excitations in supported thin films},\ }\href
	{https://doi.org/10.1103/PhysRevB.64.081402} {\bibfield  {journal} {\bibinfo
			{journal} {Phys. Rev. B}\ }\textbf {\bibinfo {volume} {64}},\ \bibinfo
		{pages} {081402} (\bibinfo {year} {2001})}\BibitemShut {NoStop}%
	\bibitem [{\citenamefont {Zhang}\ \emph {et~al.}(2003)\citenamefont {Zhang},
		\citenamefont {Sooryakumar},\ and\ \citenamefont {Bussmann}}]{ZhangPRB2003}%
	\BibitemOpen
	\bibfield  {author} {\bibinfo {author} {\bibfnamefont {X.}~\bibnamefont
			{Zhang}}, \bibinfo {author} {\bibfnamefont {R.}~\bibnamefont {Sooryakumar}},\
		and\ \bibinfo {author} {\bibfnamefont {K.}~\bibnamefont {Bussmann}},\
	}\bibfield  {title} {\bibinfo {title} {Confinement and transverse standing
			acoustic resonances in free-standing membranes},\ }\href
	{https://doi.org/10.1103/PhysRevB.68.115430} {\bibfield  {journal} {\bibinfo
			{journal} {Phys. Rev. B}\ }\textbf {\bibinfo {volume} {68}},\ \bibinfo
		{pages} {115430} (\bibinfo {year} {2003})}\BibitemShut {NoStop}%
	\bibitem [{\citenamefont {Sandercock}(1972)}]{SandercockPRL1972}%
	\BibitemOpen
	\bibfield  {author} {\bibinfo {author} {\bibfnamefont {J.~R.}\ \bibnamefont
			{Sandercock}},\ }\bibfield  {title} {\bibinfo {title} {{S}tructure in the
			{B}rillouin spectra of thin films},\ }\href
	{https://doi.org/10.1103/PhysRevLett.29.1735} {\bibfield  {journal} {\bibinfo
			{journal} {Phys. Rev. Lett.}\ }\textbf {\bibinfo {volume} {29}},\ \bibinfo
		{pages} {1735} (\bibinfo {year} {1972})}\BibitemShut {NoStop}%
	\bibitem [{\citenamefont {Sandercock}(1982)}]{Sandercock1982BOOK}%
	\BibitemOpen
	\bibfield  {author} {\bibinfo {author} {\bibfnamefont {J.~R.}\ \bibnamefont
			{Sandercock}},\ }\href {https://doi.org/10.1007/3540115137$_6$} {\emph
		{\bibinfo {title} {Light Scattering in Solids III: Recent Results}}},\ edited
	by\ \bibinfo {editor} {\bibfnamefont {M.}~\bibnamefont {Cardona}}\ and\
	\bibinfo {editor} {\bibfnamefont {G.}~\bibnamefont {G{\"u}ntherodt}}\
	(\bibinfo  {publisher} {Springer Berlin Heidelberg},\ \bibinfo {address}
	{Berlin, Heidelberg},\ \bibinfo {year} {1982})\ pp.\ \bibinfo {pages}
	{173--206}\BibitemShut {NoStop}%
	\bibitem [{\citenamefont {Gomopoulos}\ \emph {et~al.}(2009)\citenamefont
		{Gomopoulos}, \citenamefont {Cheng}, \citenamefont {Efremov}, \citenamefont
		{Nealey},\ and\ \citenamefont {Fytas}}]{GomopoulosMACRO2009}%
	\BibitemOpen
	\bibfield  {author} {\bibinfo {author} {\bibfnamefont {N.}~\bibnamefont
			{Gomopoulos}}, \bibinfo {author} {\bibfnamefont {W.}~\bibnamefont {Cheng}},
		\bibinfo {author} {\bibfnamefont {M.}~\bibnamefont {Efremov}}, \bibinfo
		{author} {\bibfnamefont {P.~F.}\ \bibnamefont {Nealey}},\ and\ \bibinfo
		{author} {\bibfnamefont {G.}~\bibnamefont {Fytas}},\ }\bibfield  {title}
	{\bibinfo {title} {{Out-of-Plane Longitudinal Elastic Modulus of Supported
				Polymer Thin Films}},\ }\href {https://doi.org/10.1021/ma901246y} {\bibfield
		{journal} {\bibinfo  {journal} {Macromolecules}\ }\textbf {\bibinfo {volume}
			{42}},\ \bibinfo {pages} {7164} (\bibinfo {year} {2009})}\BibitemShut
	{NoStop}%
	\bibitem [{\citenamefont {El~Abouti}\ \emph {et~al.}(2022)\citenamefont
		{El~Abouti}, \citenamefont {Cuffe}, \citenamefont {El~Boudouti},
		\citenamefont {Sotomayor~Torres}, \citenamefont {Chavez-Angel}, \citenamefont
		{Djafari-Rouhani},\ and\ \citenamefont {Alzina}}]{ElAboutiCRY2022}%
	\BibitemOpen
	\bibfield  {author} {\bibinfo {author} {\bibfnamefont {O.}~\bibnamefont
			{El~Abouti}}, \bibinfo {author} {\bibfnamefont {J.}~\bibnamefont {Cuffe}},
		\bibinfo {author} {\bibfnamefont {E.~H.}\ \bibnamefont {El~Boudouti}},
		\bibinfo {author} {\bibfnamefont {C.~M.}\ \bibnamefont {Sotomayor~Torres}},
		\bibinfo {author} {\bibfnamefont {E.}~\bibnamefont {Chavez-Angel}}, \bibinfo
		{author} {\bibfnamefont {B.}~\bibnamefont {Djafari-Rouhani}},\ and\ \bibinfo
		{author} {\bibfnamefont {F.}~\bibnamefont {Alzina}},\ }\bibfield  {title}
	{\bibinfo {title} {Comparison of brillouin light scattering and density of
			states in a supported layer: Analytical and experimental study},\ }\bibfield
	{journal} {\bibinfo  {journal} {Crystals}\ }\textbf {\bibinfo {volume}
		{12}},\ \href {https://doi.org/10.3390/cryst12091212} {10.3390/cryst12091212}
	(\bibinfo {year} {2022})\BibitemShut {NoStop}%
	\bibitem [{\citenamefont {Passeri}\ \emph {et~al.}(2023)\citenamefont
		{Passeri}, \citenamefont {{Di Michele}}, \citenamefont {Neri}, \citenamefont
		{Cottone}, \citenamefont {Fioretto}, \citenamefont {Mattarelli},\ and\
		\citenamefont {Caponi}}]{PasseriBioAdv2023}%
	\BibitemOpen
	\bibfield  {author} {\bibinfo {author} {\bibfnamefont {A.}~\bibnamefont
			{Passeri}}, \bibinfo {author} {\bibfnamefont {A.}~\bibnamefont {{Di
					Michele}}}, \bibinfo {author} {\bibfnamefont {I.}~\bibnamefont {Neri}},
		\bibinfo {author} {\bibfnamefont {F.}~\bibnamefont {Cottone}}, \bibinfo
		{author} {\bibfnamefont {D.}~\bibnamefont {Fioretto}}, \bibinfo {author}
		{\bibfnamefont {M.}~\bibnamefont {Mattarelli}},\ and\ \bibinfo {author}
		{\bibfnamefont {S.}~\bibnamefont {Caponi}},\ }\bibfield  {title} {\bibinfo
		{title} {Size and environment: The effect of phonon localization on
			micro-brillouin imaging},\ }\href
	{https://doi.org/10.1016/j.bioadv.2023.213341} {\bibfield  {journal}
		{\bibinfo  {journal} {Biomaterials Advances}\ }\textbf {\bibinfo {volume}
			{147}},\ \bibinfo {pages} {213341} (\bibinfo {year} {2023})}\BibitemShut
	{NoStop}%
	\bibitem [{\citenamefont {Carlotti}(2018)}]{CarlottiApplSci2018}%
	\BibitemOpen
	\bibfield  {author} {\bibinfo {author} {\bibfnamefont {G.}~\bibnamefont
			{Carlotti}},\ }\bibfield  {title} {\bibinfo {title} {Elastic characterization
			of transparent and opaque films, multilayers and acoustic resonators by
			surface {B}rillouin scattering: {A} review},\ }\href
	{https://doi.org/10.3390/app8010124} {\bibfield  {journal} {\bibinfo
			{journal} {Applied Sciences}\ }\textbf {\bibinfo {volume} {8}},\ \bibinfo
		{pages} {124} (\bibinfo {year} {2018})}\BibitemShut {NoStop}%
	\bibitem [{\citenamefont {Hosono}(1991)}]{HosonoJAP1991}%
	\BibitemOpen
	\bibfield  {author} {\bibinfo {author} {\bibfnamefont {H.}~\bibnamefont
			{Hosono}},\ }\bibfield  {title} {\bibinfo {title} {Fourier transform infrared
			attenuated total reflection spectra of ion-implanted silica glasses},\ }\href
	{https://doi.org/10.1063/1.348925} {\bibfield  {journal} {\bibinfo  {journal}
			{Journal of Applied Physics}\ }\textbf {\bibinfo {volume} {69}},\ \bibinfo
		{pages} {8079} (\bibinfo {year} {1991})}\BibitemShut {NoStop}%
	\bibitem [{\citenamefont {{de los Arcos}}\ \emph {et~al.}(2021)\citenamefont
		{{de los Arcos}}, \citenamefont {Müller}, \citenamefont {Wang},
		\citenamefont {Damerla}, \citenamefont {Hoppe}, \citenamefont {Weinberger},
		\citenamefont {Tiemann},\ and\ \citenamefont
		{Grundmeier}}]{DeLosArcosVibSpect2021}%
	\BibitemOpen
	\bibfield  {author} {\bibinfo {author} {\bibfnamefont {T.}~\bibnamefont {{de
					los Arcos}}}, \bibinfo {author} {\bibfnamefont {H.}~\bibnamefont {Müller}},
		\bibinfo {author} {\bibfnamefont {F.}~\bibnamefont {Wang}}, \bibinfo {author}
		{\bibfnamefont {V.~R.}\ \bibnamefont {Damerla}}, \bibinfo {author}
		{\bibfnamefont {C.}~\bibnamefont {Hoppe}}, \bibinfo {author} {\bibfnamefont
			{C.}~\bibnamefont {Weinberger}}, \bibinfo {author} {\bibfnamefont
			{M.}~\bibnamefont {Tiemann}},\ and\ \bibinfo {author} {\bibfnamefont
			{G.}~\bibnamefont {Grundmeier}},\ }\bibfield  {title} {\bibinfo {title}
		{Review of infrared spectroscopy techniques for the determination of internal
			structure in thin {S}i{O}$_2$ films},\ }\href
	{https://doi.org/10.1016/j.vibspec.2021.103256} {\bibfield  {journal}
		{\bibinfo  {journal} {Vibrational Spectroscopy}\ }\textbf {\bibinfo {volume}
			{114}},\ \bibinfo {pages} {103256} (\bibinfo {year} {2021})}\BibitemShut
	{NoStop}%
	\bibitem [{\citenamefont {Efthimiopoulos}\ \emph {et~al.}(2018)\citenamefont
		{Efthimiopoulos}, \citenamefont {Palles}, \citenamefont {Richter},
		\citenamefont {Hoppe}, \citenamefont {Möncke}, \citenamefont {Wondraczek},
		\citenamefont {Nolte},\ and\ \citenamefont
		{Kamitsos}}]{EfthimiopoulosJAppPhys2018}%
	\BibitemOpen
	\bibfield  {author} {\bibinfo {author} {\bibfnamefont {I.}~\bibnamefont
			{Efthimiopoulos}}, \bibinfo {author} {\bibfnamefont {D.}~\bibnamefont
			{Palles}}, \bibinfo {author} {\bibfnamefont {S.}~\bibnamefont {Richter}},
		\bibinfo {author} {\bibfnamefont {U.}~\bibnamefont {Hoppe}}, \bibinfo
		{author} {\bibfnamefont {D.}~\bibnamefont {Möncke}}, \bibinfo {author}
		{\bibfnamefont {L.}~\bibnamefont {Wondraczek}}, \bibinfo {author}
		{\bibfnamefont {S.}~\bibnamefont {Nolte}},\ and\ \bibinfo {author}
		{\bibfnamefont {E.~I.}\ \bibnamefont {Kamitsos}},\ }\bibfield  {title}
	{\bibinfo {title} {Femtosecond laser-induced transformations in ultra-low
			expansion glass: Microstructure and local density variations by vibrational
			spectroscopy},\ }\href {https://doi.org/10.1063/1.5030687} {\bibfield
		{journal} {\bibinfo  {journal} {Journal of Applied Physics}\ }\textbf
		{\bibinfo {volume} {123}},\ \bibinfo {pages} {233105} (\bibinfo {year}
		{2018})}\BibitemShut {NoStop}%
	\bibitem [{\citenamefont {Amma}\ \emph {et~al.}(2015)\citenamefont {Amma},
		\citenamefont {Luo}, \citenamefont {Pantano},\ and\ \citenamefont
		{Kim}}]{AmmaJNCS2015}%
	\BibitemOpen
	\bibfield  {author} {\bibinfo {author} {\bibfnamefont {S.}~\bibnamefont
			{Amma}}, \bibinfo {author} {\bibfnamefont {J.}~\bibnamefont {Luo}}, \bibinfo
		{author} {\bibfnamefont {C.~G.}\ \bibnamefont {Pantano}},\ and\ \bibinfo
		{author} {\bibfnamefont {S.~H.}\ \bibnamefont {Kim}},\ }\bibfield  {title}
	{\bibinfo {title} {Specular reflectance ({SR}) and attenuated total
			reflectance ({ATR}) infrared ({IR}) spectroscopy of transparent flat glass
			surfaces: {A} case study for soda lime float glass},\ }\href
	{https://doi.org/10.1016/j.jnoncrysol.2015.08.015} {\bibfield  {journal}
		{\bibinfo  {journal} {Journal of Non-Crystalline Solids}\ }\textbf {\bibinfo
			{volume} {428}},\ \bibinfo {pages} {189} (\bibinfo {year}
		{2015})}\BibitemShut {NoStop}%
	\bibitem [{\citenamefont {Tan}(2005)}]{TanOPE2005}%
	\BibitemOpen
	\bibfield  {author} {\bibinfo {author} {\bibfnamefont {C.~Z.}\ \bibnamefont
			{Tan}},\ }\bibfield  {title} {\bibinfo {title} {Determination of surface
			structure and the depth profile of silica glass by infrared spectroscopy},\
	}\href {https://doi.org/1004-924X(2005)04-0413-08} {\bibfield  {journal}
		{\bibinfo  {journal} {Optics and Precision Engineering}\ }\textbf {\bibinfo
			{volume} {13}},\ \bibinfo {pages} {413} (\bibinfo {year} {2005})}\BibitemShut
	{NoStop}%
	\bibitem [{\citenamefont {Vacher}\ \emph {et~al.}(1997)\citenamefont {Vacher},
		\citenamefont {Pelous},\ and\ \citenamefont {Courtens}}]{VacherPRB1997}%
	\BibitemOpen
	\bibfield  {author} {\bibinfo {author} {\bibfnamefont {R.}~\bibnamefont
			{Vacher}}, \bibinfo {author} {\bibfnamefont {J.}~\bibnamefont {Pelous}},\
		and\ \bibinfo {author} {\bibfnamefont {E.}~\bibnamefont {Courtens}},\
	}\bibfield  {title} {\bibinfo {title} {Mean free path of high-frequency
			acoustic excitations in glasses with application to vitreous silica},\ }\href
	{https://doi.org/10.1103/PhysRevB.56.R481} {\bibfield  {journal} {\bibinfo
			{journal} {Phys. Rev. B}\ }\textbf {\bibinfo {volume} {56}},\ \bibinfo
		{pages} {R481} (\bibinfo {year} {1997})}\BibitemShut {NoStop}%
	\bibitem [{\citenamefont {Baldi}\ \emph {et~al.}(2011)\citenamefont {Baldi},
		\citenamefont {Giordano}, \citenamefont {Monaco},\ and\ \citenamefont
		{Ruta}}]{BaldiJNCS2011}%
	\BibitemOpen
	\bibfield  {author} {\bibinfo {author} {\bibfnamefont {G.}~\bibnamefont
			{Baldi}}, \bibinfo {author} {\bibfnamefont {V.}~\bibnamefont {Giordano}},
		\bibinfo {author} {\bibfnamefont {G.}~\bibnamefont {Monaco}},\ and\ \bibinfo
		{author} {\bibfnamefont {B.}~\bibnamefont {Ruta}},\ }\bibfield  {title}
	{\bibinfo {title} {High frequency acoustic attenuation of vitreous silica:
			{N}ew insight from inelastic x-ray scattering},\ }\href
	{https://doi.org/10.1016/j.jnoncrysol.2010.05.085} {\bibfield  {journal}
		{\bibinfo  {journal} {Journal of Non-Crystalline Solids}\ }\textbf {\bibinfo
			{volume} {357}},\ \bibinfo {pages} {538} (\bibinfo {year}
		{2011})}\BibitemShut {NoStop}%
	\bibitem [{\citenamefont {Paolone}\ \emph {et~al.}(2022)\citenamefont
		{Paolone}, \citenamefont {Placidi}, \citenamefont {Stellino}, \citenamefont
		{Betti}, \citenamefont {Majorana}, \citenamefont {Mariani}, \citenamefont
		{Nucara}, \citenamefont {Palumbo}, \citenamefont {Postorino}, \citenamefont
		{Sbroscia}, \citenamefont {Trequattrini}, \citenamefont {Granata},
		\citenamefont {Hofman}, \citenamefont {Michel}, \citenamefont {Pinard},
		\citenamefont {Lemaitre}, \citenamefont {Shcheblanov}, \citenamefont
		{Cagnoli},\ and\ \citenamefont {Ricci}}]{PaoloneCOAT2022}%
	\BibitemOpen
	\bibfield  {author} {\bibinfo {author} {\bibfnamefont {A.}~\bibnamefont
			{Paolone}}, \bibinfo {author} {\bibfnamefont {E.}~\bibnamefont {Placidi}},
		\bibinfo {author} {\bibfnamefont {E.}~\bibnamefont {Stellino}}, \bibinfo
		{author} {\bibfnamefont {M.~G.}\ \bibnamefont {Betti}}, \bibinfo {author}
		{\bibfnamefont {E.}~\bibnamefont {Majorana}}, \bibinfo {author}
		{\bibfnamefont {C.}~\bibnamefont {Mariani}}, \bibinfo {author} {\bibfnamefont
			{A.}~\bibnamefont {Nucara}}, \bibinfo {author} {\bibfnamefont
			{O.}~\bibnamefont {Palumbo}}, \bibinfo {author} {\bibfnamefont
			{P.}~\bibnamefont {Postorino}}, \bibinfo {author} {\bibfnamefont
			{M.}~\bibnamefont {Sbroscia}}, \bibinfo {author} {\bibfnamefont
			{F.}~\bibnamefont {Trequattrini}}, \bibinfo {author} {\bibfnamefont
			{M.}~\bibnamefont {Granata}}, \bibinfo {author} {\bibfnamefont
			{D.}~\bibnamefont {Hofman}}, \bibinfo {author} {\bibfnamefont
			{C.}~\bibnamefont {Michel}}, \bibinfo {author} {\bibfnamefont
			{L.}~\bibnamefont {Pinard}}, \bibinfo {author} {\bibfnamefont
			{A.}~\bibnamefont {Lemaitre}}, \bibinfo {author} {\bibfnamefont
			{N.}~\bibnamefont {Shcheblanov}}, \bibinfo {author} {\bibfnamefont
			{G.}~\bibnamefont {Cagnoli}},\ and\ \bibinfo {author} {\bibfnamefont
			{F.}~\bibnamefont {Ricci}},\ }\bibfield  {title} {\bibinfo {title} {Argon and
			other defects in amorphous {S}i{O}$_2$ coatings for gravitational-wave
			detectors},\ }\href {https://doi.org/10.3390/coatings12071001} {\bibfield
		{journal} {\bibinfo  {journal} {Coatings}\ }\textbf {\bibinfo {volume}
			{12}},\ \bibinfo {pages} {1001} (\bibinfo {year} {2022})}\BibitemShut
	{NoStop}%
	\bibitem [{\citenamefont {Hirose}\ \emph {et~al.}(2006)\citenamefont {Hirose},
		\citenamefont {Saito},\ and\ \citenamefont {Ikushima}}]{HiroseJNCS2006}%
	\BibitemOpen
	\bibfield  {author} {\bibinfo {author} {\bibfnamefont {T.}~\bibnamefont
			{Hirose}}, \bibinfo {author} {\bibfnamefont {K.}~\bibnamefont {Saito}},\ and\
		\bibinfo {author} {\bibfnamefont {A.~J.}\ \bibnamefont {Ikushima}},\
	}\bibfield  {title} {\bibinfo {title} {Structural relaxation in
			sputter-deposited silica glass},\ }\href
	{https://doi.org/10.1016/j.jnoncrysol.2006.02.056} {\bibfield  {journal}
		{\bibinfo  {journal} {Journal of Non-Crystalline Solids}\ }\textbf {\bibinfo
			{volume} {352}},\ \bibinfo {pages} {2198} (\bibinfo {year}
		{2006})}\BibitemShut {NoStop}%
	\bibitem [{\citenamefont {Auld}(1973)}]{Auld1973}%
	\BibitemOpen
	\bibfield  {author} {\bibinfo {author} {\bibfnamefont {B.~A.}\ \bibnamefont
			{Auld}},\ }\href@noop {} {\emph {\bibinfo {title} {Acoustic fields and waves
				in solids}}},\ edited by\ \bibinfo {editor} {\bibnamefont {{John Wiley \&
				Sons}}}\ (\bibinfo  {publisher} {Elsevier},\ \bibinfo {year}
	{1973})\BibitemShut {NoStop}%
	\bibitem [{\citenamefont {{Heraeus Quarzglas GmbH \& Co. KG}}()}]{Heraeus}%
	\BibitemOpen
	\bibfield  {author} {\bibinfo {author} {\bibnamefont {{Heraeus Quarzglas GmbH
					\& Co. KG}}},\ }\href@noop {} {\bibinfo {title} {{F}used {Q}uartz \& {F}used
			{S}ilica for {O}ptical {A}pplications}},\ \bibinfo {howpublished}
	{\url{https://www.heraeus-covantics.com}},\ \bibinfo {note} {accessed March
		2026}\BibitemShut {NoStop}%
	\bibitem [{\citenamefont {{Crystran Ltd.}}()}]{Crystran}%
	\BibitemOpen
	\bibfield  {author} {\bibinfo {author} {\bibnamefont {{Crystran Ltd.}}},\
	}\href@noop {} {\bibinfo {title} {{O}ptical {M}aterials, {F}used {S}ilica
			{G}lass}},\ \bibinfo {howpublished}
	{\url{https://www.crystran.com/optical-materials/fused-silica-sio2/}},\
	\bibinfo {note} {accessed March 2026}\BibitemShut {NoStop}%
	\bibitem [{\citenamefont {{Corning Inc.}}()}]{Corning}%
	\BibitemOpen
	\bibfield  {author} {\bibinfo {author} {\bibnamefont {{Corning Inc.}}},\
	}\href@noop {} {\bibinfo {title} {{H}{P}{F}{S}\textregistered {F}used
			{S}ilica {I}ndustrial {G}rade: Technical data sheet}},\ \bibinfo
	{howpublished} {\url{https://www.corning.com}},\ \bibinfo {note} {accessed
		March 2026}\BibitemShut {NoStop}%
	\bibitem [{\citenamefont {Gilroy}\ and\ \citenamefont
		{Phillips}(1981)}]{GilroyPM1981}%
	\BibitemOpen
	\bibfield  {author} {\bibinfo {author} {\bibfnamefont {K.~S.}\ \bibnamefont
			{Gilroy}}\ and\ \bibinfo {author} {\bibfnamefont {W.~A.}\ \bibnamefont
			{Phillips}},\ }\bibfield  {title} {\bibinfo {title} {An asymmetric
			double-well potential model for structural relaxation processes in amorphous
			materials},\ }\href {https://doi.org/10.1080/01418638108222343} {\bibfield
		{journal} {\bibinfo  {journal} {Philosophical Magazine B}\ }\textbf {\bibinfo
			{volume} {43}},\ \bibinfo {pages} {735} (\bibinfo {year} {1981})}\BibitemShut
	{NoStop}%
	\bibitem [{\citenamefont {Cesarini}\ \emph {et~al.}(2009)\citenamefont
		{Cesarini}, \citenamefont {Lorenzini}, \citenamefont {Campagna},
		\citenamefont {Martelli}, \citenamefont {Piergiovanni}, \citenamefont
		{Vetrano}, \citenamefont {Losurdo},\ and\ \citenamefont
		{Cagnoli}}]{CesariniRSI2009}%
	\BibitemOpen
	\bibfield  {author} {\bibinfo {author} {\bibfnamefont {E.}~\bibnamefont
			{Cesarini}}, \bibinfo {author} {\bibfnamefont {M.}~\bibnamefont {Lorenzini}},
		\bibinfo {author} {\bibfnamefont {E.}~\bibnamefont {Campagna}}, \bibinfo
		{author} {\bibfnamefont {F.}~\bibnamefont {Martelli}}, \bibinfo {author}
		{\bibfnamefont {F.}~\bibnamefont {Piergiovanni}}, \bibinfo {author}
		{\bibfnamefont {F.}~\bibnamefont {Vetrano}}, \bibinfo {author} {\bibfnamefont
			{G.}~\bibnamefont {Losurdo}},\ and\ \bibinfo {author} {\bibfnamefont
			{G.}~\bibnamefont {Cagnoli}},\ }\bibfield  {title} {\bibinfo {title} {{A
				'gentle' nodal suspension for measurements of the acoustic attenuation in
				materials}},\ }\href {https://doi.org/10.1063/1.3124800} {\bibfield
		{journal} {\bibinfo  {journal} {Review of Scientific Instruments}\ }\textbf
		{\bibinfo {volume} {80}},\ \bibinfo {pages} {053904} (\bibinfo {year}
		{2009})}\BibitemShut {NoStop}%
	\bibitem [{\citenamefont {Wang}\ \emph {et~al.}(2003)\citenamefont {Wang},
		\citenamefont {Liu},\ and\ \citenamefont {Tan}}]{WangJCP2003}%
	\BibitemOpen
	\bibfield  {author} {\bibinfo {author} {\bibfnamefont {T.~B.}\ \bibnamefont
			{Wang}}, \bibinfo {author} {\bibfnamefont {Z.~G.}\ \bibnamefont {Liu}},\ and\
		\bibinfo {author} {\bibfnamefont {C.~Z.}\ \bibnamefont {Tan}},\ }\bibfield
	{title} {\bibinfo {title} {Relationship between the frequency of the main
			{LO} mode of silica glass and angle of incidence},\ }\href
	{https://doi.org/10.1063/1.1577535} {\bibfield  {journal} {\bibinfo
			{journal} {The Journal of Chemical Physics}\ }\textbf {\bibinfo {volume}
			{119}},\ \bibinfo {pages} {505} (\bibinfo {year} {2003})}\BibitemShut
	{NoStop}%
	\bibitem [{\citenamefont {Huibers}\ and\ \citenamefont
		{Shah}(1997)}]{HuibersLANG1997}%
	\BibitemOpen
	\bibfield  {author} {\bibinfo {author} {\bibfnamefont {P.~D.~T.}\
			\bibnamefont {Huibers}}\ and\ \bibinfo {author} {\bibfnamefont {D.~O.}\
			\bibnamefont {Shah}},\ }\bibfield  {title} {\bibinfo {title} {{Multispectral
				Determination of Soap Film Thickness}},\ }\href
	{https://doi.org/10.1021/la960738n} {\bibfield  {journal} {\bibinfo
			{journal} {Langmuir}\ }\textbf {\bibinfo {volume} {13}},\ \bibinfo {pages}
		{5995} (\bibinfo {year} {1997})}\BibitemShut {NoStop}%
	\bibitem [{ET()}]{ET}%
	\BibitemOpen
	\href@noop {} {\bibinfo {title} {{Einstein Telescope Italy}}},\ \bibinfo
	{howpublished} {\url{https://www.einstein-telescope.it/en/home-en/}},\
	\bibinfo {note} {accessed March 2026}\BibitemShut {NoStop}%
	\bibitem [{\citenamefont {Dobrzynski}\ and\ \citenamefont
		{Maradudin}(1976)}]{MaradudinPRB1976}%
	\BibitemOpen
	\bibfield  {author} {\bibinfo {author} {\bibfnamefont {L.}~\bibnamefont
			{Dobrzynski}}\ and\ \bibinfo {author} {\bibfnamefont {A.~A.}\ \bibnamefont
			{Maradudin}},\ }\bibfield  {title} {\bibinfo {title} {Surface contribution to
			the low-temperature specific heat of a hexagonal crystal},\ }\href
	{https://doi.org/10.1103/PhysRevB.14.2200} {\bibfield  {journal} {\bibinfo
			{journal} {Phys. Rev. B}\ }\textbf {\bibinfo {volume} {14}},\ \bibinfo
		{pages} {2200} (\bibinfo {year} {1976})}\BibitemShut {NoStop}%
\end{thebibliography}

%

\end{document}